\documentclass[fleqn,usenatbib]{mnras}

\usepackage{newtxtext,newtxmath}

\usepackage[T1]{fontenc}

\DeclareRobustCommand{\VAN}[3]{#2}
\let\VANthebibliography\thebibliography
\def\thebibliography{\DeclareRobustCommand{\VAN}[3]{##3}\VANthebibliography}

\usepackage{amsmath}	
\usepackage{subcaption}
\usepackage[utf8]{inputenc}
\usepackage{multirow}
\usepackage{booktabs}
\usepackage{hyperref}
\usepackage{url}
\usepackage{datatool}
\usepackage{lineno}
\usepackage{ulem}
\usepackage{graphicx}	

\usepackage{xcolor}

\definecolor{forestgreen}{rgb}{0.1,0.49,0.07}
\definecolor{applegreen}{rgb}{0.55,0.71,0.0}
\definecolor{cadetblue}{rgb}{0.37, 0.62, 0.63}

\DTLsetseparator{ = }

\usepackage{graphicx}
\graphicspath{{./figures/}}

\title[EPTA DR2 single-pulsar model selection]{Noise analysis in the European Pulsar Timing Array data release 2 and its implications on the gravitational-wave background search}

\author[A. Chalumeau et al.]{\parbox{\textwidth}{
	A.~Chalumeau$^{1,2,3}$\thanks{E-mail:aurelien.chalumeau@cnrs-orleans.fr},
	S.~Babak$^{1,4}$\thanks{E-mail:stas@apc.in2p3.fr},
	A.~Petiteau$^{5,1}$,
	S.~Chen$^{2,3}$,
	A.~Samajdar$^{6}$,
	R.~N.~Caballero$^{7}$,
	G.~Theureau$^{2,3,8}$,
	L.~Guillemot$^{2,3}$,
	G.~Desvignes$^{9,10}$,
	A.~Parthasarathy$^{9}$,
	K.~Liu$^{9}$,
	G.~Shaifullah$^{6,11}$, 
	H.~Hu$^{9}$,
	E.~van~der~Wateren$^{12,13}$,
	J.~Antoniadis$^{14,9,15}$,
	A.-S.~Bak~Nielsen$^{9,16}$,
	C.~G.~Bassa$^{12}$,
	A.~Berthereau$^{2,3}$,
	M.~Burgay$^{17}$,
	D.~J.~Champion$^{9}$,
	I.~Cognard$^{2,3}$,
	M.~Falxa$^{1}$,
	R.~D.~Ferdman$^{18}$,
	P.~C.~C.~Freire$^{9}$,
	J.~R.~Gair$^{19}$,
	E.~Graikou$^{9}$,
	Y.~J.~Guo$^{9}$,
	J.~Jang$^{9}$,
	G.~H.~Janssen$^{12,13}$,
	R.~Karuppusamy$^{9}$,
	M.~J.~Keith$^{20}$,
	M.~Kramer$^{9,20}$,
	K.~J.~Lee$^{7,21,9}$,
	X.~J.~Liu$^{22,20}$,
	A.~G.~Lyne$^{20}$,
	R.~A.~Main$^{9}$,
	J.~W.~McKee$^{23}$,
	M.~B.~Mickaliger$^{20}$,
	B.~B.~P.~Perera$^{24}$,
	D.~Perrodin$^{17}$,
	N.~K.~Porayko$^{9}$,
	A.~Possenti$^{17}$,
	S.~A.~Sanidas$^{20}$,
	A.~Sesana$^{6,11}$,
	L.~Speri$^{25}$,
	B.~W.~Stappers$^{20}$,
	C.~Tiburzi$^{12}$,
	A.~Vecchio$^{26}$,
	J.~P.~W.~Verbiest$^{16,9}$,
	J.~Wang$^{16}$,
	L.~Wang$^{21}$
	and H.~Xu$^{6,27,21}$}
\vspace{0.4cm} \\
Affiliations are at the end of the paper
}

\date{Accepted XXX. Received YYY; in original form ZZZ}

\pubyear{2021}

\begin{document}
\newcommand{\getval}[1]{\DTLfetch{data}{thekey}{#1}{thevalue}}
\newcommand{\geterr}[1]{\DTLfetch{data}{thekey}{#1}{errs}}

\label{firstpage}
\pagerange{\pageref{firstpage}--\pageref{lastpage}}
\maketitle

\begin{abstract}
	The European Pulsar Timing Array (EPTA) collaboration has recently released an extended data set for six pulsars (DR2) and reported evidence for a common  red noise signal.  Here we present a  
	noise analysis for each of the six pulsars. We consider several types of noise: (i)  radio frequency independent, ``achromatic'', and time-correlated red  noise; (ii) variations of dispersion measure and scattering; (iii) system and band noise; and (iv)  deterministic signals (other than gravitational waves) that could be present in the PTA data. We perform Bayesian model selection to find the optimal combination of noise components for each pulsar.
	Using these custom models we revisit the presence of the common uncorrelated red noise signal previously reported in the EPTA DR2 and show that the data still supports it with a high statistical significance. Next, we  confirm that there is no preference for or against the Hellings-Downs spatial correlations expected for the stochastic gravitational-wave background. The main conclusion of the EPTA DR2 paper remains unchanged despite a very significant change in the noise model of each pulsar. However, modelling the noise is essential  for the robust detection of gravitational waves and its impact could be significant when analysing the next EPTA data release, which will include a larger number of pulsars and more precise measurements.
\end{abstract}




\section{Introduction} \label{sec:Intro}

Millisecond pulsars are remarkable for their long-term rotational stability, comparable with 
that of the most accurate atomic clocks on Earth \citep{mat97,ver09,hob20}. The highly beamed radio emission interacts with GWs  causing deviations in the time of arrival of pulses observed by radio-telescopes. Pulsar Timing Arrays (PTAs) are used to search for ultra-low-frequency (nHz-$\mu$Hz) gravitational-waves by looking for their characteristic imprints on the times-of-arrival (ToAs) of the radio signals \citep{saz78,det79,fos90,per18}. \par

The most promising GW source in the PTA frequency band is a population of  nearby  ($z \leq 2$)  slowly
inspiralling supermassive ($\ge 10^7 \ \mathrm{M}_{\odot}$) black hole binaries (SMBHBs) with orbital periods $\sim 0.1-10$ years. Those binaries were formed as a result of galaxy collisions and their mergers will be observed in the LISA band, while PTA will see only the early and long lasting inspiral \citep{ses04}.  The most massive and closest binaries might be resolved as individual sources emitting continuous GWs \citep{ses09,bab15}, but the bulk of the SMBHB population emits rather weak GW signals which superpose and form a stochastic GW background (GWB) signal at low (nano-Hz) frequencies \citep{raj95,jaf03,wyi03}. The key feature of this noise-like signal is a very specific spatial correlation in the data which depends only on the angular separation between two pulsars and is described by the Hellings-Downs curve \citep{hel83} assuming General Relativity. \par

Several PTA collaborations are working actively on the detection and characterisation of such GWs.  
The three historical PTAs: Parkes Pulsar Timing Array \citep[PPTA;][]{ker20}, the North American Nanohertz Observatory for Gravitational-waves \citep[NANOGrav;][]{ala21} and the European Pulsar Timing Array \citep[EPTA;][]{dev16} cooperate with the emerging Indian Pulsar Timing Array (InPTA)\footnote{\url{https://inpta.gitlab.io/profile/index.html}} as the International Pulsar Timing Array \citep[IPTA;][]{ver16,per19}. 
The PPTA, NANOGrav and the EPTA have recently published results produced using with their own independently processed datasets that show consistent evidence for the presence of a  red (time-correlated) signal common among pulsars  \citep{gon21b,arz20,DR2_Chen+} but without the necessary evidence for Hellings-Downs correlations, which would confirm the detection of a GWB.
Encouraged by these promising results, the IPTA is planning to combine extended datasets from each PTA collaboration, in the hope that the GW nature of the observed signal can be confirmed (or disproved). \par

The biggest problem with PTA data is the lack of control of the \mbox{noise:} there are many potential sources which could contribute to the observed data which we model in a parametrized  (and sometimes simplified) way when it is included in the ``global fit'' (i.e. that fits all pulsars simultaneously).
The high frequency end of PTA data is usually dominated by measurement white noise. The low frequency end is expected to be dominated by red noise processes. 
Besides the GWB discussed above, some of the red noise is also expected to be correlated among pulsars as a function of angular separation of each pair of pulsars on the sky \citep{tib15}.  Such sources of correlated noise are errors in the clock time standard (causing monopolar-type correlations) or systematic errors in the Solar-system ephemeris (causing dipolar-type correlations). Moreover, we also expect the presence of spatially-uncorrelated red noise which is individual to each pulsar in the array. This is the spin noise (or timing noise) which is caused by the rotational variations of the pulsar arising from a variety of different phenomena (e.g., intrinsic processes, unmodeled objects in the vicinity of the neutron star, etc.). The red noise types described above are commonly referred to as \emph{achromatic} red noise since they are independent of the observing radio frequency. Most of the PTA data also show the presence of \emph{chromatic} red noise that depends on the radio frequency of the observations. In particular we will consider the long-term variations of dispersion measure (DM), which add time delays to the ToAs as $\Delta t \propto \nu^{-2}$, and scattering variations ($\Delta t \propto \nu^{-4}$), which are both caused by the time-varying electron column density between the pulsar and the radio telescope. \par

The sensitivity of PTA data to GWs is significantly affected by the levels of red noise
and our ability to both detect and to characterize GW signals strongly depends on
the faithfulness of the pulsar noise model \citep{cll16,len16,haz20,gon21}, which can vary significantly from pulsar to pulsar.
Due to the large choice of possible noise 
components (we see it as various possible models of the noise) and their description/parametrisation, the 
search for a GWB usually assumes a common (and simplified) noise model that is the same for each 
pulsar. The parameters of that simplified model are then inferred together
with the parameters characterizing the GWB. It was shown \citep[see, for example,][and the references 
therein]{gon21} that the actual noise model
could vary significantly from pulsar to pulsar, which could influence the detectability of 
a GWB \citep{haz20}.
This is the main motivation behind this paper.  We consider this paper as a companion of  \citet{DR2_Chen+}, where the main focus was on finding and characterizing the common red signal (CRS) using two independent pipelines. 

Based on previous exploratory investigations for each pulsar in the DR2, 
we suggest a finite set of noise models and use Bayes 
factors as a ranking statistic to choose between them, assuming that all models are equally probable 
\textit{a priori}. When the Bayes factor is not informative (close to one) we make the selection 
based either on the simplicity of the model or on the basis of computational efficiency with very few 
exceptions which we explicitly discuss below. Once we had customized the noise model for each pulsar, we reproduced the main results of \citet{DR2_Chen+},
namely we confirmed the high statistical significance of a common red signal, but without sufficient evidence for a GWB induced Hellings-Downs correlation.  We want to emphasize that the final noise model used for each pulsar was quite different from the simpler model assumed in \citet{DR2_Chen+} for all pulsars. The noise model selection method described in this paper will be used on the extended (25 pulsars) EPTA data. 

The rest of the paper is organized as follows. We give a brief description of DR2 data in the section~\ref{S:dr2}. Section \ref{sec:Sig} gives a detailed description of each noise process that will be used in building the noise model. Section \ref{sec:bayes} summarizes the Bayesian framework used in this paper. The details of the single-pulsar noise model selection are given in  Section \ref{sec:Models}.  In Section  \ref{sec:Com} we consider the presence of a common red noise and we summarize our results in Section \ref{sec:Conc}.

\section{Brief description of EPTA DR2}
\label{S:dr2}

\par
The EPTA Data Release 2 (DR2) -- 6-pulsars dataset \citep{DR2_Chen+} -- comprises  up to $24$ years of high cadence observations of PSRs J0613$-$0200, J1012+5307, J1600$-$3053, J1713+0747, J1744$-$1134 and J1909$-$3744. These pulsars are observed in single-dish mode at four European radio telescopes: the Effelsberg 100-m radio telescope (EFF), the Nan\c{c}ay Radio telescope (NRT), the Lovell Telescope at the Jodrell Bank Observatory (JBO) and the Westerbork Synthesis Radio Telescope (WSRT).
In addition,EPTA DR2 includes data from the Large European Array of Pulsars (LEAP), which is based on the combination of these four telescopes with the Sardinia Radio Telescope (SRT), forming a tied-array telescope \citep{bas15}.
The five radio telescopes contribute to all pulsars except PSR~J1909$-$3744, which has a dataset that contains only  NRT observations because of its low-declination.
Upgrades of telescopes, including improvements to or changes of receivers/backends,  have been applied during the observational period, which makes the dataset heterogeneous in timing precision and radio frequency coverage. We label the data by the telescope (or observatory) and the system that collected it, followed by the radio frequency in MHz (e.g., EFF.P200.1400). 
Having multiple systems in a PTA dataset is both a curse, as we need to combine the data from all systems together taking into account possible systematics, and a blessing as the multiband observations are required to disentangle and characterize the chromatic noise and the system specific instrumental red noise \citep[e.g., system noise,][]{len16}.\par

A characteristic ToA is computed from the time and frequency averaged profile of each observation, except for JBO.ROACH and NRT.NUPPI backends, which use respectively 2 and 4 (radio-frequency) sub-band ToAs per epoch. The ToAs of each pulsar are assembled together and used to fit the timing model (TM) parameters that describe the pulsars's sky position and proper-motion, its spin frequency and corresponding derivative, and the DM and its two first derivatives. For pulsars in binary systems, the timing model also accounts  for the orbital motion including Keplerian and post-Keplerian parameters. Phase jumps are included in the TM for each system and also for each of the JBO.ROACH \& NRT.NUPPI.1484 sub-bands.
The fits for the TM parameters were obtained using the \textsc{Tempo2} package \citep{hob06} with the JPL Solar-system ephemeris DE438  (to transform the local observatory ToAs to the Solar-system barycentre) and with the clock corrections TT(BIPM2019) (time conversion from the observatory time standard to the Terrestrial Time (TT) given by the Bureau International des Poids et Mesures (BIPM)). The end result are the \emph{timing residuals}, i.e. the differences between the observed ToAs and those  predicted by the TM, which are then analysed to search for GW signals. \par

\section{Modelling noise in PTA data} \label{sec:Sig}
In this section we describe the basic noise models. We will use an elegant description based on the Gaussian Process (GP) introduced in \citet{van14}. Later, we will use those basic noise components to build a complete noise model for each pulsar in DR2 using Bayesian techniques. \par

Let us very briefly introduce the likelihood for Gaussian processes following \citet{van14}. We assume that all noise components are Gaussian and stationary and we separate the white noise component $ \mathbf{N}$ from the rest. The Gaussian process can be introduced in two equivalent ways: (i) As a sum of deterministic basis functions $\sum_i F_i(t) w_i$, where $w_i$ are weights -- random Gaussian distributed variable $\mathcal{N}(w_i^0, \Sigma_{ij})$,  where $w_i^0$ is a mean value for each weight, $\Sigma_{ij}$ is a covariance matrix, and  $F_i(t)$ are the basis functions. This is the weight-space view. (ii) As a continuous function such that the ensemble average is $\mathbb{E}[f(t)] = m(t)$ and the covariance is $\mathbb{E}[(f(t) - m(t))(f(t') -m(t'))] = C(t, t')$. This is the function space view.  Those two descriptions are related via
\begin{equation}
	C(t, t') = \sum_{a,b} F_a(t) \Sigma_{ab} F_b(t'),
	\label{Eq:GPequiv}
\end{equation}
with $a,b = 1, ..., N$.
The red noise covariance matrix $C(t,t')$ was introduced in \cite{van12} and it was approximated using an incomplete Fourier basis 
in \cite{len13}.  Applying the Gaussian process approach to the PTA likelihood function we get \citep{van14}:
\begin{align}
	p(\delta t| w_a, \text{GP}) &= \frac{
		e^{-\frac1 2 \sum_{ij}
			(\delta t_i - \sum_a F_a(t_i) w_a) (\mathbf{N}_{ij})^{-1}
			(\delta t_j - \sum_a F_a(t_j) w_a)}
	}
	{\sqrt{(2\pi)^n {\rm{det}}(\mathbf{N})}} \nonumber \\
	&\times
	\frac{
		e^{-\frac1{2}  \sum_{a,b} w_a (\Sigma_{ab})^{-1}
			w_b }
	}
	{\sqrt{(2\pi)^m {\rm{det}}(\Sigma)}},
	\label{Eq:weightGP}
\end{align}
where $\delta t_i$ the $i$-th observed residuals with $i,j = 1, ..., n$.  The equivalent representation is given by
\begin{align}
	p(\delta t| \text{GP}) =
	\frac{
		e^{-\frac1 2 .\sum_{ij}
			\delta t_i  (\mathbf{N} + C_{ij})^{-1}
			\delta t_j}
	}
	{\sqrt{(2\pi)^n {\rm{det}}(\mathbf{N} + \mathbf{C})}},
	\label{Eq:MargGP}
\end{align}
where  $ C_{ij}^{rn} = \sum_{a,b} F_a(t_i) \Sigma_{ab} F_b(t_j)$. The convenience of the latter description is that it can be computed efficiently using the Woodbury equality:

\begin{align}
	(\mathbf{N}+\mathbf{C})^{-1} &\simeq ( \mathbf{N}+\mathbf{F} \mathbf{\Sigma} \mathbf{F}^T )^{-1} \nonumber \\
	&= \mathbf{N}^{-1} - \mathbf{N}^{-1}\mathbf{F} (\mathbf{\Sigma}^{-1} + \mathbf{F}^T \mathbf{N}^{-1} \mathbf{F})^{-1} \mathbf{F}^T \mathbf{N}^{-1}
\end{align}
In what follows we consider  $\mathbf{C}$ as a combination of several (chromatic and achromatic) red noise  components each decomposed in its own set of basis functions.

\subsection{Marginalization over timing model}
Before we introduce the noise components, we should explain how we treat the timing model. We assume that an initial fit of the timing model obtained with \textsc{Libstempo} \citep{val20} reduces it to a linear model where the coefficients are given by a design matrix.  We analytically marginalize the likelihood over the TM parameter errors described by that linear model. The analytic marginalization was first demonstrated in \cite{van09}, but \cite{van14}
have shown that it is equivalent to the marginalization of a corresponding Gaussian process with an improper prior.

The implementation of this marginalization in \textsc{Enterprise} \citep{ell19}  (the package that we use throughout this project) uses the equivalence of
weight space and function space description of a Gaussian process.
The design matrices ($M_a(t_i)$) are used as basis functions, the covariance for the TM process is given as
$C^{\mathrm{TM}} = \sum_{a,b} M_a(t_i) \Sigma^{\mathrm{TM}}_{ab} M_b(t_j)$,  where the prior on the parameter errors is modelled as $\Sigma = \lambda I$ with
$I$ being a diagonal unit matrix and $\lambda$ is a large numerical number \citep[see][for details]{van14}. In the limit $\lambda \rightarrow \infty$ this prior becomes improper, but in this analysis the values of $\lambda$ are fixed but large so the prior is formally proper. The marginalization over timing errors
(``weights'') of Eqn.  (\ref{Eq:weightGP}) leads to Eqn. (\ref{Eq:MargGP}) (this is a manifestation of the duality of the two descriptions).

The use of a very wide or improper prior in Bayesian model selection should be taken with great caution especially when comparing two models where only one of them uses marginalization over the improper prior (this was also discussed in \cite{DR2_Chen+}). The penalization which is embedded in the prior (for being too wide) and propagates into the computation of the evidence is lost and reliable results from evidence-based model selection cannot be guaranteed. However in \emph{all} noise models described below we perform marginalization over the TM parameters which brings them all to a common starting point for further comparison.

\subsection{White noise}

As mentioned earlier, the white noise dominates the high frequency end of the PTA sensitivity  band.  The ToA errors ($\sigma_{\mathrm{ToA}}$) are estimated within the template-matching method \citep{tay92} that is used to compute the ToAs.
This method is based on the Fourier domain cross-correlation of a template profile with the integrated pulse profile at the corresponding epoch.
The uncertainties of each ToA are further modified as
$$\sigma = \sqrt{\mathrm{E_f^2 \ \sigma^2_{\mathrm{ToA}}} + \mathrm{E_q}^2}.$$
\textsc{EFAC} ($\mathrm{E_f}$) is a  multiplicative factor that takes  into account  ToA measurement errors (or radiometer noise). \textsc{EQUAD}  ($\mathrm{E_q}$)  is added in quadrature to account for any other white noise \citep[such as stochastic profile variations][]{lkl+11,sha14,lcc+16} and for possible systematic errors. The white noise model, therefore, is given as
\begin{equation}
	\mathbf{N} = \left(\mathrm{E_f}^2 \ \sigma^2_{\mathrm{ToA}}(t_i) + \mathrm{E_q}^2\right) \delta_{i,j},
\end{equation}
where $i$ and $j$ indexing the ToAs of the corresponding backend.   \textsc{EFAC} and  \textsc{EQUAD} are phenomenological parameters that characterize the white noise for each system and for each pulsar.


\subsection{Stochastic red signals} \label{ssec:modredsig}
It is essential for PTA analysis to properly describe  the intrinsic red noise because of its possible correlation with low-frequency GW signals \citep{sha10}. Results from simulations in \cite{haz20} have clearly demonstrated the impact of inaccurate red noise modelling on GWB results.  \par

The single-pulsar stochastic red noise is a time-correlated signal modelled as a stationary Gaussian process. In this work we adopted the "weight-space" representation of the Gaussian Process. The timing residuals due to red noise at  each epoch $t_i$ are approximated as:
\begin{align}
	\delta t^{\mathrm{SRS}}(t_i) &= \sum_{l=1}^{N} X_{l} \ \mathrm{cos} \left( 2 \pi  t_i f_l \right) + Y_l \ \mathrm{sin} \left(2 \pi  t_i f_l \right),
	\label{Eq:RNasGPweights}
\end{align}
where $X_l$ and $Y_l$ are playing the role of weights and the basis functions are
\begin{equation}
	\begin{aligned}
		F_{2l-1}(t_i)&=\mathrm{cos} \left( 2 \pi \ t_i f_l \right),\\
		F_{2l}(t_i)&=\mathrm{sin} \left( 2 \pi \ t_i f_l \right),
	\end{aligned}
\end{equation}
where $l = 1, ..., N$.  This representation would correspond to the usual Fourier transform
if $f_{l} = l/T$ (where $T$ is the total timespan) and we had regularly spaced epochs, $t_i$. However the radio observations are quite irregular (besides maybe the last 5 years or so) which makes the Fourier basis not exactly orthogonal. In addition, we do not use a complete set: we usually truncate it at some low frequency as we are interested in modelling  the red noise. The optimal choice of frequencies was considered in \cite{van15}, however for all results presented here, we have used an evenly spaced  $\Delta f = 1/T$ set of frequencies, starting at $f = 1/T$ and truncating at $N/T$ where $N$ is one of the parameters in the model selection. \par

The covariance matrix $\mathbf{\Sigma}$ for the Fourier coefficients (weights $X_l, Y_l$) is defined by the power spectral density (PSD), $S(f)$.  The simplest model for the PSD of a stochastic red process in a single pulsar data set is a power-law:

\begin{equation} \label{eq:RN_powerlaw}
	S_{\mathrm{P}}(f ; A,\gamma) = \frac{A^2}{12 \pi^2} \left( \frac{f}{{\mathrm{yr}^{-1}}} \right)^{-\gamma} \ \mathrm{yr}^3,
\end{equation}
where the amplitude $A$ is the normalized value at the frequency of one over one year  ($f=1/\mathrm{yr}$).  The covariance matrix is given in the frequency domain by
\begin{equation}
	\Sigma_{k\alpha, l\beta} = S_P(f_k; A_{\alpha}, \gamma_{\alpha}) \delta_{kl} \delta_{\alpha\beta} \ / \ T
\end{equation}
where $k, l= 1,...,N$, and $\alpha, \beta$ are pulsar indices -- we consider spatially uncorrelated red noise -- therefore we have placed the Kronecker delta-function  on the right hand side.

An alternative description takes into account that the data is
dominated by white noise at high frequencies, and that
is captured in the broken power law \citep{arz20} :
\begin{equation}
	S_{\mathrm{BPL}}(f ; A, \gamma, \delta, f_b, \kappa) = \frac{A^2}{12 \pi^2} \left( \frac{f}{{\mathrm{yr}^{-1}}} \right)^{-\gamma} \left( 1 + \left( \frac{f}{f_b} \right)^{1 / \kappa} \right)^{\kappa \ (\gamma - \delta)} \mathrm{yr}^3
\end{equation}
in which $A$ is the amplitude at frequency $f=\frac{1}{\mathrm{yr}}$, $f_b$ is the transition frequency, $\gamma$ and $\delta$ are respectively the slopes below and above $f_b$ and $\kappa$ defines the smoothness of the transition. Jumping a bit ahead, we will perform Bayesian analysis of the data with priors on $\kappa$ and $f_b$ as uniform $\mathcal{U}(0.01, 0.5)$ and log-uniform $\mathrm{log}_{10}\mathcal{U}(10^{-10}, 10^{-6})$, and the priors on $A$ and $\gamma$ are the same as for the simple power-laws models given in Table \ref{tab:priors}. The high-frequency spectral index $\delta$ is fixed at $0$. \par

A completely different approach is not to impose any particular spectral shape but rather estimate it from the data itself: this is the free-spectrum method \citep{len16} in which
\begin{equation}
	S_{\mathrm{FS}}(f_i ; \rho_i) = \rho_i^2 \ T,
\end{equation}
where $\rho_i$ is the spectral amplitude at frequency $f_i=i / T$, with $i = 1, ..., N$, in units of residuals. This modelling is particularly useful for understanding the spectral content and for interpreting the results for the red noise models given above. The number of parameters in the free-spectrum approach is equal to the number of Fourier bins and it is therefore computationally more expensive. The priors used for each  $\rho_i$ will be log-uniform:  $\mathrm{log}_{10}\mathcal{U}(10^{-10}, 10^{-4})$. \par

In the rest of this subsection, we give more details on each specific type of red noise that will be included in the total noise budget for each pulsar.

\subsubsection{Achromatic red noise}
The achromatic red-noise (which we henceforth denote as RN) is commonly used in single-pulsar noise models in order to characterize the long-term variability of the pulsar spin. Also referred to as ``timing noise'' or ``spin noise'', RN is a dominant feature in the ToAs of younger pulsars,
and several physical processes have been suggested to explain it, such as magnetospheric variability \citep[e.g.,][]{lyn10,ts13} or interactions between the pulsar's superfluid core and solid crust \citep{cor10}.  The origins of RN in MSPs may differ from that of young pulsars: due to their much weaker magnetic fields, 
superfluid turbulence has been suggested as a possible contributor to the RN in MSPs \citep{mel14}. \par

We model RN using the descriptions given above: the power law model will be our standard approach, however we will also use the free spectrum and broken power law (red noise becomes white after some frequency) to guide the selection of the truncation frequency
in the sum given by Eq.~(\ref{Eq:RNasGPweights}). This noise component is unique, independent of the observational radio frequency and uncorrelated between different pulsars.

\subsubsection{Chromatic red noise}
During its propagation, the pulsar radio emission passes through and interacts with
the ionized interstellar medium (IISM), the Solar-system interplanetary medium and the Earth's ionosphere, which leads to frequency-dependent delays on the observed signal. \par

An important effect is the interstellar dispersion that induces a delay in the arrival time $\Delta t^{\mathrm{DM}} \propto \nu^{-2} \; \mathrm{DM} $, where $\nu$ is the radio observing frequency and DM is the dispersion measure, which is the path integral of the free-electron density~\citep{lor04}. This effect is taken into account during the observations and inside the timing model which considers its value at a reference epoch together with its first and second derivatives. 
However, the inhomogenious and turbulent nature of the IISM also induces chromatic (i.e., dependent on the observing radio frequency $\nu$) red noise that are important
on the decade-long timescales of PTA data \citep{you07,kei13}. In addition, the orbital motion of the Earth around the Sun may induce an additional deterministic chromatic signal. \par

Another result of the radio signal's interaction with the IISM are scattering variations (Sv), corresponding to the multi-path propagation of the radio signal due to diffraction and refraction in the IISM \citep{lor04,lyn10}. This causes a chromatic pulse broadening and a time delay with $\sim \nu^{-4}$ chromaticity. The scattering variations are described as a stochastic red signal such that $\Delta t^{\mathrm{Sv}} \propto \nu^{-4}$.\par

We describe phenomenologically any general chromatic red noise 
using the basis functions
\begin{equation}
	\mathbf{F}_j^{\mathrm{chrom.}}(t_i) = \mathbf{F}(t_i) * \left( \frac{\nu_j}{1.4 \ \mathrm{GHz}} \right)^{-\chi},
\end{equation}
where $ \mathbf{F}$ is the incomplete set of sin/cos basis functions, $\nu_j$ is an observational radio frequency for a corresponding residual at the epoch $t_i$, and $\chi$
is  the chromatic index.  We use the same covariance matrices for chromatic red noise
as for achromatic red noise (power-law, broken power-law, free spectrum).  It is essential to have multiband radio observations to disentangle chromatic from achromatic red noise \citep[otherwise they are completely degenerate, see for example][]{cll16}. \par

During model selection, we will consider the following chromatic red processes: (i) dispersion measurement variations
(DMv) with $\chi=2$; (ii) scattering variations  (Sv) with  $\chi=4$; and (iii) a phenomenological ``free chromatic noise'' model (FCN) with $\chi$ taken to be a free parameter with prior $\mathcal{U}(0,7)$. The FCN was first introduced in \citep{gon21}
and is used here as a diagnostic to verify the combined noise model.

\subsubsection{System and Band noise}
The EPTA DR2 dataset is a combination of ToAs produced by five radio telescopes which use different systems observing at radio frequencies ranging from $\sim 300$ MHz to $\sim 5$ GHz.  Following \cite{len16} we introduce ``system'' and ``band'' red noise.
The system noise (SN) term is a stochastic red signal specific to a single receiver system. Such a signal could, for example, arise from a miscalibration of polarizations or specific radio frequency interferences. We model this process as a stochastic red noise applied to the ToAs of only one considered system at a time. This noise is considered to be achromatic for every system except for NRT.NUPPI.1484 that is divided into 4 sub-bands and will be probed for the presence of both chromatic red process SN and DMv (labelled as DMv-SN). \par

The band noise (BN) is a stochastic red noise assigned to a specific radio frequency band. This is to account either for a possible  frequency-dependent DM in the amplitude (additional to the overall $\nu^{-2}$ factor)
caused by  multi-path propagation of radio emission \citep{cor16} or by frequency-dependent calibration errors \citep{van13}. Given the frequency coverage of the EPTA DR2 dataset \citep{DR2_Chen+}, we consider four radio bands for the BN:
\begin{itemize}
	\item Band.1 : $< 1$ GHz
	\item Band.2 : $[1, 2]$ GHz
	\item Band.3 : $[2, 3]$ GHz
	\item Band.4 : $> 3$ GHz.
\end{itemize}

\subsection{Common red noise} \label{ssec:SigCRS}
In this subsection we describe the red noise common to all pulsars and
differentiate between two groups of common stochastic red signals: (i) a common spatially-uncorrelated red noise (CURN) signal; and (ii) a correlated common red noise signal.
The CURN shares spectral properties across all pulsars but does not appear with any particular spatial correlation (random) for each pair of pulsars, its covariance matrix is described as
\begin{equation}
	\Sigma_{k\alpha, l\beta} = S_P(f_k; A_{\rm CURN}, \gamma_{\rm CURN}) \delta_{kl} \delta_{\alpha\beta} \ / \ T,
\end{equation}
where the amplitude and spectral index ($A_{\rm CURN}, \gamma_{\rm CURN}$) are the same for all pulsars.  

On the other hand, the stochastic GWB, and the clock  and ephemerides errors  are examples of truly spatially correlated red signals. We describe in detail the GWB, which is the only correlated red process considered in this paper (clock and ephemerides errors in the EPTA DR2 were investigated in \cite{DR2_Chen+}  and their presence was not 
supported by the data). The dimensionless characteristic strain spectrum of the GWB is given as a power-law \citep{maj2000,jen06} with a reference frequency at $1 \ \mathrm{yr}^{-1}$:
\begin{align}
	h_c(f) = A_{\mathrm{GWB}} \ \left( \frac{f}{\mathrm{yr}^{-1}} \right)^{\alpha_{\mathrm{GWB}}}
\end{align}
with $A_{\mathrm{GWB}}$ and $\alpha_{\mathrm{GWB}}$ respectively the GWB strain amplitude and spectral index. The corresponding  PSD $S_P^{\mathrm{GWB}}(f)$
can then be written as
\begin{align}
	S_P^{\mathrm{GWB}}(f) &= \frac{1}{12 \pi^2} \frac{1}{f^3} \ h^2_c(f) \nonumber \\
	&= \frac{A^2_{\mathrm{GWB}}}{12 \pi^2} \ \left( \frac{f}{\mathrm{yr}^{-1}} \right)^{-\gamma_{\mathrm{GWB}}} \mathrm{yr}^3
\end{align}
with $\gamma_{\mathrm{GWB}} = 3 - 2 \alpha_{\mathrm{GWB}}$. \\

The spectral slope of a GWB generated by a population of SMBHBs on circular and GW-driven orbits \citep{jaf03,che17} is expected to be $\alpha_{\mathrm{GWB}} = -2/3$, or $\gamma_{\mathrm{GWB}} = 13/3$.  We use the same incomplete set of Fourier basis function
as for the achromatic red noise described above but with a covariance matrix with spatial correlation coefficients $\Gamma(\theta_{ab})$ corresponding to the Hellings-Downs curve
\citep{lee08}:
\begin{equation}
	\Sigma_{k\alpha, l\beta} = S_{\mathrm{P}}(f_k; A_{\mathrm{GWB}}, \gamma_{\mathrm{GWB}}) \delta_{kl} \Gamma(\theta_{\alpha\beta}) \ / \ T,
\end{equation}
where $\theta_{\alpha\beta}$ is the angular separation of a pair of pulsars. This model describes an isotropic component of the GWB. \par

Note that it is the spatial correlation that distinguishes a GWB from the CURN, therefore to clearly identify this it is necessary to infer $\Gamma(\theta_{ab})$. 
This is not possible with six pulsars and with the present EPTA timing precision as shown in \cite{DR2_Chen+}.

In addition, a CURN could be mimicked by the RN of the most sensitive
pulsars, so it is crucial to study in detail the red processes in each pulsar in order to understand the significance of the CURN reported in  \cite{DR2_Chen+}. 

\subsection{Deterministic signals} \label{ssec:detsig}
In addition to stochastic processes we also consider two types of deterministic signals
of non-GW nature. We have used prior information about the (possible) presence of these signals in the data of some pulsars.


\subsubsection*{Exponential dips}
Several pulsars have displayed exponential timing-residual dips (E), in which there is a sudden frequency-dependent advance in the ToAs. It is relevant to our analysis that such events have been observed at least twice for PSR~J1713+0747 over our observational timespan, in $2008$ ($\sim$ MJD $54757$) and in $2016$ ($\sim$ MJD $57510$). \par

The first event, reported in \cite{col15}, \cite{zhu15} and \cite{dev16}, is interpreted as a ``DM event'', i.e., a drop in the electron column density along the line-of-sight  producing a sudden reduction of DM that returns to the previously observed level exponentially over time.  \par

The second event, reported in \cite{lam18}, was accompanied by a pulse shape change and corresponding chromatic index lower than $2$, so it is not compatible with a  DM-related process. It was proposed instead \citep{gon21} that this event is related to processes in the pulsar's magnetosphere.\par

We model the exponential dip delay at epoch $t_i$ and radio frequency $\nu_k$ as :

\begin{align}
	&d^{\mathrm{E}}(t_i, \nu_k ; A_E, \tau, t_0, \chi_E) = \nonumber\\
	&\left\{
	\begin{aligned}
		&0, \mathrm{if} \ t_i < t_0\\
		&A_E \ \left( \frac{\nu_k}{1.4 \ \mathrm{GHz}} \right)^{-\chi_E} \ \mathrm{exp} \left( - \frac{t_i - t_0}{\tau} \right), \mathrm{if} \ t_i \geq t_0
	\end{aligned}
	\right.
\end{align}
where $A_E$ is the amplitude in residual units, $t_0$ is the reference epoch of the event, $\tau$ is the relaxation time and $\chi_E$ is the chromatic index, either fixed or being a free parameter with prior $\mathcal{U}(0,7)$.

\subsubsection*{Annual chromatic signals}
The second deterministic signal which could be present in the data is an annual chromatic process (which  we label as ``Y'') that results from electron density variations as the line-of-sight to the pulsar changes during the annual Earth motion around the Sun.\par

Previous investigations \citep{kei13,mai20} indicated such a signal is present in PSR~J0613$-$0200, which we model as \citep{len16,gon21}:
\begin{equation}
	d^{\mathrm{Y}}(t_i, f_l; A_Y, \phi, \chi_y) = A_Y \ \left( \frac{f_l}{1.4 \ \mathrm{GHz}} \right)^{-\chi_y} \mathrm{sin} \left( 2 \pi \ \frac{t_i}{\mathrm{yr}} + \phi \right)
\end{equation}
where $A_Y$ is the characteristic amplitude in residual units, $\chi_y$ is the chromatic index and $\phi$ is the initial phase. We consider either annual DM variations or annual scattering variations, with a chromatic index fixed at $2$ or $4$ respectively. \\

\begin{table}
	\centering
	\caption{Models and priors used for the single-pulsar model selection.}
	\label{tab:priors}
	\begin{tabular}{lcc}
		\hline
		Model & \multirow{2}{*}{Parameters} & \multirow{2}{*}{Priors (or fixed val.)}\\
		(abrev.) & & \\
		\hline\hline
		White-noise & EFAC & $\mathcal{U}(0.1, 5)$ \\
		(WN) & EQUAD [s] & $\mathrm{log}_{10}\mathcal{U}(10^{-9}, 10^{-5})$ \\
		\hline
		Achromatic red-noise & $A_{\mathrm{RN}}$ & $\mathrm{log}_{10}\mathcal{U}(10^{-18}, 10^{-10})$ \\
		(RN) & $\gamma_{\mathrm{RN}}$ & $\mathcal{U}(0,7)$ \\
		\hline
		DM variations & $A_{\mathrm{DM}}$ & $\mathrm{log}_{10}\mathcal{U}(10^{-18}, 10^{-10})$ \\
		(DMv) & $\gamma_{\mathrm{DM}}$ & $\mathcal{U}(0,7)$ \\
		\hline
		Scattering variations & $A_{\mathrm{Sv}}$ & $\mathrm{log}_{10}\mathcal{U}(10^{-18}, 10^{-10})$ \\
		(Sv) & $\gamma_{\mathrm{Sv}}$ & $\mathcal{U}(0,7)$ \\
		\hline
		\multirow{2}{*}{Free-chromatic noise} & $A_{\mathrm{FCN}}$ & $\mathrm{log}_{10}\mathcal{U}(10^{-18}, 10^{-10})$ \\
		\multirow{2}{*}{(FCN)} & $\gamma_{\mathrm{FCN}}$ & $\mathcal{U}(0,7)$ \\
		& $\chi_{\mathrm{FCN}}$ & $\mathcal{U}(0,7)$ \\
		\hline
		\multirow{2}{*}{System-noise} & $A_{\mathrm{SN}}$ & $\mathrm{log}_{10}\mathcal{U}(10^{-18}, 10^{-10})$ \\
		\multirow{2}{*}{(SN or DMv-SN)} & $\gamma_{\mathrm{SN}}$ & $\mathcal{U}(0,7)$ \\
		& $\chi_{\mathrm{SN}}$ & $0$ or $2$ \\
		\hline
		Band-noise & $A_{\mathrm{BN}}$ & $\mathrm{log}_{10}\mathcal{U}(10^{-18}, 10^{-10})$ \\
		(BN) & $\gamma_{\mathrm{BN}}$ & $\mathcal{U}(0,7)$ \\
		\hline
		\multirow{4}{*}{DM events} & $A_{\mathrm{E}}$ [s] & $\mathrm{log}_{10}\mathcal{U}(10^{-10}, 10^{-2})$ \\
		\multirow{4}{*}{($\mathrm{E}$)} & $\tau_{\mathrm{E}}$ [day] & $\mathrm{log}_{10}\mathcal{U}(1, 10^{2.5})$ \\
		& $t_0$ [MJD] & $\mathcal{U}(54650, 54850)$ or \\
		&  & $\mathcal{U}(57490, 57530)$  \\
		& $\chi_{\mathrm{E}}$ & $1$, $2$, $4$ or $\mathcal{U}(0,7)$\\
		\hline
		\multirow{2}{*}{Annual chrom.} & $A_{\mathrm{Y}}$ [s] & $\mathrm{log}_{10}\mathcal{U}(10^{-10}, 10^{-2})$ \\
		\multirow{2}{*}{($\mathrm{Y}$)} & $\phi_{\mathrm{Y}}$ & $\mathcal{U}(0, 2\pi)$ \\
		& $\chi_{\mathrm{y}}$ & $2$ or $\mathcal{U}(0,7)$\\
		\hline
	\end{tabular}
\end{table}



\section{Bayesian inference framework}\label{sec:bayes}
In this section, we briefly describe the Bayesian framework that will be applied to the model selection (see, for example, \cite{siv06} for further reading). 
The main purpose of this section is to introduce notation that will be used in the following sections.

We will consider a set of models, $\mathcal{M}_a$, each characterized by parameters $\vec{\theta}_a$, where the subscript $a$ enumerates the models.  The probability of a given model $\mathcal{M}_a$ given the observed residuals  $\vec{\delta t}$ can be written using Bayes theorem:

\begin{align}
	\label{Eq:BayesModelSelection}
	\mathrm{P}(\mathcal{M}_a | \vec{\delta t}) &= \frac{\mathrm{P}(\vec{\delta t} | \mathcal{M}_a) \ \ \pi_{\mathcal{M}_a}}{\mathrm{P}(\vec{\delta t})},
\end{align}
where $\pi_{\mathcal{M}_a}$ is the prior probability of model $\mathcal{M}_a$, $\mathrm{P}(\vec{\delta t} | \mathcal{M}_a)$ is the probability of observing $\vec{\delta t}$ assuming that model $\mathcal{M}_a$ is the correct one (this is the evidence of model $\mathcal{M}_a$ and we denote it as $\mathcal{Z}_{\mathcal{M}_a}$), and $ \mathrm{P}(\vec{\delta t}) = \sum_b \mathrm{P}(\vec{\delta t} | \mathcal{M}_b) \pi_{\mathcal{M}_b}$ is the probability of the observed data set, which we consider as a normalization factor.  We have used previously published results as a guide for selecting models for a given pulsar, assuming that all considered models have equal prior probabilities unless otherwise specified. 
The model selection is based on the odds ratio:
\begin{align}
	\frac{\mathrm{P}(\mathcal{M}_a | \vec{\delta t})}{\mathrm{P}(\mathcal{M}_b | \vec{\delta t})} = \frac{\mathcal{Z}_{\mathcal{M}_a}}{\mathcal{Z}_{\mathcal{M}_b}} \  \frac{\pi_{\mathcal{M}_a}}{\pi_{\mathcal{M}_b}}
\end{align}
Since  we use equal priors, the odds ratio reduces to the  \emph{Bayes factor} $\mathcal{B}^{\mathcal{M}_a}_{\mathcal{M}_b} = \mathcal{Z}_{\mathcal{M}_a} / \mathcal{Z}_{\mathcal{M}_b}$ .\par

For a given model the posterior on the parameters $\vec{\theta}_a$ is given again by Bayes theorem:
\begin{align}
	\label{eq:bayes}
	\mathrm{P}(\vec{\theta}_a | \vec{\delta t}, \mathcal{M}_a) &= \frac{\mathrm{P}(\vec{\delta t} | \vec{\theta}_a, \mathcal{M}_a) \ \ \mathrm{P}(\vec{\theta}_a | \mathcal{M}_a)}{\mathrm{P}(\vec{\delta t} | \mathcal{M}_a)} \nonumber \\
	&= \frac{\mathcal{L} (\vec{\delta t} | \vec{\theta}_a, \mathcal{M}_i ) \ \ \pi(\vec{\theta}_a | \mathcal{M}_a)}{\mathcal{Z}_{\mathcal{M}_a}},
\end{align}
where we will use the likelihood $\mathcal{L} (\vec{\delta t} | \vec{\theta}_a, \mathcal{M}_a ) $ marginalized over the timing model parameters (see discussion in the previous section) and
$\pi(\vec{\theta}_a | \mathcal{M}_a)$ are priors on the model parameters. The evidence of a given model  is computed from the fully marginalized posterior
\citep{siv06}:
\begin{align} \label{eq:ev}
	\mathcal{Z}_{\mathcal{M}_a} =  \int d \vec{\theta}_a \ \ \mathcal{L} (\vec{\delta t} | \vec{\theta}_a, \mathcal{M}_a) \ \ \pi(\vec{\theta}_a | \mathcal{M}_a).
\end{align}
\par
Our decisions are based on the scale proposed in \cite{jef61},  that is $\mathcal{B}^{\mathcal{M}_a}_{\mathcal{M}_b} > 100$ indicates a preference for the model ${\mathcal{M}_a}$ against ${\mathcal{M}_b}$ with ``decisive'' evidence. This interpretation criteria has been set phenomenologically, and revised in \cite{kas95}, which suggests using the threshold value  of $150$ ($\log_{10} {\mathcal{B}^{\mathcal{M}_a}_{\mathcal{M}_b}} \gtrsim 2.2$). Therefore, we use the range \mbox{$2 < | \mathrm{log}_{10}\mathcal{B}^{\mathcal{M}_a}_{\mathcal{M}_b} | < 2.2$} as a selection criteria. In the case of a non-conclusive Bayes factor  we follow the Occam principle and select the model with the lowest prior volume (or computational cost). \par

The dimensionality of models that we consider varies from $16$ to  $75$ parameters. To infer the parameter posterior for each model, we have used several numerical tools:  (i) a parallel tempering Markov Chain Monte-Carlo (MCMC) sampler \textsc{PTMCMCSAMPLER} \citep{ell17}; and (ii) \textsc{$\textrm{MC}^3$}  (\url{https://gitlab.in2p3.fr/stas/samplermcmc}), both based on the Metropolis-Hastings algorithm \citep{met53,has70}. For computation of the evidence we have used the \textsc{Dynesty} \citep{spe20} package based on the  nested sampling algorithm \citep{ski04,ski06}. In addition, we have used the hyper-model  method (proposed in \cite{car95}, extended  in \cite{hee15} and applied to PTA in \cite{tay20}) to obtain Bayes factors without evidence evaluation. In that approach models and their corresponding parameters are sampled using  a hyper-parameter that switches between the models. We compare both approaches whenever possible.   For multi-chain MCMC runs, we check convergence by computing the Gelman-Rubin ratio \citep{gel92,bro98}. Finally, we use the  Jensen-Shannon divergence \citep[JSD;][]{man99} to compare marginalized posteriors across models. This is the symmetric
version of the Kullback-Leibler divergence \citep{kul59} which is bound to the range $0 < \mathrm{JSD}(A || B) < \mathrm{ln}(2) \sim 0.69$, with zero corresponding to two identical probability distribution functions. \par

In Table \ref{tab:priors}, we summarize all noise models, their parameters and the prior range used in this paper.

\section{Customising the noise model for each pulsar} \label{sec:Models}

In this section, we describe the selection of the optimised noise model for each pulsar.
The selection is performed in several steps in a partially iterative way.
We start with the base model
used in \citep{DR2_Chen+} that contains only  achromatic red noise  and DM variation components. Both these models depend on the number of Fourier frequency bins we use for the
basis functions, or, in other words, on the high frequency cut-off that roughly corresponds to the transition from the red-noise to the white-noise dominated region. The analysis in \citet{DR2_Chen+} used the 30 and 100 lowest Fourier frequencies ($k/T$, where
$k=1...30(100)$) for RN and DMv respectively for \emph{all} 6 pulsars, 
as a reasonable balanced choice given exploratory tests. \par

We start with these two sources of noise and apply Bayesian model selection to find the number of frequency bins for each pulsar. We use a simple power-law model for each red noise, and we use the broken power-law and  free-spectrum models for guidance to minimize the set of model trials. Next, we include  stochastic chromatic noise and deterministic signals to the noise budget, and, finally we test for the presence of system and/or band noise. \par

Each pulsar's noise model always includes white noise and we marginalize over the TM parameter errors (as implemented in \textsc{Enterprise}).


\subsection{Selection on number of Fourier modes for the achromatic red noise and DM variation} \label{ssec:Nbins}

The importance of the choice of spectral binning has been discussed in \cite{van15}, where the authors show
limitations of the usual Fourier-sum approach with $f_k = \{1/T, ..., N/T\}$ in the presence of linear and/or quadratic signals, or if the stochastic red process spectral index is relatively high
(i.e., $\gamma \geq 7$). 
Earlier RN measurements have shown that such steeply rising noise is highly unlikely in MSPs and absent in the six pulsars considered here \citep[see][]{ver09,cll16,rea16}.  We, therefore, use a prior on the spectral index
$\mathcal{U}(0,7)$, and, as we will see later, this prior range is sufficiently broad.
We have chosen to use Fourier frequencies $f_k$ in our analysis.

We begin by identifying the most favoured number of Fourier bins for each pulsar. To do this we assume  that only the most commonly used noise components,  RN and DMv, are present in each  pulsar's data. This is the noise model that was assumed in  Chen et al. 2021. We reassess the presence of those noise components in Section \ref{ssec:Stoch_det}. We extend the short-hand notation for RN and DMv by appending the number of bins (basis functions) used in its description. For example,   \textit{RN30\_DMv100} refers to a model marginalized over the TM parameter errors, including white-noise parameters, and both RN and DMv with $30$ and $100$ Fourier modes, respectively.

The  RN and DMv components could be highly correlated if we lack multi-band observations. The RN dominates in the several lowest Fourier bins but is covered by the white noise at high frequencies. DMv, on other hand, depends on the observational radio frequency and, therefore, weakly (if at all) correlates with the white noise. Therefore we should try to use a large number of Fourier bins to accommodate the dispersion information stored at high frequencies. 

For the RN, we expect that a relatively small number of bins contribute to the analysis before the white noise becomes dominant. For example, we have found that the use of \textit{RN30\_DMv30} (which used to be the default choice)  is disfavoured by a Bayes factor of more than $10^5$ compared to the most favourable model \textit{RN30\_DMv100}
for PSR~J1744$-$1134.  In addition, the use of 30  bins  for both noise components shows a very strong cross-model ``leakage''
in the posterior of the RN and DMv parameters, which disappears completely in the favourable model.

\begin{figure}
	\centering
	\hspace*{-.3cm}\includegraphics[keepaspectratio=true,scale=0.21]{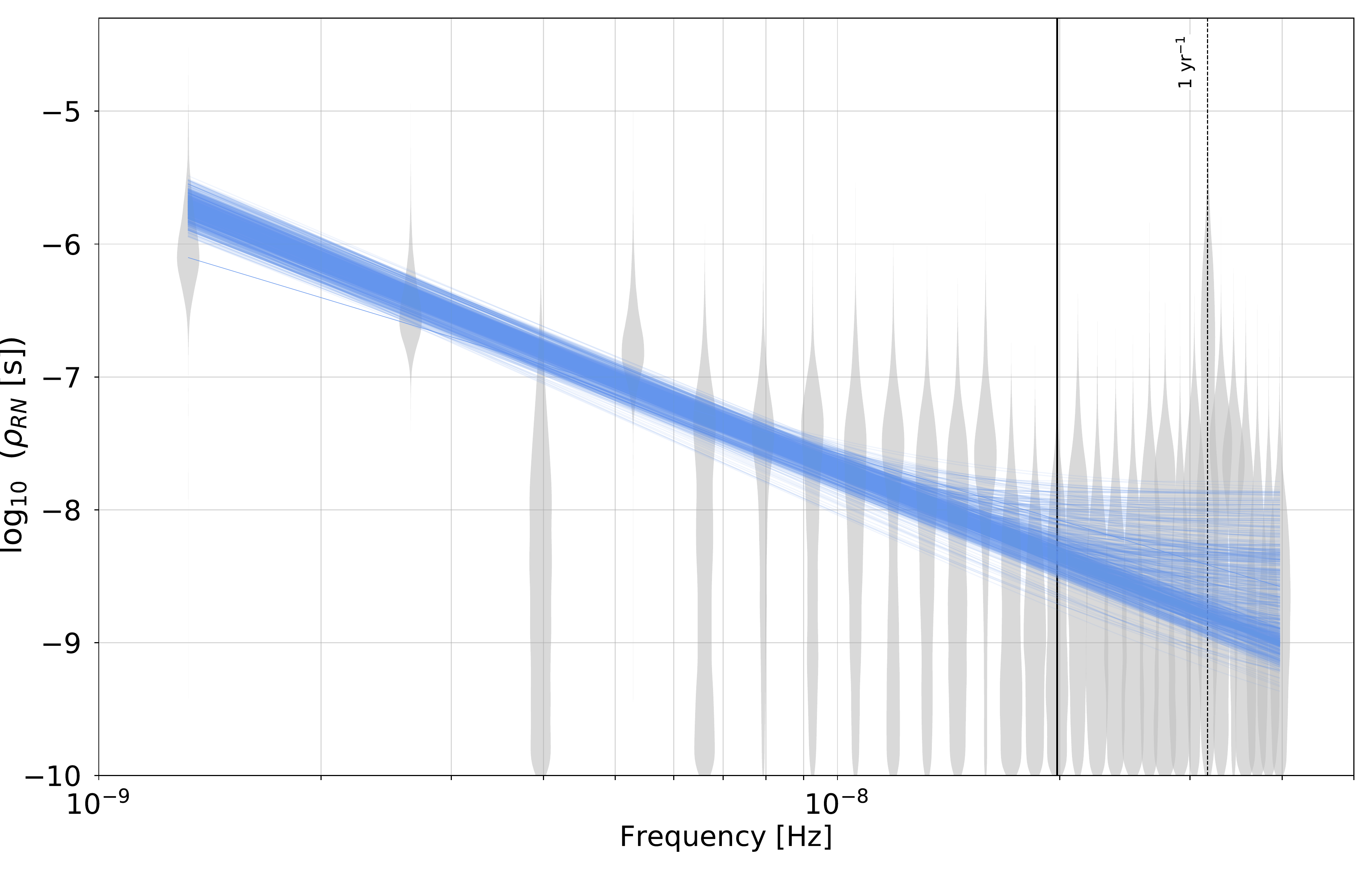}
	\caption{Achromatic red-noise spectrum of PSR~J1713+0747 using the model \textit{RN30\_DMv100}, with the free-spectrum PSD (grey violin plot) and broken power-law (blue) PSDs, showing $1000$ random realisations of the posterior distribution. The vertical thin dotted line displays the $1 \mathrm{yr}^{-1}$ frequency and the solid vertical line show the $15$th bin.}
	\label{fig:spectrum_J1713}
\end{figure}

We have analysed each pulsar using the broken power-law and free-spectrum models in order to get a rough indication on the expected range for the number of Fourier modes.
Typical results of such an analysis (for RN)  are given in  Fig.~\ref{fig:spectrum_J1713} for PSR~J1713+0747
where the estimation of the power in each free-spectrum bin is given by the grey violin-type histograms and we have over-plotted $1000$ realizations of the broken power-law randomly drawn from the posterior as blue solid lines. The broken power-law model suggests that the transitional frequency $f_b$ should be  above 15 bins (as indicated by a vertical dashed line in Fig.~\ref{fig:spectrum_J1713}). Therefore for this pulsar we try 15, 20 and 30 Fourier modes for the RN.  Similar analysis was performed for  DMv and we decided to use  $30$, $50$, $70$, $100$ and $150$ modes (to choose from) for every pulsar, which allowed us to consider frequencies up to $f_{\mathrm{max}} \simeq 1 / (2 \ \mathrm{months})$ for the pulsar with the longest data set (PSR~J1713+0747).  \par

We have performed Bayesian model selection across the pre-selected number of modes for both RN and DMv. The favourable models are summarized in Table \ref{tab:NBins}.
It shows that the data supports frequencies higher than $100/\mathrm{T_{span}}$ for DMv, except for J0613$-$0200 which displays no difference, either in evidence or in the posterior distribution of parameters, across the range of frequency-bin numbers for DMv that we have tried. As expected,  for the RN, we  require no more than $20$ frequency bins for all pulsars except PSR~J1012+5307, which requires $30$.
Where there was no clear preference between two (or more) models (Bayes factor less than 10) we preferred the model with the smallest number of bins for reasons of computational efficiency. \par

Let us give a few comments on the results presented in Table \ref{tab:NBins}. Each quoted Bayes factor compares the selected model for the number of Fourier modes with the model used in \cite{DR2_Chen+}. One can see a significant gain in the Bayes factor for PSRs J1012+5307 and J1713+0747 which is mainly due to the extension of DMv to higher frequencies. In fact all pulsars give a slight preference to 150 bins but with the Bayes factor close to one, and following our logic we have chosen to use 100 (30 for PSR~J0613$-$0200) bins to reduce the computational cost. The negative, but close to zero, log-Bayes factor for  PSRs J1600$-$3053, J1744$-$1134,
J1909$-$3744 indicates that the use of 30 modes for RN has only a slight preference (what we would call
\emph{inconclusive}) and we have chosen the lowest allowed number of bins.
The residuals in pulsar J0613$-$0200 data did not favour any particular number of modes. As we are mainly interested in the red noise, we have quantified the difference in the posterior of the red noise between two models by computing the Jensen-Shannon divergence. The last column indicates that there has been a statistically significant change in the RN relative to 
the base \citep{DR2_Chen+} models only for J1012+5307: the increase in the frequency range of the DMv process has constrained the RN to lower frequencies (the amplitude of the RN has slightly dropped while the spectral index has increased). We will revisit model selection for this pulsar in Section \ref{ssec:Stoch_det}.

\begin{table}
	\centering
	\caption{Favored number of Fourier modes for RN and DMv for the $6$ pulsars. The third and fourth columns compare the favored model with \textit{RN30\_DMv100}, respectively showing the Bayes factors and the Jensen-Shannon divergences for the RN amplitude (top) and spectral index (bottom). The models for PSR~J1713+0747 also include the $2$ exponential dips.}
	\label{tab:NBins}
	\begin{tabular}{cccc}
		\hline
		\multirow{2}{*}{Pulsar} & \multirow{2}{*}{Favored NBins} & \multirow{2}{*}{$\mathrm{log}_{10} \mathcal{B}^{\mathrm{Fav. NBins}}_{\mathrm{\textit{RN30\_DMv100}}}$} & \multicolumn{1}{|r|}{\multirow{2}{*}{J-S div.} $A_{\mathrm{RN}}$}\\
		& & & \multicolumn{1}{|r|}{$\gamma_{\mathrm{RN}}$}\\
		\hline \hline
		\multirow{2}{*}{J0613$-$0200} & \multirow{2}{*}{\textit{RN10\_DMv30}} & \multirow{2}{*}{$0.0$} & $2.31 \times 10^{-3}$\\
		& & & $1.32 \times 10^{-3}$\\
		\hline
		\multirow{2}{*}{J1012+5307} & \multirow{2}{*}{\textit{RN30\_DMv150}} & \multirow{2}{*}{$3.1$} & $2.68 \times 10^{-2}$\\
		& & & $2.99 \times 10^{-2}$\\
		\hline
		\multirow{2}{*}{J1600$-$3053} & \multirow{2}{*}{\textit{RN20\_DMv100}} & \multirow{2}{*}{$-0.3$} & $4.44 \times 10^{-3}$\\
		& & & $4.03 \times 10^{-3}$\\
		\hline
		\multirow{2}{*}{J1713+0747} & \multirow{2}{*}{\textit{RN15\_DMv150}} & \multirow{2}{*}{$6.3$} & $2.71 \times 10^{-3}$\\
		& & & $1.04 \times 10^{-3}$\\
		\hline
		\multirow{2}{*}{J1744$-$1134} & \multirow{2}{*}{\textit{RN10\_DMv100}} & \multirow{2}{*}{$-0.3$} & $4.85 \times 10^{-3}$\\
		& & & $4.66 \times 10^{-3}$\\
		\hline
		\multirow{2}{*}{J1909$-$3744} & \multirow{2}{*}{\textit{RN10\_DMv100}} & \multirow{2}{*}{$-0.1$} & $3.72 \times 10^{-3}$\\
		& & & $1.98 \times 10^{-3}$\\
		\hline
	\end{tabular}
\end{table}

\subsubsection*{Red noise free-spectrum of PSR~J1909$-$3744}
\begin{figure}
	\centering
	\hspace*{-.3cm}\includegraphics[keepaspectratio=true,scale=0.22]{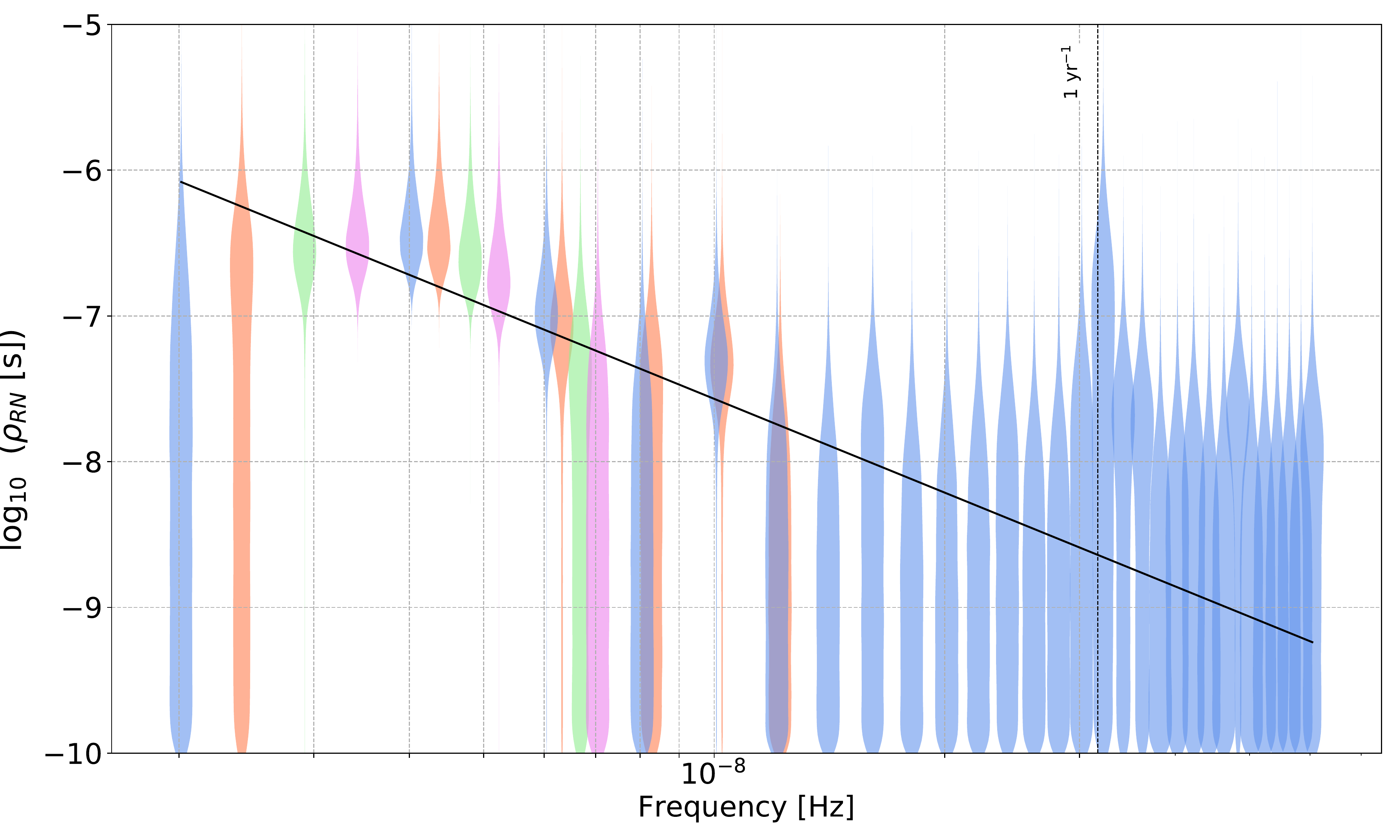}
	\caption{Achromatic red-noise spectrum of J1909$-$3744 using the model \textit{RN30\_DMv100}, with a power-law (black solid line drawn with the maximum a posteriori), and four free-spectrum (violin plots) PSDs computed with four different minimum frequencies : $i/\mathrm{T_{span}}$ with $i =$ $1$ (blue), $1.2$ (orange), $1.45$ (green) and $1.7$ (pink). Here, $30$ bins are drawn for the first one, $6$ for the second and $3$ for the two others.}
	\label{fig:spectrum_J1909}
\end{figure}

In this paragraph we discuss PSR~J1909$-$3744. This is one of the best timers, but
it has the shortest observational span (about 11 years) and the data was only acquired by NRT.
The free spectrum and the power-law (corresponding to the maximum a posteriori parameters) of the RN are plotted in
Fig.~\ref{fig:spectrum_J1909} . The blue violin plot shows the power distribution using our standard Fourier modes ($i/\mathrm{T_{span}}$).
We performed additional runs (given by different colours)  
with scaled-down timespan values as, $\mathrm{T_{span}} \to (\mathrm{T_{span}}/1.2, \mathrm{T_{span}}/1.45, \mathrm{T_{span}}/1.7$), to get better resolution at low frequencies.
Note that those frequency bins are not
independent as we have used over-sampling in the frequency domain, and the corresponding basis functions are not orthogonal even for evenly spaced data.  One can clearly see that the spectrum flattens out and probably bends downwards at the lowest frequency bin. This bend is not very conclusive: the posterior at the lowest bin is poorly constrained, and could be caused by the gaps in the data
($214$ epochs from $2004$ to $2011$ with BON vs. $695$ from $2011$ to $2020$ with NUPPI backends).
If this downturn is real, it could be related to processes intrinsic to the neutron star \citep{gon20}, or to a putative GWB produced by eccentric SMBHBs \citep{che17}. We require a longer timespan to better constrain the lowest frequencies and refine our interpretation.

\subsection{Extending model selection to stochastic and deterministic signals} \label{ssec:Stoch_det}
Let us now investigate the different red signals presented in Section \ref{sec:Sig}.
We start with the pulsar stochastic processes by (i) inspecting the presence of the chromatic signals in the data; and
(ii) performing model selection to obtain the most favoured signal combination with RN, DMv and Sv.
The selection of the optimal number of modes for Sv was done in a manner similar to the one described in the previous section. We have also checked that the selected number of basis functions for RN and DMv is still optimal and
found that to be the case for all pulsars except PSR~J1600$-$3053, which we will discuss separately.

Next, we fit for the presence of an annual chromatic signal in PSR~J0613$-$0200 data and exponential dips in PSR~J1713+0747 (as discussed in Section~\ref{ssec:detsig}). Finally, we search for the presence of system and band noise in each pulsar.

\subsubsection{Stochastic chromatic signals}

In the previous subsection we assumed the presence of RN and DMv and concentrated on choosing the number of basis functions  to describe the noise by a Gaussian Process. Now we fix the number of modes and check  
whether the data supports RN, DM and Sv noise components. The probed models and the Bayes factors (with respect to the most favourable model indicated by the bold zeros) are summarized in Table \ref{tab:BFchrom}.   
Below, we outline the procedure that we have followed. \par

\begin{figure}
	\centering
	\includegraphics[keepaspectratio=true,scale=0.22]{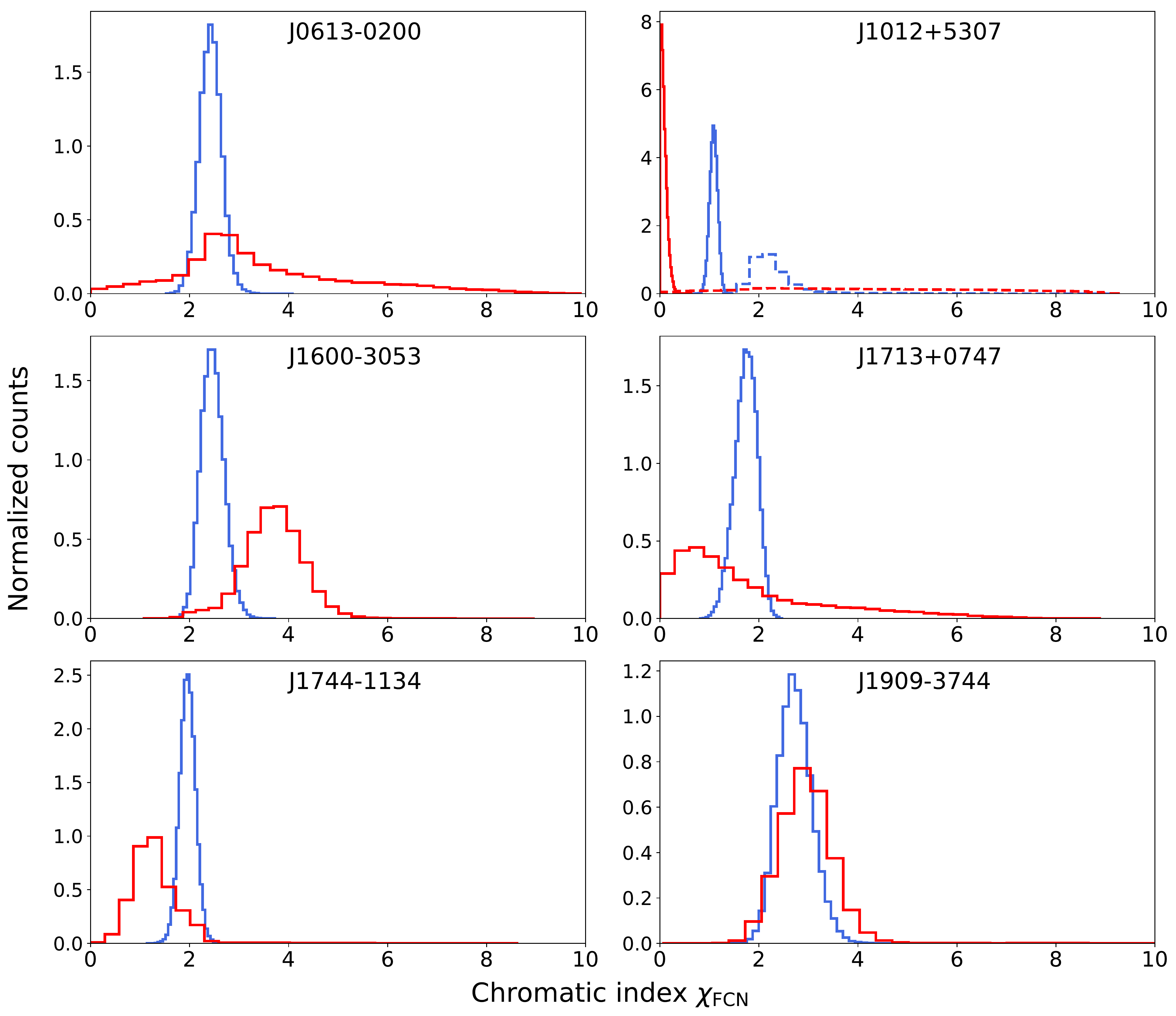}
	\caption{Marginalized posterior distributions of the chromatic index $\chi_{\mathrm{FCN}}$ in \textit{RN\_FCN} (solid blue) and \textit{RN\_DMv\_FCN} (solid red) for the $6$ pulsars. The number of frequency bins for the RN and DMv power-law are taken from Table \ref{tab:NBins}. For PSR~J1012+5307 we performed additional analyses  with models \textit{RN150\_FCN150} (dashed blue) and \textit{RN150\_DMv150\_FCN150} (dashed red).}
	\label{fig:CNidx}
\end{figure}

In parallel to the direct computation of the evidence for each model, we have also conducted a noise diagnostic by using a noise model with RN and FCN (free chromatic index), which covers RN (chromatic index $\chi_{\mathrm{FCN}} = 0$),
DMv ($\chi_{\mathrm{FCN}} = 2$) and Sv ($\chi_{\mathrm{FCN}} = 4$). The posterior on the chromatic index with \textit{RN\_FCN} is shown as blue histograms in Fig. ~\ref{fig:CNidx} and in most cases it is centred around 2, indicating the presence of DMv, with two exceptions: PSR~J1012+5307 (centred at 1.1) and J1909$-$3744 (centred at 3). We add DMv in our model and repeat the analysis with \textit{RN\_DMv\_FCN}. The FCN now captures the remaining noise not covered by DMv and is indicated by the red histograms in Fig. ~\ref{fig:CNidx}. PSR~J1600$-$3053 shows the presence of scattering noise ($\chi_{\mathrm{FCN}} = 4$). For this pulsar, variable and clumpy scintillation arcs in the secondary spectrum \citep[i.e., the power-spectrum of the dynamic spectrum, see e.g.,][]{cor86} are also seen in the L-band LEAP data (Main et al. in prep.), with power extending up to $16\,\mu$s in delay, and averaged time delays at the $100$ns level. These results add more confidence for the inclusion of this process, and indeed its presence is confirmed by the Bayes factor. However, including Sv into the noise model absorbs most of the red noise, making its presence inconclusive (as indicated in the table). PSRs~J0613$-$0200, J1713+0747 and J1744$-$1134 show no sign of scattering variation noise. The chromatic index remains unchanged for J1909$-$3744, and the model selection indicates (though not very strongly) the presence of both DMv and Sv. \par

\begin{figure*}
	\centering
	\begin{subfigure}{.4\textwidth}
		\centering
		\includegraphics[keepaspectratio=true,scale=0.4]{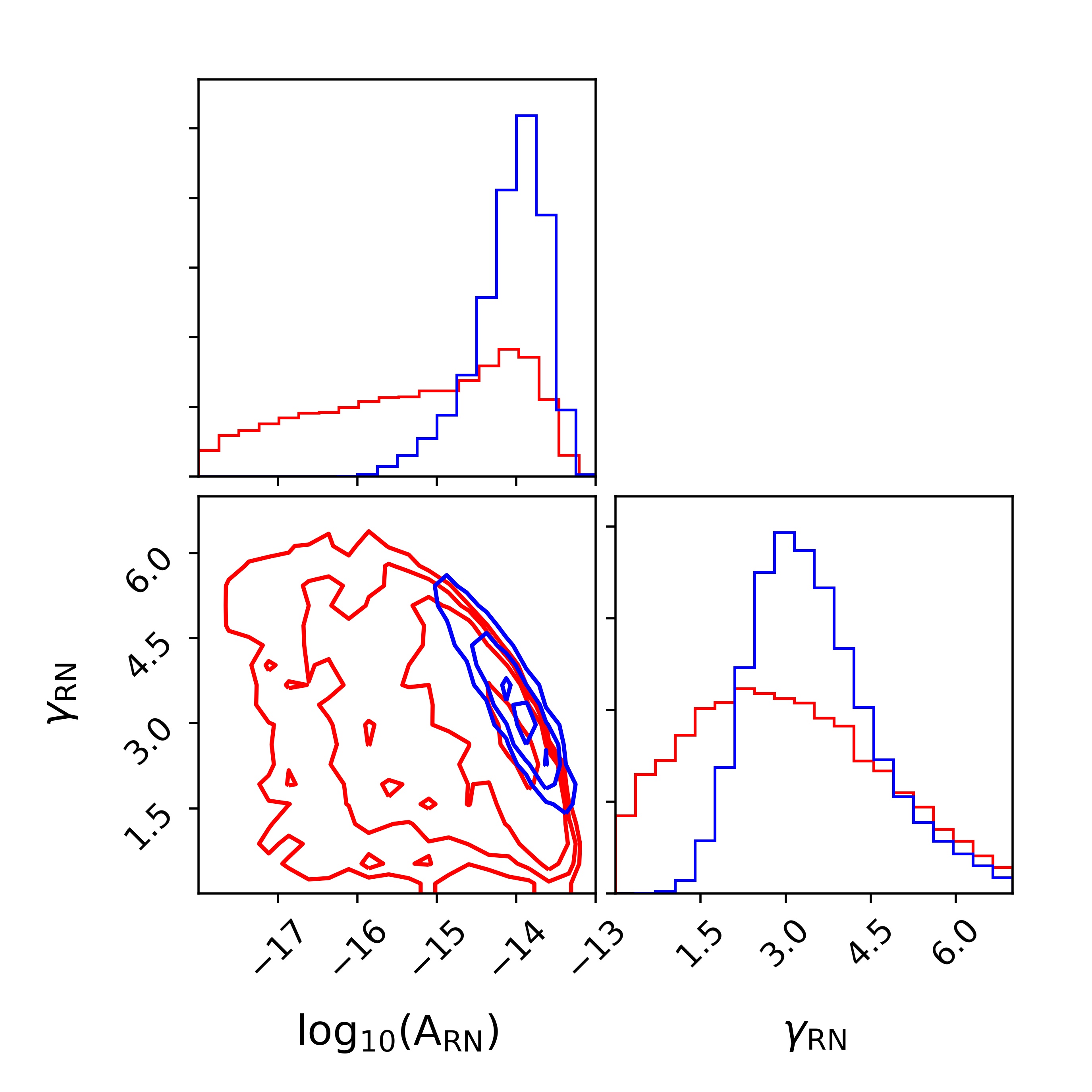}
	\end{subfigure}
	\begin{subfigure}{.4\textwidth}
		\centering
		\includegraphics[keepaspectratio=true,scale=0.4]{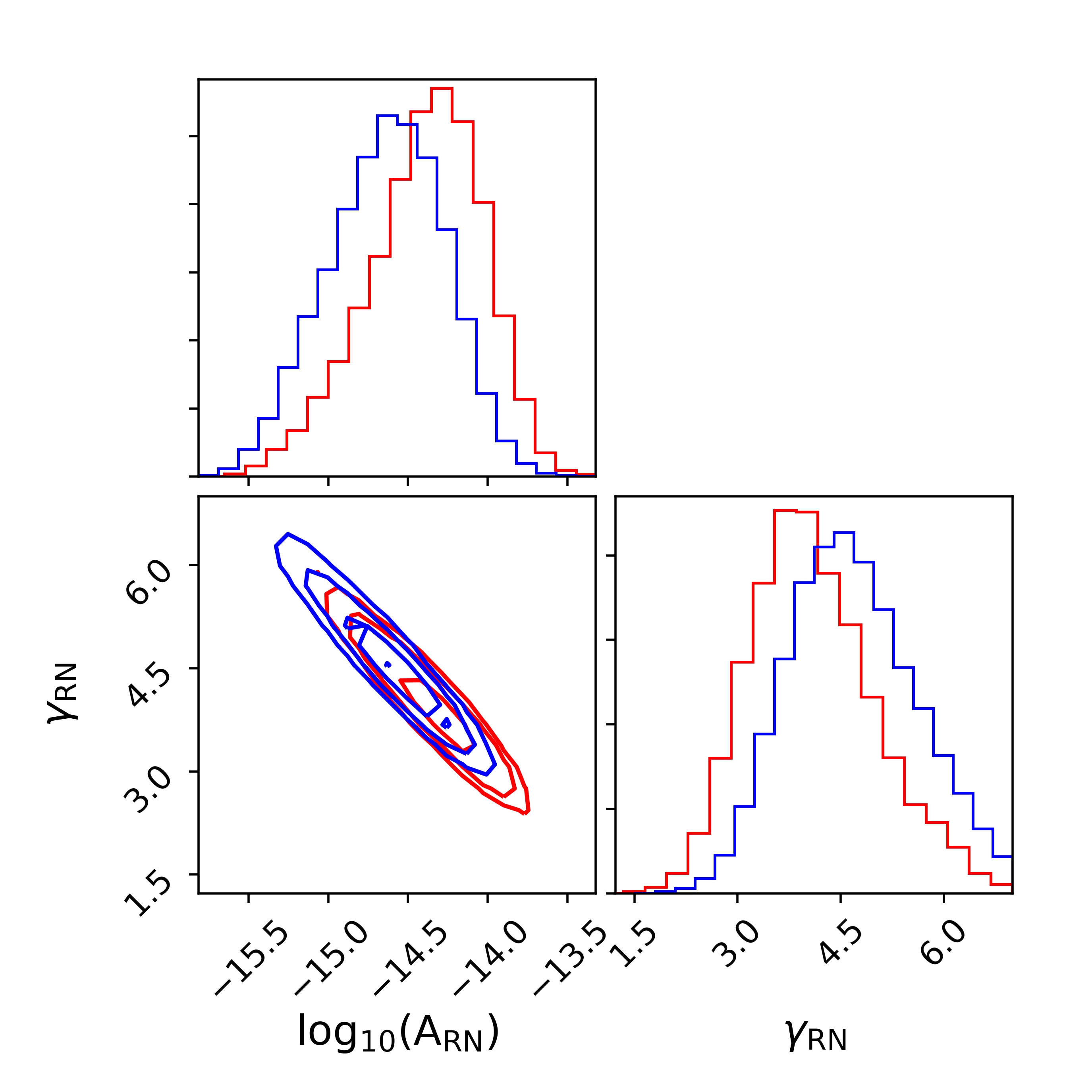}
	\end{subfigure}
	\caption{2D distributions of achromatic red-noise amplitude and spectral index.
		Left plot:  J1600$-$3053, the model \textit{RN\_DMv} is in blue and \textit{RN\_DMv\_Sv} is red ; Right  plot:  J1909$-$3744,  with \textit{RN\_DMv} in blue and \textit{RN\_DMv\_Sv} (favourable) in red.}
	\label{fig:cornerchrom_1}
\end{figure*}

It is important to see how the inclusion of Sv changes the RN properties.
In Fig.~\ref{fig:cornerchrom_1} we show the corner plot of the RN parameters for the model
\textit{RN\_DMv} in blue and the model \textit{RN\_DMv\_Sv} in red, for the two pulsars that favour Sv.
As mentioned above, the data is non-informative on the presence of RN in PSR~J1600$-$3053 if we add Sv to the model and this can be seen by the poorly  constrained posterior (red) in the left panel. For J1909$-$3744, the inclusion of Sv has a less drastic effect: it absorbs a small part of the RN at very low frequencies, which reduces the spectral index but pushes the amplitude slightly up.

\begin{table*}
	\centering
	\caption{Model selection for the stochastic chromatic signals. This table contains the log-10 Bayes factors of the highest evidence model ($\mathrm{log_{10}} \ \mathcal{B} = 0.0$) over the set of other models that we have tried (given in columns). The selected model is indicated in bold. We have used $150$ Fourier modes  for RN in PSR~J1012+5307 and 2 exponential dips are  always included in the analysis of PSR~J1713+0747.}
	\label{tab:BFchrom}
	\begin{tabular}{cccccccc}
		\hline
		Pulsar & \textit{RN} & \textit{DMv} & \textit{Sv} & \textit{RN\_DMv} & \textit{RN\_Sv} & \textit{DMv\_Sv} & \textit{RN\_DMv\_Sv} \\
		\hline \hline
		J0613$-$0200 & $-12.5$ & $-10.3$ & $-37.7$ & $\mathbf{0.0}$ & $-2.2$ & $-1.7$ & $-0.3$ \\
		\hline
		J1012+5307 & $-25.0$ & $-63.0$ & $-143.7$ & $\mathbf{0.0}$ & $-2.0$ & $-47.5$ & $0.4$\\
		\hline
		J1600$-$3053 & $-146.1$ & $-10.2$ & $-59.3$ & $-6.5$ & $-9.1$ & $\mathbf{0.0}$ & $0.0$ \\
		\hline
		J1713+0747 & $-36.8$ & $-42.6$ & $-125.0$ & $\mathbf{0.0}$ & $-30.0$ & $-28.5$ & $-0.8$ \\
		\hline
		J1744$-$1134 & $-12.1$ & $-3.1$ & $-27.9$ & $\mathbf{0.0}$ & $-10.7$ & $-2.4$ & $-1.8$ \\
		\hline
		J1909$-$3744 & $-66.7$ & $-82.1$ & $-244.4$ & $-2.1$ & $-3.4$ & $-21.0$ & $\mathbf{0.0}$ \\
		\hline
	\end{tabular}
\end{table*}

\subsubsection*{Peculiar red noise in PSR~J1012+5307}
\label{S:peculiar1012}

PSR~J1012+5307 required a special investigation given the rather strong signal in model \textit{RN\_DMv\_FCN} displayed in Fig.~\ref{fig:CNidx}, with chromaticity index close to zero. It turned out to be unaccounted RN at high frequencies. In our initial analysis for the number of Fourier components we concentrated on low frequencies (up to 30 bins) for the RN and the FCN in  \textit{RN\_DMv\_FCN}  picks up excess red noise which extends also to high frequencies. \par

\begin{figure}
	\centering
	\includegraphics[keepaspectratio=true,scale=0.45]{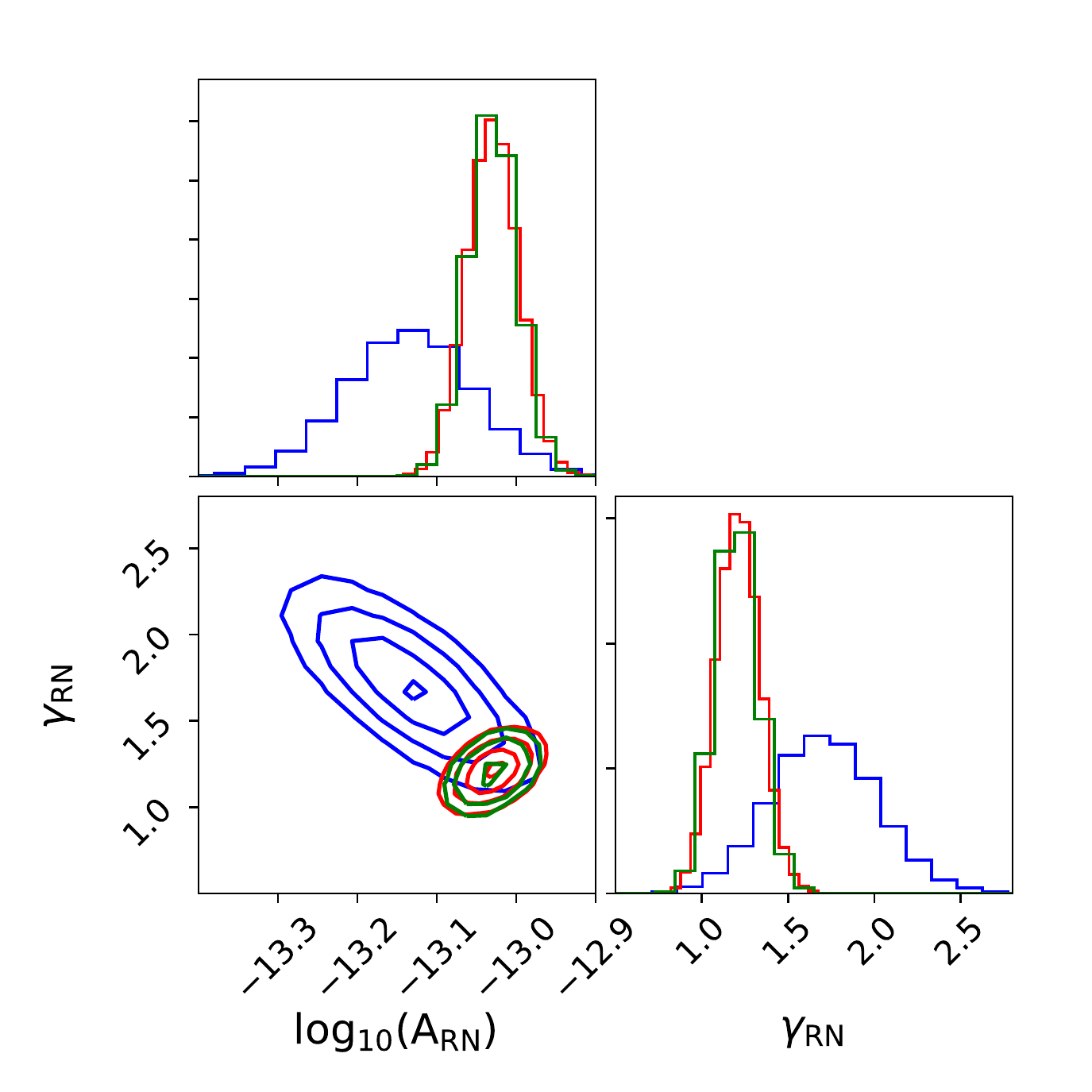}
	\caption{2D distribution of achromatic red-noise amplitude and spectral index for J1012+5307  with the noise models \textit{RN30\_DMv150} (blue), \textit{RN150\_DMv30} (red) and \textit{RN150\_DMv30\_SV150} (green).}
	\label{fig:cornerchrom_2}
\end{figure}

Based on these findings we revisited  the  selection of Fourier modes done in Section \ref{ssec:Nbins} for this pulsar by allowing the RN to go up to $150$ Fourier bins. The  favoured model \textit{RN150\_DMv30} has a Bayes factor  $\mathrm{log}_{10} \mathcal{B}^{\mathrm{\textit{RN150\_DMv30}}}_{\mathrm{\textit{RN30\_DMv150}}} = 29.8$ over the previous one, and therefore we adopt it in further investigations.
In Fig.~\ref{fig:cornerchrom_2} we show the evolution of the red noise parameters (amplitude and spectral index) as we move from \textit{RN30\_DMv150} (blue) to the new model  \textit{RN150\_DMv30} (red). The parameters of the RN process are better constrained, the red noise is significantly shallower (to accommodate the high frequency contribution) but the amplitude is slightly higher.  The free spectrum estimation for this pulsar can be seen in the second panel of the right hand column of Fig.~\ref{fig:timedom_freespec}. One can clearly see the low frequency red noise (well constrained power at the three lowest Fourier bins), but we also observe significant fluctuations at higher frequencies. The high frequency red noise is also evident in the time realization of this signal in the corresponding plot in the left hand column. The red noise at high frequencies flattens out the power-law of the overall RN process.  However, we do not exclude the possibility that the red noise comprises two components of different origin. \par

Using this number of modes, we have repeated the analysis of the data with the \textit{RN150\_FCN150} and \textit{RN150\_DMv30\_FCN150} models. The results are presented in Fig.~\ref{fig:CNidx} as dashed lines and confirm that  \textit{RN150\_DMv30} is sufficient to describe the data. The green histogram in Fig.~\ref{fig:cornerchrom_2} shows that adding Sv does not change the properties of the RN process. \par

\subsubsection{Deterministic chromatic signals}

The scattering and scintillation effects for J0613$-$0200 have been studied by \cite{mai20}, who also
discuss the presence of annual variations of the arc  curvatures. Furthermore, \cite{kei13} has shown the presence of annual chromatic signals which were reported in \cite{gon21},
suggesting the need for a noise model that includes an annual DM process (i.e., $\chi_{\mathrm{y}} =2$).
However, we did not find conclusive evidence for the presence of such a signal, with Bayes factors
$\mathcal{B}^{\mathrm{\textit{RN\_DMv\_AnnualDM}}}_{\mathrm{\textit{RN\_DMv}}} = 8.3$ and
$\mathcal{B}^{\mathrm{\textit{RN\_DMv\_AnnualSv}}}_{\mathrm{\textit{RN\_DMv}}} = 7.0$,
which we do not consider sufficiently significant to justify its inclusion in the noise model.
Note that (at least part of) the annual variations in the timing residuals might be absorbed by the TM parameter fit (through the pulsar sky position and proper motion). \\

\begin{figure*}
	\centering
	\begin{subfigure}{.45\textwidth}
		\centering
		\includegraphics[keepaspectratio=true,scale=0.3]{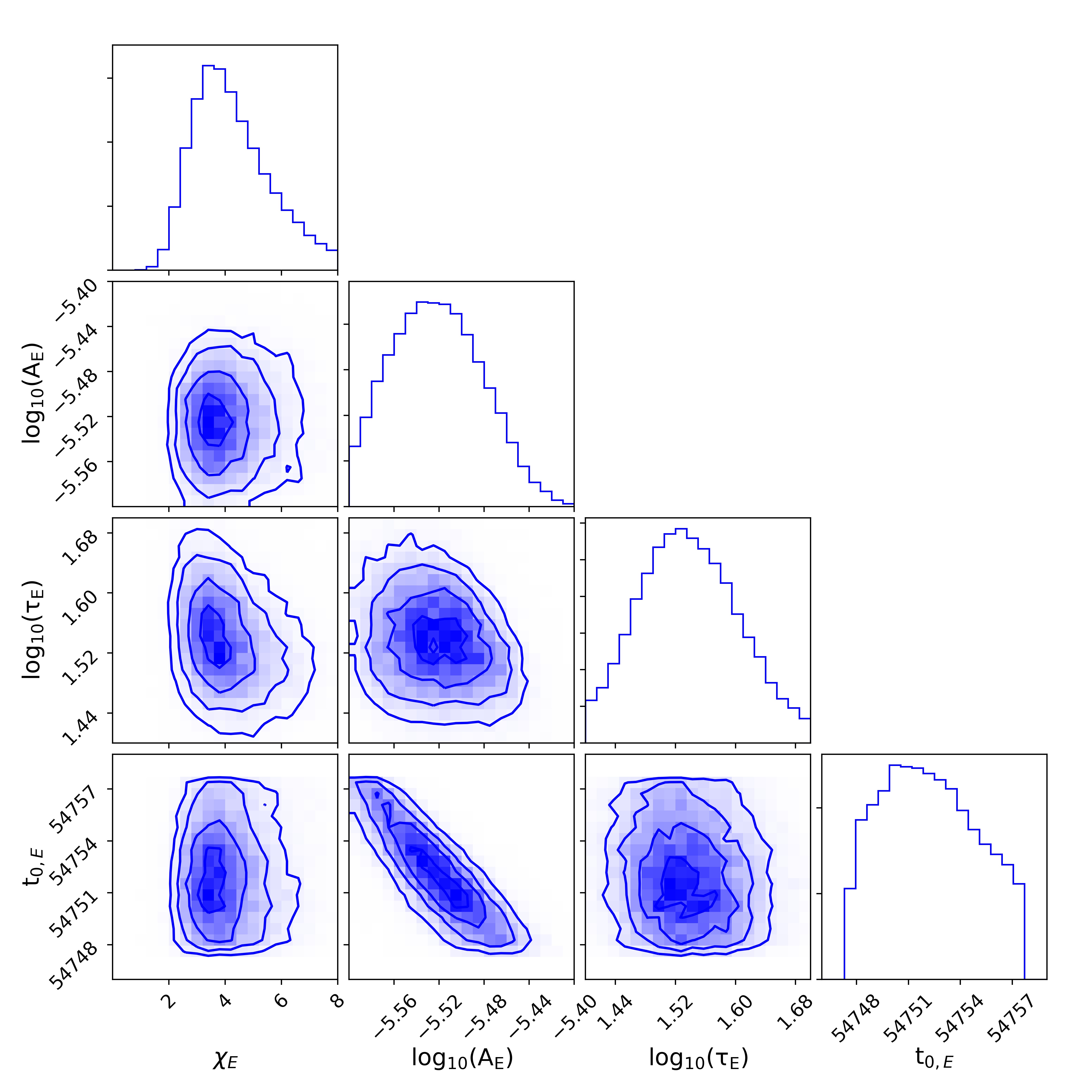}
	\end{subfigure}
	\begin{subfigure}{.45\textwidth}
		\centering
		\includegraphics[keepaspectratio=true,scale=0.3]{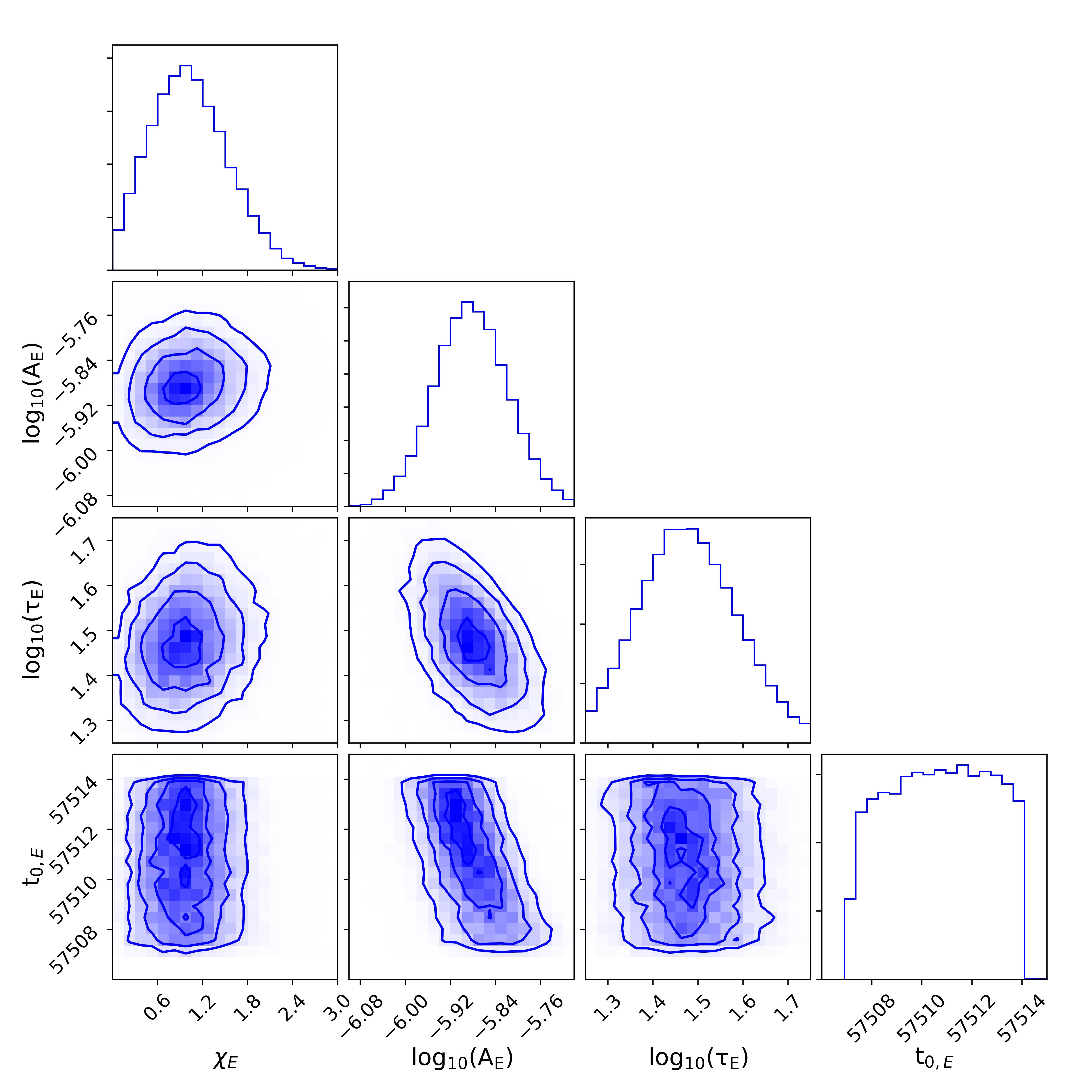}
	\end{subfigure}
	\caption{Posterior distributions  for the exponential dips found in J1713+0747
		at  MJD 54757 (left) and at MJD 57510 (right). The chromatic index $\chi_E$ was used as
		a model parameter.}
	\label{fig:cornerexpd}
\end{figure*}

As for the exponential  dip events, we found their presence in PSR~J1713+0747 data with high statistical confidence: the log-10 Bayes factors were  30.5, 13.9 and 46.8 favouring models that include respectively one single event at MJD 54757, one single event at MJD 57510 and both those events together. We found a chromatic index very consistent with scattering variations ($\chi_{\mathrm{E}_1} = 4.07^{+1.77}_{-1.13}$, with errors corresponding to the $68\%$ confidence interval) for the first event (left panel  of Fig. \ref{fig:cornerexpd}), and an index
$\chi_{\mathrm{E}_2} = 1.00^{+0.56}_{-0.49}$  for the second event.
The index for the second event is consistent with \cite{gon21}, who reported a profile change for this event and proposed a cause linked with the pulsar's magnetosphere instead of an IISM process.  The posteriors for both events are shown in Fig.~\ref{fig:cornerexpd}.
Note that the posteriors of $t_0$ for both epochs are sharply constrained between the two consecutive ToAs that surround the actual event dates and it is not railing against the prior range (see Table \ref{tab:priors}). For the rest of this work, we fix the chromatic indices of both events at $4$ and $1$ as discussed above.

\subsubsection{System and Band noise}
The dataset for each pulsar is made from the combination of ToAs produced with different receiver systems integrated in different radio telescopes. It might happen that one (or several)
of these systems unexpectedly introduce extra noise. The idea of the system-noise model is to check this hypothesis.\par

The EPTA data have a large number of systems, and checking all of them at the same time is computationally prohibitive.
Instead, we have used an approach based on the hyper-model selection framework \citep{hee15} to check for the presence of a noise in each system in turn.   We introduce a switch (hyper) parameter which regulates the prior on the amplitude of the system  noise being either significant/detectable or negligible.  Those two models for each system imply that SN is always present but it could be at a detectable level or not.  Note that the prior on the amplitude (log-uniform) is always taken into account and it has the same range for both models and therefore the main difference is in the likelihood, which gives some similarity to the dropout analysis \citep{arz20}. The posterior mean, $P_i$, of any given hyper-parameter indicates the probability of having red noise in the corresponding system. The ratio $P_i/(1-P_i)$ is the Bayes factor for the model including system noise in that component over the model without.\par

For this analysis, we exclude systems that have less than $3$ years of time span for any of the $6$ pulsars. The excluded systems are JBO.DFB.1400, JBO.DFB.1520, WSRT.P1.323.C, WSRT.P1.367.C, WSRT.P1.840.C, WSRT.P1.1380.C, WSRT.P1.1380.1, WSRT.P1.1380.2.C, NRT.BON.1600 and NRT.NUPPI.1854. We also do not investigate NRT.NUPPI.2154, which contributes $\sim 2.4$ years for J1600$-$3053 and about $4.7$ years for J1909$-$3744, but with $16$ epochs distributed in $1.1$ year, and only $2$ epochs $4.6$ years after.

The dataset of PSR~J1909$-$3744 is composed by ToAs produced only from NRT observations, with three systems (BON backend) before MJD $\sim 55812$ and four systems (NUPPI backend) after that date.
This means that any possible system red noise will be totally correlated with RN, which will absorb it. \par

In Table~\ref{tab:SN} we report the log-10 of the Bayes factor in favour of including red noise in that component, computed as described above.

We observe three systems NRT.NUPPI.1484, JBO.ROACH.1520 and LEAP.1396 with a significant  inclusion factor ($10^5$, $10^4$ and $10^4$ respectively). We notice that these are three L-band systems active between $2011$ and $2020$, which corresponds to the major part of the data sets. The  first two of the above mentioned systems (especially NRT.NUPPI.1484) are the largest contributors to the EPTA data and LEAP.1396 produces the ToAs with the lowest uncertainties (mean at $1.86 \ \mu s$ for PSR~J1012+5307 and lower than $0.55 \ \mu s$ for the others). For PSR~J1713+0747 two additional systems (EFF.S110.2639 and NRT.BON.2000) show signs of system noise, and it  might also be present in  NRT.NUPPI.2539 for PSRs J1012+5307 and J1744$-$1134.

\begin{table*}
	\centering
	\caption{Switch hyper-parameter  (in log10-scale) for the achromatic system noise. Considered systems are given in the second column. The values above 1 (shown in bold), indicate evidence for 
			the presence of red noise in that system. These systems were selected for a detailed analysis.}
	\label{tab:SN}
	\begin{tabular}{cccccccc}
		\hline
		Radio band & System & J0613$-$0200 & J1012+5307 & J1600$-$3053 & J1713+0747 & J1744$-$1134 & J1909$-$3744 \\
		\hline\hline
		\multirow{7.5}{*}{Band.1} & WSRT.P1.328 & $-0.18$ & $-0.23$ & - & - & - & - \\
		\cmidrule(l){2-8}
		\multirow{7.7}{*}{<$1$ GHz} & WSRT.P1.328.C & $-0.23$ & $-0.07$ & - & - & - & - \\
		\cmidrule(l){2-8}
		& WSRT.P1.382 & $-0.07$ & $-0.10$ & - & - & - & - \\
		\cmidrule(l){2-8}
		& WSRT.P1.382.C & $-0.02$ & $-0.01$ & - & - & - & - \\
		\cmidrule(l){2-8}
		& WSRT.P1.840 & - & - & - & $-0.15$ & - & - \\
		\cmidrule(l){2-8}
		& WSRT.P1.840.C & - & - & - & $-0.07$ & - & - \\
		\cmidrule(l){2-8}
		& WSRT.P2.350 & - & $-0.11$ & - & $-0.16$ & - & - \\
		\hline\hline
		\multirow{18.5}{*}{Band.2} & EFF.EBPP.1360 & $-0.16$ & $-0.15$ & - & $0.24$ & $-0.23$ & - \\
		\cmidrule(l){2-8}
		\multirow{18.7}{*}{[$1,2$] GHz} & EFF.EBPP.1410 & $0.41$ & $-0.16$ & - & $-0.18$ & $-0.36$ & - \\
		\cmidrule(l){2-8}
		& EFF.P200.1380 & - & $0.54$ & - & - & - & - \\
		\cmidrule(l){2-8}
		& EFF.P200.1400 & $-0.08$ & - & $-0.28$ & - & - & - \\
		\cmidrule(l){2-8}
		& EFF.P200.1400.np & - & - & - & $-0.44$ & - & - \\
		\cmidrule(l){2-8}
		& EFF.P217.1380 & - & $0.26$ & - & - & - & - \\
		\cmidrule(l){2-8}
		& EFF.P217.1400 & $-0.09$ & - & $-0.29$ & $0.32$ & $-0.14$ & - \\
		\cmidrule(l){2-8}
		& EFF.P217.1400.np & - & - & - & $-0.09$ & - & - \\
		\cmidrule(l){2-8}
		& JBO.ROACH.1520 (JBO\_1.5) & $\mathbf{1.93}$ & $\mathbf{2.48}$ & $0.01$ & $\geq \mathbf{5.00}$ & $\geq \mathbf{5.00}$ & - \\
		\cmidrule(l){2-8}
		& LEAP.1396 (LEAP\_1.4) & $\geq \mathbf{5.00}$ & $0.42$ & $\geq \mathbf{5.00}$ & $\geq \mathbf{5.00}$ & $\geq \mathbf{5.00}$ & - \\
		\cmidrule(l){2-8}
		& NRT.BON.1400 & $0.66$ & $-0.23$ & $-0.27$ & $-0.13$ & $0.62$ & $-0.33$ \\
		\cmidrule(l){2-8}
		& NRT.NUPPI.1484 (NUP\_1.4) & $\geq \mathbf{5.00}$ & $\mathbf{2.65}$ & $\mathbf{3.80}$ & $\geq \mathbf{5.00}$ & $\geq \mathbf{5.00}$ & $-0.41$ \\
		\cmidrule(l){2-8}
		& WSRT.P1.1380 & $-0.21$ & - & - & - & - & - \\
		\cmidrule(l){2-8}
		& WSRT.P1.1380.2 & - & $-0.18$ & - & $-0.22$ & - & - \\
		\cmidrule(l){2-8}
		& WSRT.P2.1380 & $-0.01$ & $-0.13$ & $-0.28$ & $0.56$ & $-0.16$ & - \\
		\hline\hline
		\multirow{7.5}{*}{Band.3} & EFF.EBPP.2639 & $0.05$ & $-0.29$ &  & $-0.20$ & $-0.19$ & - \\
		\cmidrule(l){2-8}
		\multirow{7.7}{*}{[$2,3$] GHz} & EFF.S110.2487 & - & $-0.30$ & - & - & $-0.17$ & - \\
		\cmidrule(l){2-8}
		& EFF.S110.2639 & $0.08$ & - & $-0.25$ & $\mathbf{2.63}$ & - & - \\
		\cmidrule(l){2-8}
		& NRT.BON.2000 (BON\_2.0) & $-0.24$ & $-0.23$ & $-0.01$ & $\geq \mathbf{5.00}$ & $-0.31$ & $-0.30$ \\
		\cmidrule(l){2-8}
		& NRT.NUPPI.2539 (NUP\_2.5) & $-0.23$ & $\mathbf{1.53}$ & $-0.34$ & $-0.24$ & $\mathbf{1.48}$ & $-0.40$ \\
		\cmidrule(l){2-8}
		& WSRT.P1.2273.C & - & - & - & $-0.10$ & - & - \\
		\cmidrule(l){2-8}
		& WSRT.P2.2273 & - & - & - & $-0.03$ & - & - \\
		\hline\hline
		\multirow{1.2}{*}{Band.4}& EFF.S60.4850 & - & - & - & $-0.06$ & - & - \\
		\cmidrule(l){2-8}
		\multirow{1.3}{*}{>$3$ GHz} & EFF.S60.4857 & - & $-0.16$ & - & - & $0.13$ & - \\
		\hline
	\end{tabular}
\end{table*}
The systems with an inclusion factor above $10$ (presented in bold in Table~\ref{tab:SN}) were selected for a detailed analysis
of  all possible combinations of system noise.
We have found that DM-type chromatic system noise (DM-SN) is always favoured (in terms of Bayes factor) over the achromatic SN for the NRT.NUPPI.1484 system
and it is included in the total noise budget for  PSRs J0613$-$0200, J1012+5307, J1713+0747 and J1744$-$1134. It is not entirely clear why the data does not support its presence in J1600$-$3053 (where we have identified SN only in
LEAP.1396). One plausible explanation is that the NRT.NUPPI.1484 ToAs dominate the data for this pulsar and to clearly disentangle SN from RN we would need to include data from other PTAs to test this assumption (that is a plan for the future IPTA data combination, repeating the previous such effort in \cite{len16}).

\begin{figure}
	\centering
	\begin{subfigure}{.4\textwidth}
		\centering
		\includegraphics[keepaspectratio=true,scale=0.5]{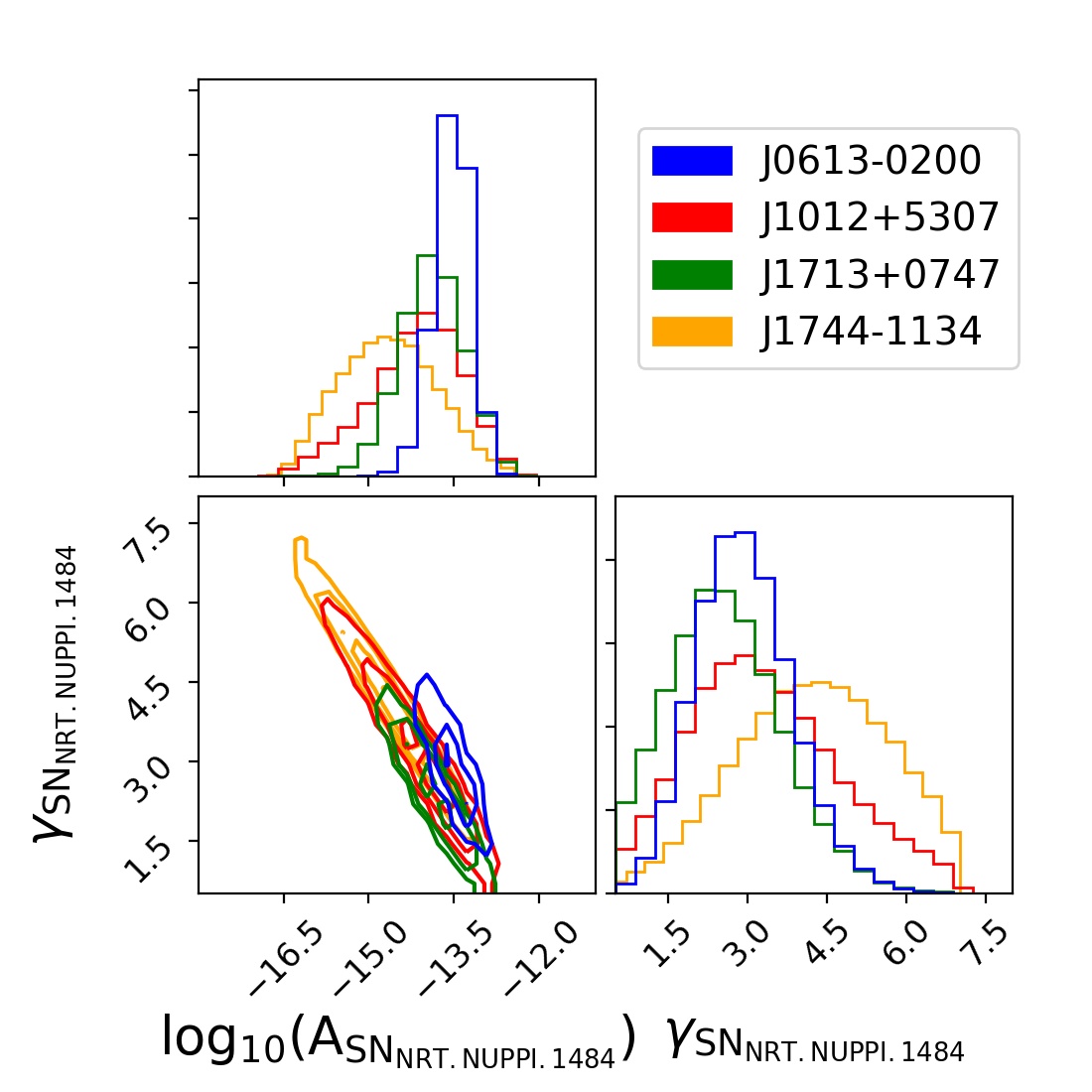}
	\end{subfigure}
	\begin{subfigure}{.4\textwidth}
		\centering
		\includegraphics[keepaspectratio=true,scale=0.5]{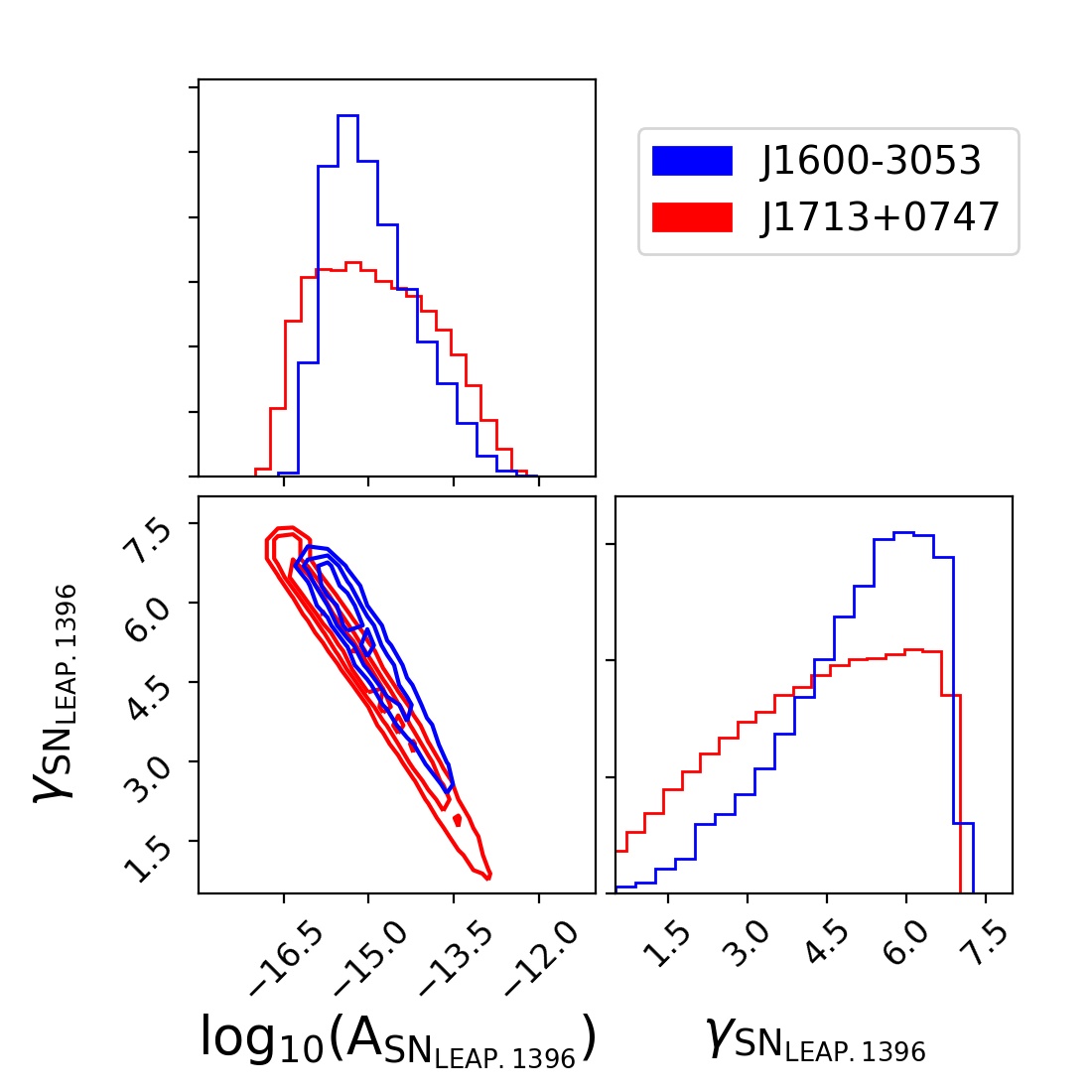}
	\end{subfigure}
	\caption{2D distributions of amplitude and spectral slope of NRT.NUPPI.1484 DM-SN (top) and LEAP.1396 SN (bottom), using the final favored model for each pulsar.}
	\label{fig:cornerSNBN}
\end{figure}

Polarization calibration errors and radio frequency interference are possible causes for SN. Parameter posteriors for the SN  of NRT.NUPPI.1484 and LEAP.1396 (Figure \ref{fig:cornerSNBN}) display overall consistency across pulsars, which corroborates the 
assumption of a red noise specific to these systems. We should emphasize again the presence of data from other systems (as expected in the IPTA data set) should greatly help to identify the system noise and disentangle it from the RN as was demonstrated in  \cite{len16} for PSR~J1730$-$2304 using the IPTA DR1 dataset.

The results of the SN selection are presented in Table~\ref{tab:finalMS}: inclusion of the SN leads to log-10 Bayes factors  ($15.9$, $7.4$, $13.8$, $118.1$ \& $18.6$) for PSRs J0613$-$0200, J1012+5307, J1600$-$3053, J1713+0747 and J1744$-$1134 respectively . \\

We switch now to the band noise investigation. The main radio frequency bands (cf. Table \ref{tab:SN}) in the datasets are Band.2 and Band.3, which contain the bulk of observations for all pulsars. Band.1 and Band.4 are only covered by one telescope:  
the WSRT and Effelsberg, respectively. As for SN, 
we do not investigate for BN if the corresponding time span of
observations is less than three years. Unfortunately, this made investigations in Band.1 and Band.4 inconclusive due to lack of sufficient data. \par

We found evidence of BN only in Band.3 of  PSR~J1713+0747 and Band.2 of PSR~J1744$-$1134,  with corresponding log-10 Bayes factors of $6.5$ and $6.2$. This result is somewhat
consistent with \cite{gon21}, where Band-noise for both Band.2 and Band.3 were reported for these two pulsars. However, we have found that the inclusion of Band.2 for PSR~J1713+0747 and Band.3 for PSR~J1744$-$1134 results in insignificant log-10 Bayes factors, $0.6$ in both cases. 
This points again to the importance of the IPTA data combination for the band noise analysis. Our results for BN are summarized in Table \ref{tab:finalMS}.

\subsubsection*{Poor constraint on RN for PSR~J1744$-$1134}
\label{S:peculiar1744}

\begin{figure}
	\centering
	\includegraphics[keepaspectratio=true,scale=0.5]{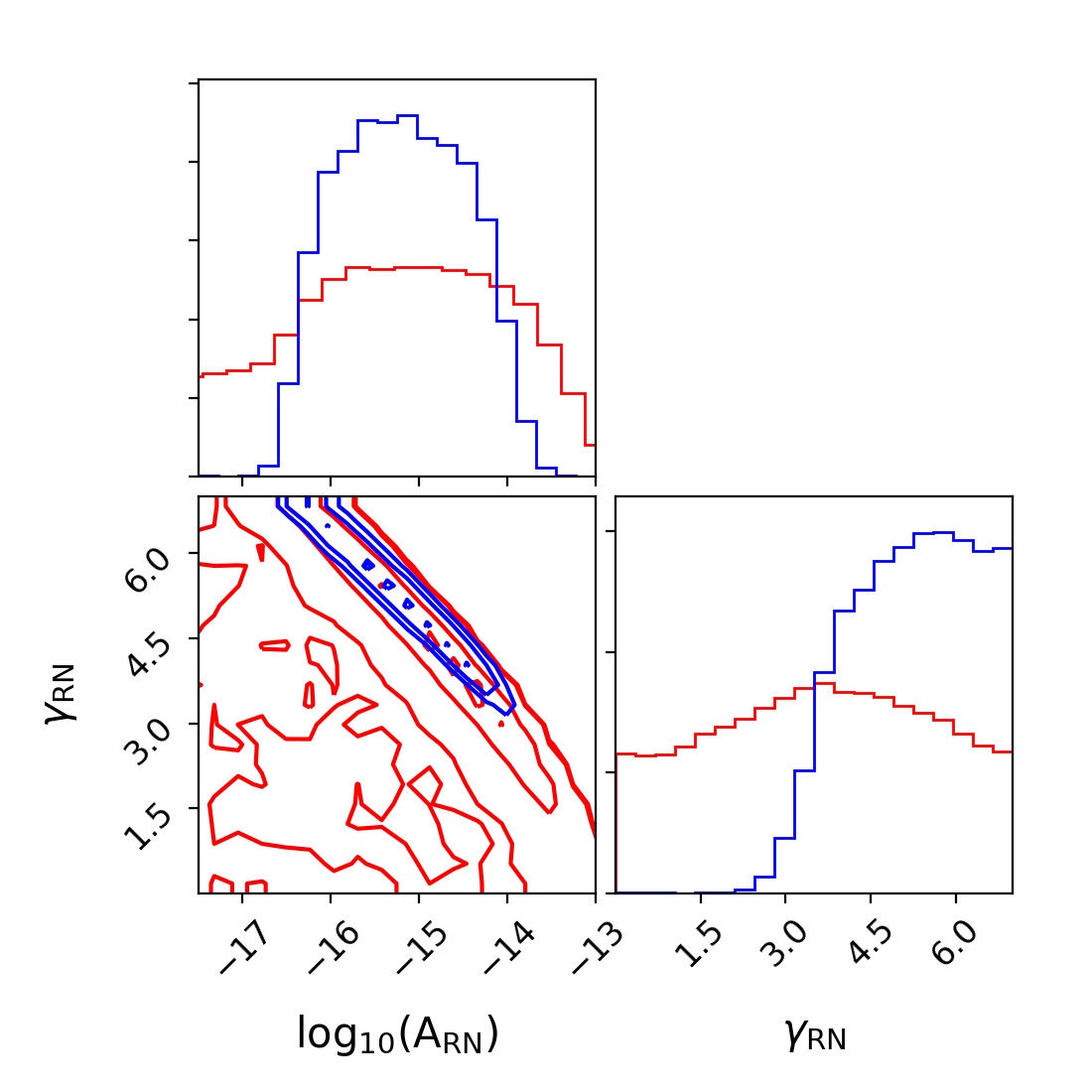}
	\caption{2D posterior distribution of the red-noise parameters for PSR~J1744$-$1134 for model \textit{RN\_DMv} (blue) and the most favored model, \textit{RN\_DMv\_SN\_BN} (red).}
	\label{fig:cornerJ1744}
\end{figure}

We revisited Table \ref{tab:BFchrom} after finding and fixing the set of noise sources included for each pulsar. 
We noticed that the RN for PSR~J1744$-$1134 becomes poorly constrained using the \textit{RN\_DMv\_SN\_BN} model (see red histograms in Fig. \ref{fig:cornerJ1744}).  After further investigation, we found that the Bayes factor $\mathcal{B}^{\mathrm{\textit{RN\_DMv\_SN\_BN}}}_{\mathrm{\textit{DMv\_SN\_BN}}} = 2$ hardly supports the presence of the RN. A similar result was found in \cite{gon21}, where the RN of this pulsar does not enter the favoured noise model. As another confirmation, the time-domain noise realizations of the red noise signal (see Fig. \ref{fig:timedom_freespec}) are quite reduced for the \textit{RN\_DMv\_SN\_BN} model (light grey) as compared to \textit{RN30\_DMv100} (red).  We have decided to keep the  \textit{RN\_DMv\_SN\_BN} model (as a conservative assumption), however we will address its impact when we discuss the common red noise.

\subsection{Summary of the model selection} \label{ssec:Sum}

The results  of the noise model selection for each pulsar are summarized in Table \ref{tab:finalMS}.  We report the RN parameters for each model, giving the
median and 95\% confidence interval.  We also compare the
custom-build noise model (M2) with the default noise model (M1) used in~\cite{DR2_Chen+}.
We quote the log-10 Bayes factors showing a strong preference for the custom noise model (ranging from 2 to 195). The biggest impact on the Bayes factor was caused by inclusion of the SN and BN components. Finally we have  evaluated the Anderson-Darling diagnostic $A^2$ \citep{and52} for the whitened residuals, obtained after subtracting the time domain noise realization drawn from the maximum likelihood from the residuals and dividing by the ToA uncertainties.
While the Bayes factor
measures the relative closeness of the whitened residuals to a Gaussian distribution, it does not tell us if the final result is actually Gaussian (in other words, if the favoured model is actually good). The Anderson-Darling statistic addresses precisely this question. $A^2=2.5$ is a value at which one fails to reject the null hypothesis (i.e., following a Gaussian distribution) at $95\%$ confidence level. A lower value of the statistic corresponds to a better agreement with a Gaussian distribution.  We see the overall improvement for the M2 model.
The high values for PSR~J1713+0747 could be caused by a few outliers in the whitened residuals. To test this assumption, we recompute the statistic after removing outliers found with the Grubbs' test \citep{gru50}, obtaining $3.2$ and $2.1$ respectively for M1 (with $9$ outliers) and M2 (with $6$ outliers). The Anderson-Darling statistic for PSR~J1744$-$1134 reduces to 1.0 if we exclude RN from the favourable (M2) model.
\par

\begin{table*}
	\centering
	\caption{Final noise models for the six pulsars in EPTA DR2. The third and fourth columns show the median of the RN amplitude ($\mathrm{A_{RN}}$) and spectral slope ($\mathrm{\gamma_{RN}}$), with corresponding $95\%$ confidence interval. The fifth column displays the $\mathrm{log}_{10}$ Bayes factors of the custom model over the model \textit{RN30\_DMv100}. The sixth and seventh columns show the Anderson-Darling statistic applied to the whitened residuals for (i) \textit{RN30\_DMv100}; and (ii) the selected noise model. The last column shows the Jensen-Shannon divergences of the RN amplitude and spectral index posteriors of the selected noise model relative to the same model with fixed white-noise parameters.}
	\label{tab:finalMS}
	\begin{tabular}{cccccccc}
		\hline
		\multirow{2}{*}{Pulsar} & \multirow{2}{*}{Sel. model} & \multirow{2}{*}{$\mathrm{A_{RN}}$} & \multirow{2}{*}{$\mathrm{\gamma_{RN}}$} & \multirow{2}{*}{$\mathrm{log}_{10} \ \mathcal{B}^{\mathrm{M2}}_{\mathrm{M1}}$} & \multirow{2}{*}{$\mathrm{A^2_{M1}}$} & \multirow{2}{*}{$\mathrm{A^2_{M2}}$\texttt{}} & \multicolumn{1}{|r|}{\multirow{2}{*}{$\mathrm{J}$-$\mathrm{S \ div_{WNf}}$} $A_{\mathrm{RN}}$}\\
		& & & & & & & \multicolumn{1}{|r|}{$\gamma_{\mathrm{RN}}$}\\
		\hline \hline
		\multirow{2}{*}{J0613$-$0200} & \textit{RN10 DMv30} & \multirow{2}{*}{$-14.93^{+1.26}_{-1.17}$} & \multirow{2}{*}{$5.07^{+1.83}_{-2.34}$} & \multirow{2}{*}{$15.9$} & \multirow{2}{*}{$0.4$} & \multirow{2}{*}{$0.3$} & $2.36 \times 10^{-3}$\\
		& \textit{DMv-SN\_NUP\_1.4} & & & & & & $1.82 \times 10^{-3}$\\
		\hline
		\multirow{3}{*}{J1012+5307} & \textit{RN150 DMv30} & \multirow{3}{*}{$-13.03^{+0.08}_{-0.08}$} & \multirow{3}{*}{$1.16^{+0.32}_{-0.29}$} & \multirow{3}{*}{$40.7$} & \multirow{3}{*}{$1.8$} & \multirow{3}{*}{$1.4$} & \multirow{2}{*}{$6.60 \times 10^{-4}$}\\
		& \textit{DMv-SN\_NUP\_1.4} & & & & & & \multirow{2}{*}{$3.80 \times 10^{-4}$}\\
		& \textit{SN\_NUP\_2.5} & & & & & &\\
		\hline
		\multirow{2}{*}{J1600$-$3053} & \textit{DMv30 Sv150} & \multirow{2}{*}{-} & \multirow{2}{*}{-} & \multirow{2}{*}{$20.0$} & \multirow{2}{*}{$0.3$} & \multirow{2}{*}{$0.2$} & -\\
		& \textit{SN\_LEAP\_1.4} & & & & & & -\\
		\hline
		\multirow{7}{*}{J1713+0747} & \textit{RN15 DMv150} & \multirow{7}{*}{$-14.50^{+0.51}_{-0.86}$} & \multirow{7}{*}{$3.94^{+1.82}_{-1.13}$} & \multirow{7}{*}{$195.4$} & \multirow{7}{*}{$5.5$} & \multirow{7}{*}{$4.1$} & \multirow{6}{*}{$1.76 \times 10^{-3}$}\\
		& \textit{$2$ Exp. dips} & & & & & & \multirow{6}{*}{$1.92 \times 10^{-3}$}\\
		& \textit{DMv-SN\_NUP\_1.4} & & & & & & \\
		& \textit{SN\_JBO\_1.5} & & & & & &\\
		& \textit{SN\_LEAP\_1.4} & & & & & &\\
		& \textit{SN\_BON\_2.0} & & & & & &\\
		& \textit{BN\_Band.3} & & & & & &\\
		\hline
		\multirow{3}{*}{J1744$-$1134} & \textit{RN10 DMv100} & \multirow{3}{*}{$-15.31^{+2.03}_{-2.50}$} & \multirow{3}{*}{$3.68^{+3.13}_{-3.46}$} & \multirow{3}{*}{$22.6$} & \multirow{3}{*}{$1.0$} & \multirow{3}{*}{$1.2$} & \multirow{2}{*}{$4.39 \times 10^{-3}$}\\
		& \textit{DMv-SN\_NUP\_1.4} & & & & & & \multirow{2}{*}{$1.79 \times 10^{-3}$}\\
		& \textit{BN\_Band.2} & & & & & &\\
		\hline
		\multirow{2}{*}{J1909$-$3744} & \multirow{2}{*}{\textit{RN10 DMv100 Sv150}} & \multirow{2}{*}{$-14.45^{+0.66}_{-0.85}$} & \multirow{2}{*}{$4.22^{+2.16}_{-1.65}$} & \multirow{2}{*}{$2.1$} & \multirow{2}{*}{$0.8$} & \multirow{2}{*}{$0.6$} & $3.10 \times 10^{-4}$\\
		& & & & & & & $3.20 \times 10^{-4}$\\
		\hline
	\end{tabular}
\end{table*}

In preparation for the next section where we consider a common red signal, we investigated how much the white noise parameters impact the measurement of the RN. Following \cite{len15}, we have performed a noise analysis with all white noise parameters fixed to the maximum-likelihood values and compared the RN posteriors to the previously obtained results.
The results are quoted as Jensen-Shannon divergences (last column of Table \ref{tab:finalMS}) and show a very good consistency ( $\mathrm{J}$-$\mathrm{S} \ div< 3\times 10^{-3}$). This confirmed that we can safely fix the white noise parameters for further investigations. 


\section{Impact on the search for a common red noise} \label{sec:Com}
In this section, we investigate how the custom single-pulsar noise model  affects the results of the common red signal 
analyses reported in \cite{DR2_Chen+}. We consider here the CRS either with Hellings-Downs spatial correlations  \citep[GWB; ][]{hel83} or without (CURN). \par

As in the previous section, we denote the default base noise model 
used for all pulsars in \cite{DR2_Chen+}, $RN30\_DMv100$, as M1, and label the custom models (summarized in Table~\ref{tab:finalMS}) for each pulsar as M2, and fix the white-noise parameters 
to their maximum-likelihood values (in the corresponding models). 
We model the common red noise using 30 uniformly spaced Fourier modes
$f_k = k/\mathrm{T_{tot}},\;\; k=1...30$, where $\mathrm{T_{tot}}$ is the time span between the lowest and highest epoch from the combined data of all pulsars.


\subsection{Contribution of each pulsar to the common red signal}
Following \cite{arz20}, \cite{gon21b} and \cite{DR2_Chen+}, we study the contribution of each pulsar dataset to the inferred presence of a common red process using dropout analysis. One additional parameter is added to the model for each pulsar, with a uniform prior, and these are sampled as part of the model. When the parameter is one, the common red signal is included in the model for that pulsar used in the likelihood and when it is zero it is not. The dropout factor is the ratio of the fraction of posterior samples when the CRS is included to the fraction when it is not included. 

\begin{figure}
	\centering
	\includegraphics[keepaspectratio=true,scale=0.4]{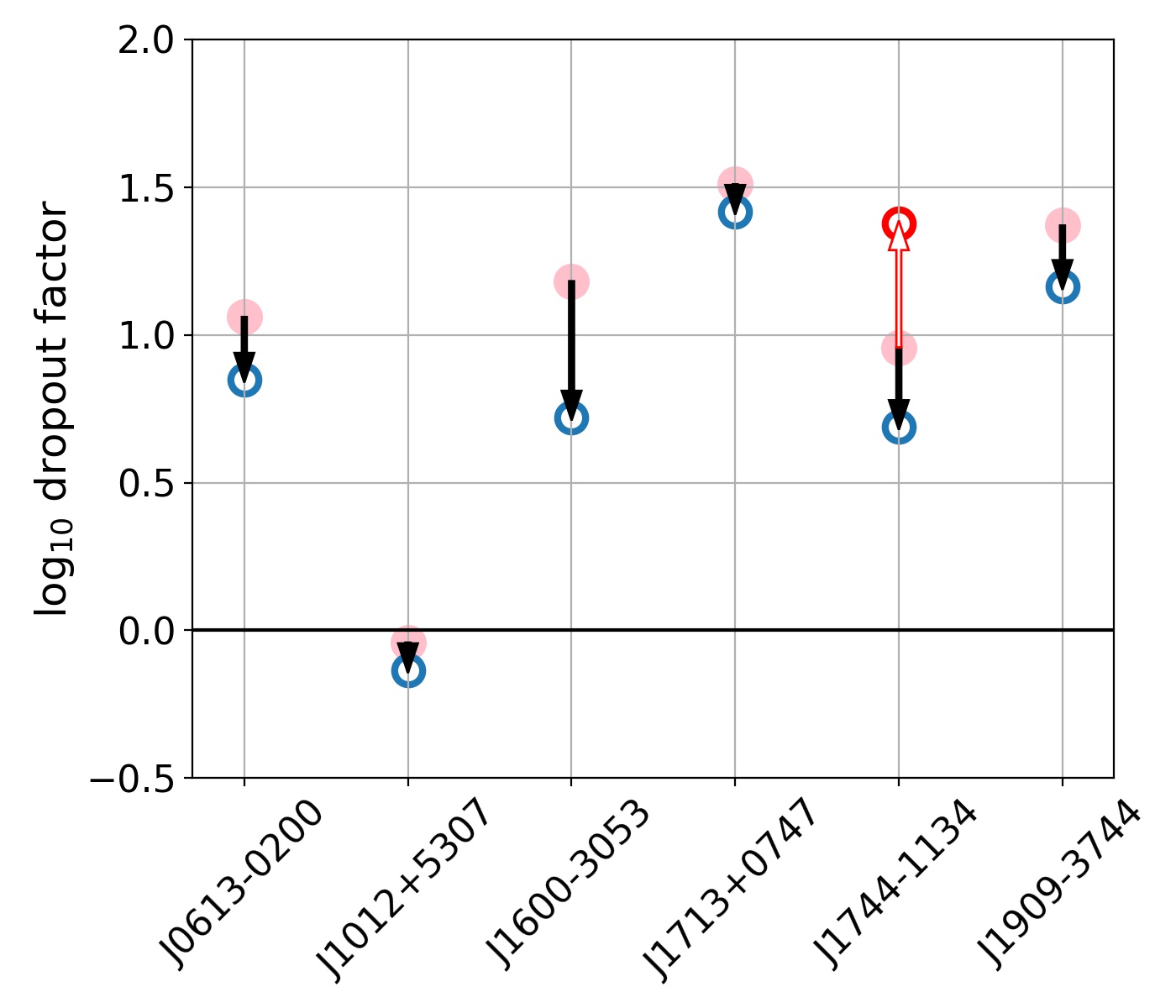}
	\caption{Dropout score for the contribution of each pulsar to the CURN model with M1 (pink dots) or M2 (empty blue dots). The same analysis for PSR~J1744$-$1134 but without intrinsic red-noise for this pulsar is also shown (empty red dot).}
	\label{fig:CRSDrop}
\end{figure}

The resulting dropout factors (see Fig. \ref{fig:CRSDrop}, blue hollow circles) with  M1+CURN are very similar to those presented in Figure $5$ in \cite{DR2_Chen+}. The dropout factor for PSR~J1012+5307  is around 1 (consistent with the results of \citet{arz20}) and most likely caused by abnormal red noise at high frequency
(see discussion in \ref{S:peculiar1012}) . \par

The contribution of each pulsar to the common red noise has decreased for the custom model M2+CURN with the biggest drop shown for PSR~J1600$-$3053 (the one which did not support RN).
Despite that the overall result remains: these pulsars support the presence of a CURN.
Interestingly, if we discard RN from the M2+CURN model of PSR~J1744$-$1134  (see discussion  in subsection~\ref{S:peculiar1744}), it is picked up by the CURN leading to an increase in the drop-out factor (see hollow red circle in Fig.~\ref{fig:CRSDrop}).  Note that this poorly constrained (and poorly understood) signal could potentially affect the sensitivity to the GWB. Our choice to keep RN inside the M2 model for PSR~J1744$-$1134 was a conservative choice.

\subsection{Spectral properties of common red signal}
\begin{figure}
	\centering
	\caption{2D posterior distributions of the CURN (top) and Hellings-Down correlated GWB (bottom) power-law parameters with M1 (blue) and M2 (red) single-pulsar noise models.}
	\label{fig:cornerCRS}
	\begin{subfigure}{.45\textwidth}
		\centering
		\includegraphics[keepaspectratio=true,scale=0.5]{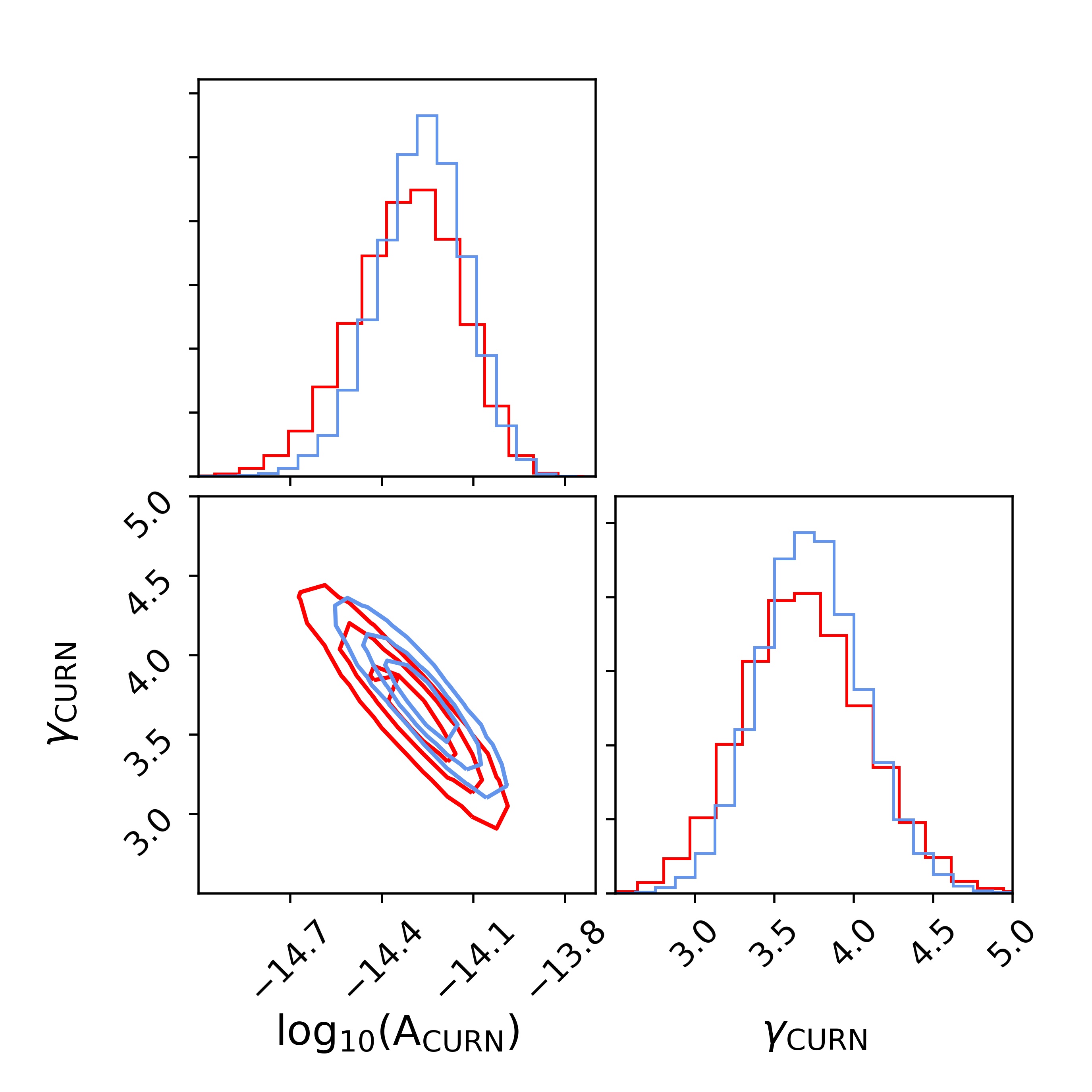}
	\end{subfigure}
	\begin{subfigure}{.45\textwidth}
		\centering
		\includegraphics[keepaspectratio=true,scale=0.5]{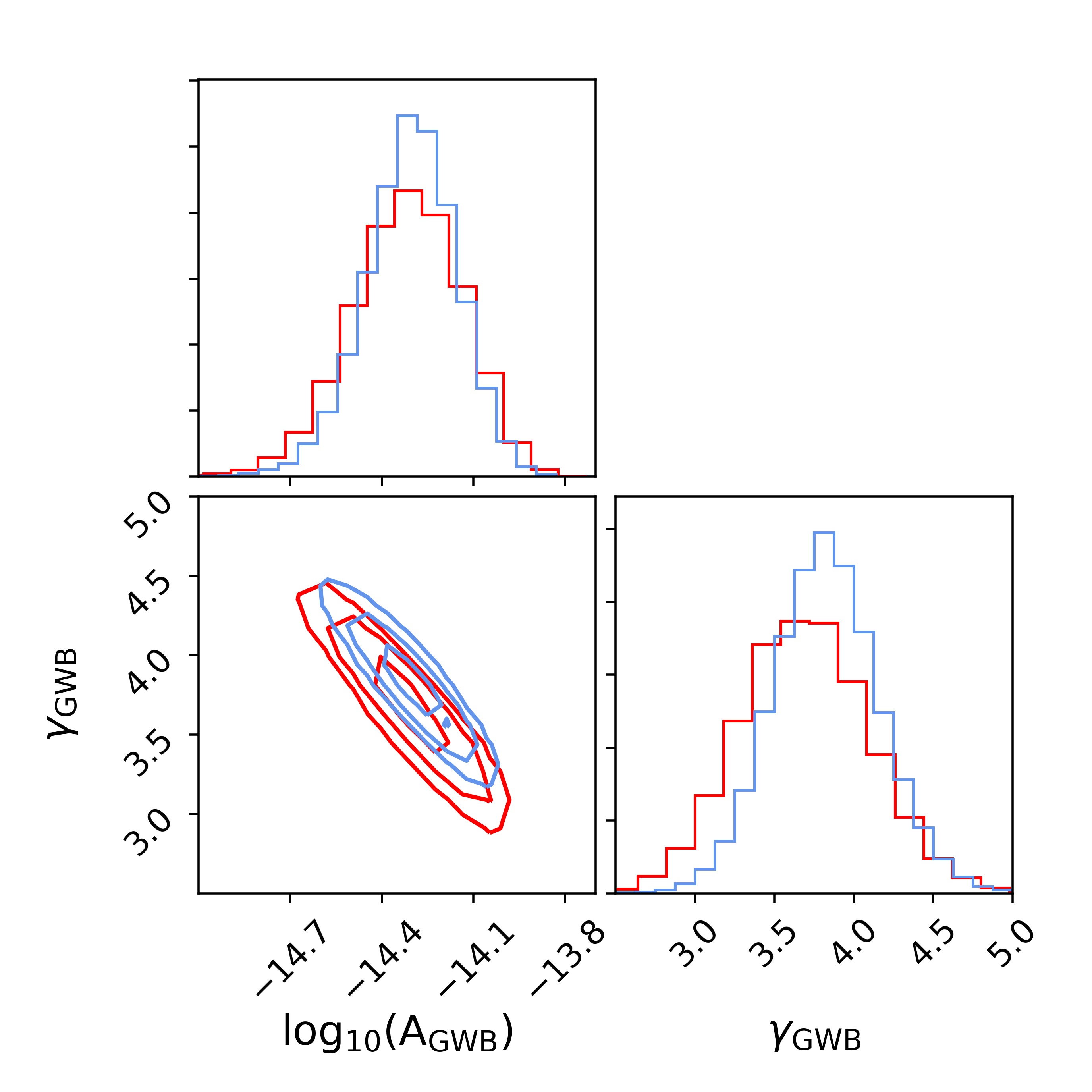}
	\end{subfigure}
\end{figure}

Figure \ref{fig:cornerCRS} and Table~\ref{tab:valCURN} summarize the spectral properties of the CURN and GWB. We do not see significant changes in the new results (red) from those found previously~\citep[blue,][]{DR2_Chen+}. The amplitudes of both CURN and GWB are slightly lower and the spectral indices are a bit shallower with the M2 models.
The median amplitude is reduced from \mbox{$A_{\mathrm{CURN, M1}} = 5.42^{+4.48}_{-2.81} \times 10^{-15}$} to \mbox{$A_{\mathrm{CURN, M2}} = 4.88^{+4.94}_{-2.85} \times 10^{-15}$} ($95\%$ credible interval) and the uncertainties of the spectral index are somewhat larger. The shift in the amplitude of the CURN  is most likely due to partial absorption of the  noise components which are unmodelled in M1, and which are accounted for in the extended set of M2.
Similarly for the GWB we have \mbox{$A_{\mathrm{GWB, M1}} = 5.01^{+4.34}_{-2.63} \times 10^{-15}$} and  \mbox{$A_{\mathrm{GWB, M2}} = 4.87^{+5.26}_{-2.88} \times 10^{-15}$}.  Note that the amplitude of the CRS in M2 is the same for the CURN and GWB models.
\par

We observe no changes in the CURN amplitude and spectral index posteriors using M2 with or without RN for PSR~J1744$-$1134, with corresponding Jensen-Shannon divergences $2.95 \times 10^{-3}$ and $1.84 \times 10^{-3}$. \\

\begin{figure*}
	\centering
	\caption{Free-spectrum of the CURN (left) and the GWB (right) signals, either with M1 (empty black violin) or M2 (filled blue violin) single-pulsar noise models.}
	\label{fig:specCRS}
	\includegraphics[keepaspectratio=true,scale=0.32]{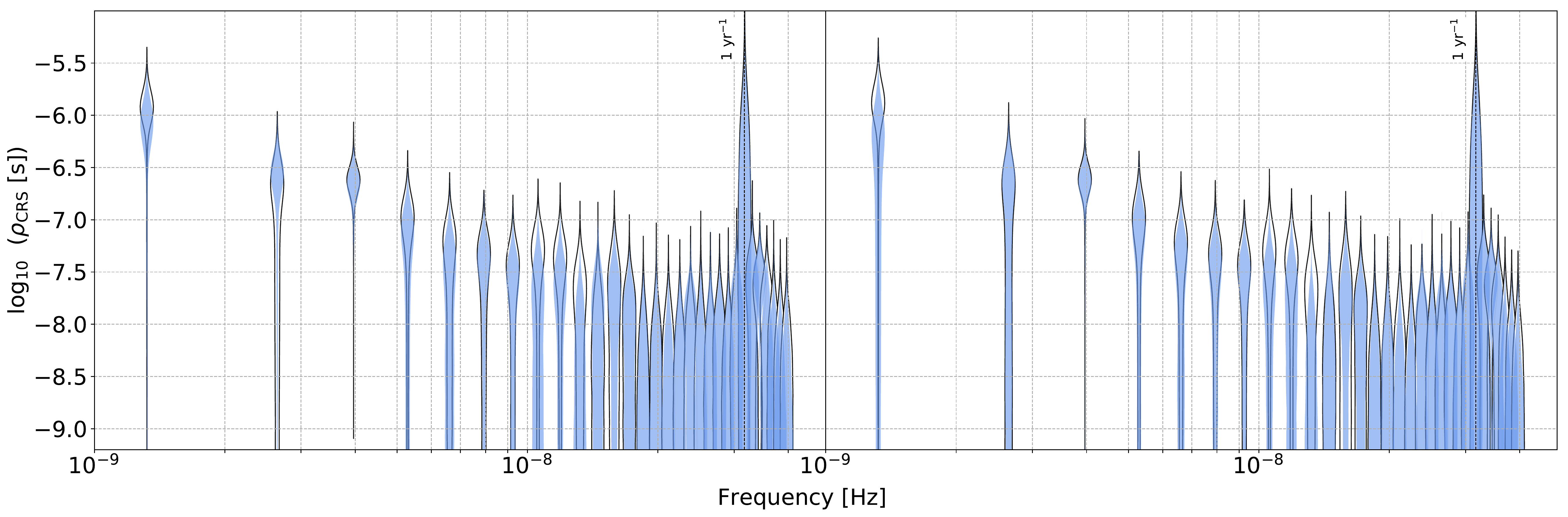}
\end{figure*}

We have also considered a free-spectrum model for CRS and plot it in Fig. \ref{fig:specCRS}.
The M1+CRS model is presented by the hollow violins and we plot M2+CRS by filled blue violins. The left plot shows the addition of CURN to the M1 and M2 noise models. We notice the slight drop in the amplitude at the  lowest frequency and slightly better constraint of amplitude at the second frequency bin. The free spectrum for GWB (right plot) shows a slight drop in power at the lowest frequency. These results nicely confirm our main findings with the power-law model of a decreased amplitude and shallower spectral index with M2 compared to M1.

\begin{table}
	\centering
	\caption{Median of amplitude and spectral index posterior  (and 95\% credible interval) of both uncorrelated (CURN) and Hellings-Down (GWB) common processes computed with M1 and M2. The last two columns contain Jensen-Shannon divergences values that compare the posteriors of a CRS  added to M1 and M2.}
	\label{tab:valCURN}
	\begin{tabular}{ccccc}
		\hline
		Model & $\mathrm{log}_{10} A$ & $\gamma$ & JS div. $A$ & JS div. $\gamma$\\
		\hline\hline
		M1 + CURN & $-14.27^{+0.26}_{-0.32}$ & $3.73^{+0.68}_{-0.61}$ & $0$ & $0$\\
		\hline
		M2 + CURN & $-14.31^{+0.30}_{-0.38}$ & $3.68^{+0.81}_{-0.72}$ & $1.55 \times 10^{-2}$ & $9.40 \times 10^{-3}$\\
		\hline\hline
		M1 + GWB & $-14.30^{+0.27}_{-0.32}$ & $3.82^{+0.70}_{-0.65}$ & $0$ & $0$\\
		\hline
		M2 + GWB & $-14.31^{+0.32}_{-0.39}$ & $3.69^{+0.84}_{-0.78}$ & $8.73 \times 10^{-3}$ & $2.60 \times 10^{-2}$\\
		\hline
	\end{tabular}
\end{table}


\subsection{Effects on the statistical significance of the common red signal}
Finally, we evaluate Bayes factors
considering M1/M2+CURN/GWB against pulsar noise models without any common process (PSRN), M1/M2.\par

The results are summarized in Table~\ref{tab:BFCRS}. We have used two methods to compute Bayes factors (Dynesty, indicated by ``Dyn.'' and the hyper-model (presented in Section \ref{sec:bayes}), indicated by ``Hyp'') for cross checking.
First of all we have re-derived the results of \cite{DR2_Chen+} for the M1 model: we observe strong evidence for the presence of a CRS but the data is not informative 
about its nature (could equally be CURN or GWB). 
Analysis with the custom noise model built 
in this work \emph{confirms} these findings. It is quite remarkable
that even though the custom noise models  are very different from the standard ones, we can still confirm the presence of a CRS. This gives additional confidence in the presence of a CRS in the EPTA DR2 data.

 The results presented in Table~\ref{tab:BFCRS} should be considered more as a comparison of the nature of the CRS: PSRN + GWB vs PSRN + CURN. It is worth mentioning that in this view the custom noise model M2 actually slightly prefers the GWB:  $\mathrm{log}_{10} {\mathcal{B}^{\rm PSRN + GWB}_{\rm PSRN + CURN}} = 0.3$, whereas it is slightly negative (-0.2) for M1 as estimated with the hypermodel method. The observed signal could also be caused by the spatially-uncorrelated component of the GWB, which is expected to dominate over the cross-correlated terms.

\begin{table}
	\centering
	\caption{Bayes factors in favor of the inclusion of CRS signals (CURN or GWB) in the M1 or M2 single-pulsar noise models. Estimations are performed either through evidence ratios (Dyn.) or using the product-space method (Hyp.).}
	\label{tab:BFCRS}
	\begin{tabular}{ccccc}
		\hline
		\multirow{2}{*}{CRS Model} & \multicolumn{4}{c}{$\mathrm{log}_{10}$ BF}\\
		& $\mathrm{M1_{Dyn.}}$ & $\mathrm{M1_{Hyp.}}$ & $\mathrm{M2_{Dyn.}}$ & $\mathrm{M2_{Hyp.}}$\\
		\hline \hline
		PSRN & - & - & - & -\\
		\hline
		PSRN + CURN & $3.6$ & $3.8$ & $2.9$ & $3.3$\\
		\hline
		PSRN + GWB & $3.4$ & $3.6$ & $3.0$ & $3.6$\\
		\hline
	\end{tabular}
\end{table}


\section{Conclusion} \label{sec:Conc}
We used a general Bayesian inference approach to select the most favoured noise model for each pulsar of EPTA DR2.
These models are summarized in  Table \ref{tab:finalMS} and show a significant improvement (in terms of Bayes factor) over the default base model used in \cite{DR2_Chen+}.  In addition to conventional stochastic processes such as achromatic red noise and DM variations, we have considered scattering variations, system noise (both chromatic and achromatic), band noise and  deterministic signals (annual DM variations, annual scattering variations and non-stationary DM event).  Our model selection was guided by previously published investigations or using auxiliary runs that helped to identify the list of models for further consideration.

The main result of this paper is that we confirm the presence of a common red noise in EPTA DR2 despite the use of much more complicated single-pulsar noise models. The data is not informative on the presence of Hellings-Downs spatial correlations, which is not surprising given that we have analysed only the 6 best EPTA pulsars. However, the use of custom noise models led to a marginal increase in the evidence for the presence of a GWB against an uncorrelated CRS.

The second main result of this paper is that it lays the practical scheme (protocol)  for
choosing a custom noise model which will be applied to a larger set of EPTA (and IPTA)
pulsars on the path to confirming the nature of the observed common red noise process. Even though we only found small differences, both in posteriors and Bayes factors, the analysis of noise in each pulsar data is essential for the detection and interpretation of the common red signal.

There are also several interesting side results obtained during this investigation.

(i) PSR~J1909$-$3744 shows signs of a downturn in its spectrum at low frequencies. This feature could be confirmed or disproved with a longer observational span.

(ii) PSR~J1012+5307 indicates the presence of a high level of red noise at high frequencies of unclear origin.

(iii) PSR~J1600$-$3053 and J1744$-$1134 are not very informative about the presence of achromatic red noise, giving only a small preference to the models with RN.

(iv) We did not observe strong evidence for the annual DM variations signal previously reported for
PSR~J0613$-$0200.

(v) We found that the first exponential dip of PSR~J1713+0747 has a chromatic index consistent with scattering variations, and confirmed the low chromatic index  for the second event.

In this work we have emphasized several times the need to combine the data from multiple systems (to disentangle SN and RN processes) and multi-band observations (to disentangle
achromatic and chromatic red noise). This gives a strong reason for a joint analysis of combined IPTA data.

In the Appendix we
show the red noise free spectrum with the default model \textit{RN30\_DM100} and its time-domain realization for each pulsar.
We also provide a Table~\ref{tab:NoiseParams} with the parameters of all noise components.


\section*{Acknowledgements}
The European Pulsar Timing Array (EPTA) is a collaboration between European and partner institutes,
namely ASTRON (NL), INAF/Osservatorio di Cagliari (IT), Max-Planck-Institut f\"{u}r
Radioastronomie (GER), Nan\c{c}ay/Paris Observatory (FRA), the University of Manchester
(UK), the University of Birmingham (UK), the University of East Anglia (UK), the
University of Bielefeld (GER), the University of Paris (FRA), the University of
Milan-Bicocca (IT) and Peking University (CHN), with the aim to provide high precision
pulsar timing to work towards the direct detection of low-frequency gravitational waves. An Advanced Grant of the European Research Council to implement the Large European Array for Pulsars (LEAP) also provides funding. The EPTA is part of the International Pulsar Timing Array (IPTA), we would like to thank our IPTA colleagues for their help with this paper.

Part of this work is based on observations with the 100-m telescope of the Max-Planck-Institut f\"{u}r Radioastronomie (MPIfR) at Effelsberg in Germany. Pulsar research at the Jodrell Bank Centre for Astrophysics and the observations using the Lovell Telescope are supported by a Consolidated Grant (ST/T000414/1) from the UK's Science and Technology Facilities Council. The Nan{\c c}ay radio Observatory is operated by the Paris Observatory, associated to the French Centre National de la Recherche Scientifique (CNRS), and partially supported by the Region Centre in France. We acknowledge financial support from ``Programme National de Cosmologie and Galaxies'' (PNCG), and ``Programme National Hautes Energies'' (PNHE) funded by CNRS/INSU-IN2P3-INP, CEA and CNES, France. We acknowledge financial support from Agence Nationale de la Recherche (ANR-18-CE31-0015), France. The Westerbork Synthesis Radio Telescope is operated by the Netherlands Institute for Radio Astronomy (ASTRON) with support from the Netherlands Foundation for Scientific Research (NWO). The Sardinia Radio Telescope (SRT) is funded by the Department of University and Research (MIUR), the Italian Space Agency (ASI), and the Autonomous Region of Sardinia (RAS) and is operated as National Facility by the National Institute for Astrophysics (INAF).

The work is supported by National SKA program of China 2020SKA0120100, Max-Planck Partner Group, NSFC 11690024, CAS Cultivation Project for FAST Scientific. This work is supported as part of the ``LEGACY'' MPG-CAS collaboration on low-frequency gravitational wave astronomy. AC acknowledges support from the Paris \^{I}le-de-France Region. GS, AS and AS acknowledge financial support provided under the European Union's H2020 ERC Consolidator Grant ``Binary Massive Black Hole Astrophysic'' (B Massive, Grant Agreement: 818691). JA acknowleges support by the Stavros Niarchos Foundation (SNF) and the Hellenic Foundation for Research and Innovation (H.F.R.I.) under the 2nd Call of ``Science and Society'' Action Always strive for excellence -- ``Theodoros Papazoglou'' (Project Number: 01431). GD, RK and MK acknowledge support from European Research Council (ERC) Synergy Grant ``BlackHoleCam'' Grant Agreement Number 610058 and ERC Advanced Grant ``LEAP'' Grant Agreement Number 337062. JWM is a CITA Postdoctoral Fellow: This work was supported by the Natural Sciences and Engineering Research Council of Canada (NSERC), (funding reference \#CITA 490888-16). AV acknowledges the support of the Royal Society and Wolfson Foundation. JPWV acknowledges support by the Deutsche Forschungsgemeinschaft (DFG) through the Heisenberg programme (Project No. 433075039). Finally, we would like to thank B. Goncharov for reading and commenting on the manuscript.

We used the following software packages to produce results and figures in this work : \textsc{Libstempo} \citep{val20}, \textsc{Enterprise} \citep{ell19}, \textsc{Enterprise\_extensions} (\url{https://github.com/nanograv/enterprise\_extensions}), \textsc{PTMCMC} \citep{ell17}, \textsc{Dynesty} \citep{spe20}, \textsc{$\textrm{MC}^3$} (\url{https://gitlab.in2p3.fr/stas/samplermcmc}), \textsc{numpy} \citep{van11}, \textsc{scipy} \citep{oli07}, \textsc{matplotlib} \citep{hun07}, \textsc{corner} \citep{for16}, \textsc{la forge} \citep{haz20b} and \textsc{outlier-utils} (\url{https://github.com/c-bata/outlier-utils}).


\section*{Data Availability}
The timing data used in this article shall be shared on reasonable request to the corresponding authors.


\bibliographystyle{mnras}
\bibliography{model_selection} 

\begin{thebibliography}{}
\makeatletter
\relax
\def\mn@urlcharsother{\let\do\@makeother \do\$\do\&\do\#\do\^\do\_\do\%\do\~}
\def\mn@doi{\begingroup\mn@urlcharsother \@ifnextchar [ {\mn@doi@}
  {\mn@doi@[]}}
\def\mn@doi@[#1]#2{\def\@tempa{#1}\ifx\@tempa\@empty \href
  {http://dx.doi.org/#2} {doi:#2}\else \href {http://dx.doi.org/#2} {#1}\fi
  \endgroup}
\def\mn@eprint#1#2{\mn@eprint@#1:#2::\@nil}
\def\mn@eprint@arXiv#1{\href {http://arxiv.org/abs/#1} {{\tt arXiv:#1}}}
\def\mn@eprint@dblp#1{\href {http://dblp.uni-trier.de/rec/bibtex/#1.xml}
  {dblp:#1}}
\def\mn@eprint@#1:#2:#3:#4\@nil{\def\@tempa {#1}\def\@tempb {#2}\def\@tempc
  {#3}\ifx \@tempc \@empty \let \@tempc \@tempb \let \@tempb \@tempa \fi \ifx
  \@tempb \@empty \def\@tempb {arXiv}\fi \@ifundefined
  {mn@eprint@\@tempb}{\@tempb:\@tempc}{\expandafter \expandafter \csname
  mn@eprint@\@tempb\endcsname \expandafter{\@tempc}}}

\bibitem[\protect\citeauthoryear{{Alam} et~al.,}{{Alam} et~al.}{2021}]{ala21}
{Alam} M.~F.,  et~al., 2021, \mn@doi [\apjs] {10.3847/1538-4365/abc6a0}, \href
  {https://ui.adsabs.harvard.edu/abs/2021ApJS..252....4A} {252, 4}

\bibitem[\protect\citeauthoryear{Anderson \& Darling}{Anderson \&
  Darling}{1952}]{and52}
Anderson T.~W.,  Darling D.~A.,  1952, The Annals of Mathematical Statistics,
  23, 193

\bibitem[\protect\citeauthoryear{Arzoumanian et~al.,}{Arzoumanian
  et~al.}{2020}]{arz20}
Arzoumanian Z.,  et~al., 2020, \mn@doi [The Astrophysical Journal Letters]
  {10.3847/2041-8213/abd401}, 905, L34

\bibitem[\protect\citeauthoryear{Babak et~al.,}{Babak et~al.}{2015}]{bab15}
Babak S.,  et~al., 2015, \mn@doi [Monthly Notices of the Royal Astronomical
  Society] {10.1093/mnras/stv2092}, 455, 1665–1679

\bibitem[\protect\citeauthoryear{Bassa et~al.,}{Bassa et~al.}{2015}]{bas15}
Bassa C.~G.,  et~al., 2015, \mn@doi [Monthly Notices of the Royal Astronomical
  Society] {10.1093/mnras/stv2755}, 456, 2196

\bibitem[\protect\citeauthoryear{Brooks \& Gelman}{Brooks \&
  Gelman}{1998}]{bro98}
Brooks S.~P.,  Gelman A.,  1998, \mn@doi [Journal of Computational and
  Graphical Statistics] {10.1080/10618600.1998.10474787}, 7, 434

\bibitem[\protect\citeauthoryear{{Caballero} et~al.,}{{Caballero}
  et~al.}{2016}]{cll16}
{Caballero} R.~N.,  et~al., 2016, \mn@doi [\mnras] {10.1093/mnras/stw179},
  \href {http://cdsads.u-strasbg.fr/abs/2016MNRAS.457.4421C} {457, 4421}

\bibitem[\protect\citeauthoryear{Carlin \& Chib}{Carlin \& Chib}{1995}]{car95}
Carlin B.~P.,  Chib S.,  1995, \mn@doi [Journal of the Royal Statistical
  Society: Series B (Methodological)]
  {https://doi.org/10.1111/j.2517-6161.1995.tb02042.x}, 57, 473

\bibitem[\protect\citeauthoryear{Chen, Sesana  \& Del~Pozzo}{Chen
  et~al.}{2017}]{che17}
Chen S.,  Sesana A.,   Del~Pozzo W.,  2017, \mn@doi [Monthly Notices of the
  Royal Astronomical Society] {10.1093/mnras/stx1093}, 470, 1738

\bibitem[\protect\citeauthoryear{Chen et~al.,}{Chen et~al.}{2021}]{DR2_Chen+}
Chen S.,  et~al., 2021, \mn@doi [Monthly Notices of the Royal Astronomical
  Society] {10.1093/mnras/stab2833}, 508, 4970

\bibitem[\protect\citeauthoryear{Coles et~al.,}{Coles et~al.}{2015}]{col15}
Coles W.~A.,  et~al., 2015, \mn@doi [The Astrophysical Journal]
  {10.1088/0004-637x/808/2/113}, 808, 113

\bibitem[\protect\citeauthoryear{Cordes \& Shannon}{Cordes \&
  Shannon}{2010}]{cor10}
Cordes J.~M.,  Shannon R.~M.,  2010, A Measurement Model for Precision Pulsar
  Timing (\mn@eprint {arXiv} {1010.3785})

\bibitem[\protect\citeauthoryear{{Cordes} \& {Wolszczan}}{{Cordes} \&
  {Wolszczan}}{1986}]{cor86}
{Cordes} J.~M.,  {Wolszczan} A.,  1986, \mn@doi [\apjl] {10.1086/184722}, \href
  {https://ui.adsabs.harvard.edu/abs/1986ApJ...307L..27C} {307, L27}

\bibitem[\protect\citeauthoryear{Cordes, Shannon  \& Stinebring}{Cordes
  et~al.}{2016}]{cor16}
Cordes J.~M.,  Shannon R.~M.,   Stinebring D.~R.,  2016, \mn@doi [The
  Astrophysical Journal] {10.3847/0004-637x/817/1/16}, 817, 16

\bibitem[\protect\citeauthoryear{Desvignes et~al.,}{Desvignes
  et~al.}{2016}]{dev16}
Desvignes G.,  et~al., 2016, \mn@doi [Monthly Notices of the Royal Astronomical
  Society] {10.1093/mnras/stw483}, 458, 3341

\bibitem[\protect\citeauthoryear{{Detweiler}}{{Detweiler}}{1979}]{det79}
{Detweiler} S.,  1979, \mn@doi [\apj] {10.1086/157593}, \href
  {http://cdsads.u-strasbg.fr/abs/1979ApJ...234.1100D} {234, 1100}

\bibitem[\protect\citeauthoryear{Ellis \& van Haasteren}{Ellis \& van
  Haasteren}{2017}]{ell17}
Ellis J.,  van Haasteren R.,  2017, jellis18/PTMCMCSampler: Official Release,
  \mn@doi{10.5281/zenodo.1037579}, \url
  {https://doi.org/10.5281/zenodo.1037579}

\bibitem[\protect\citeauthoryear{{Ellis}, {Vallisneri}, {Taylor}  \&
  {Baker}}{{Ellis} et~al.}{2019}]{ell19}
{Ellis} J.~A.,  {Vallisneri} M.,  {Taylor} S.~R.,   {Baker} P.~T.,  2019,
  {ENTERPRISE: Enhanced Numerical Toolbox Enabling a Robust PulsaR Inference
  SuitE} (\mn@eprint {ascl} {1912.015})

\bibitem[\protect\citeauthoryear{Foreman-Mackey}{Foreman-Mackey}{2016}]{for16}
Foreman-Mackey D.,  2016, \mn@doi [The Journal of Open Source Software]
  {10.21105/joss.00024}, 1, 24

\bibitem[\protect\citeauthoryear{{Foster} \& {Backer}}{{Foster} \&
  {Backer}}{1990}]{fos90}
{Foster} R.~S.,  {Backer} D.~C.,  1990, \mn@doi [\apj] {10.1086/169195}, \href
  {http://cdsads.u-strasbg.fr/abs/1990ApJ...361..300F} {361, 300}

\bibitem[\protect\citeauthoryear{Gelman \& Rubin}{Gelman \&
  Rubin}{1992}]{gel92}
Gelman A.,  Rubin D.~B.,  1992, \mn@doi [Statistical Science]
  {10.1214/ss/1177011136}, 7, 457

\bibitem[\protect\citeauthoryear{Goncharov, Zhu  \& Thrane}{Goncharov
  et~al.}{2020}]{gon20}
Goncharov B.,  Zhu X.-J.,   Thrane E.,  2020, \mn@doi [Monthly Notices of the
  Royal Astronomical Society] {10.1093/mnras/staa2081}, 497, 3264

\bibitem[\protect\citeauthoryear{{Goncharov} et~al.,}{{Goncharov}
  et~al.}{2021a}]{gon21}
{Goncharov} B.,  et~al., 2021a, \mn@doi [\mnras] {10.1093/mnras/staa3411},
  \href {https://ui.adsabs.harvard.edu/abs/2021MNRAS.502..478G} {502, 478}

\bibitem[\protect\citeauthoryear{Goncharov et~al.,}{Goncharov
  et~al.}{2021b}]{gon21b}
Goncharov B.,  et~al., 2021b, \mn@doi [The Astrophysical Journal Letters]
  {10.3847/2041-8213/ac17f4}, 917, L19

\bibitem[\protect\citeauthoryear{Grubbs}{Grubbs}{1950}]{gru50}
Grubbs F.~E.,  1950, \mn@doi [The Annals of Mathematical Statistics]
  {10.1214/aoms/1177729885}, 21, 27

\bibitem[\protect\citeauthoryear{Hastings}{Hastings}{1970}]{has70}
Hastings W.~K.,  1970, \mn@doi [Biometrika] {10.1093/biomet/57.1.97}, 57, 97

\bibitem[\protect\citeauthoryear{Hazboun}{Hazboun}{2020}]{haz20b}
Hazboun J.~S.,  2020, La Forge, \mn@doi{10.5281/zenodo.4152550}, \url
  {https://doi.org/10.5281/zenodo.4152550}

\bibitem[\protect\citeauthoryear{{Hazboun}, {Simon}, {Siemens}  \&
  {Romano}}{{Hazboun} et~al.}{2020}]{haz20}
{Hazboun} J.~S.,  {Simon} J.,  {Siemens} X.,   {Romano} J.~D.,  2020, \mn@doi
  [\apjl] {10.3847/2041-8213/abca92}, \href
  {https://ui.adsabs.harvard.edu/abs/2020ApJ...905L...6H} {905, L6}

\bibitem[\protect\citeauthoryear{Hee, Handley, Hobson  \& Lasenby}{Hee
  et~al.}{2015}]{hee15}
Hee S.,  Handley W.~J.,  Hobson M.~P.,   Lasenby A.~N.,  2015, \mn@doi [Monthly
  Notices of the Royal Astronomical Society] {10.1093/mnras/stv2217}, 455, 2461

\bibitem[\protect\citeauthoryear{{Hellings} \& {Downs}}{{Hellings} \&
  {Downs}}{1983}]{hel83}
{Hellings} R.~W.,  {Downs} G.~S.,  1983, \mn@doi [\apjl] {10.1086/183954},
  \href {https://ui.adsabs.harvard.edu/abs/1983ApJ...265L..39H} {265, L39}

\bibitem[\protect\citeauthoryear{{Hobbs}, {Edwards}  \& {Manchester}}{{Hobbs}
  et~al.}{2006}]{hob06}
{Hobbs} G.~B.,  {Edwards} R.~T.,   {Manchester} R.~N.,  2006, \mn@doi [\mnras]
  {10.1111/j.1365-2966.2006.10302.x}, \href
  {https://ui.adsabs.harvard.edu/abs/2006MNRAS.369..655H} {369, 655}

\bibitem[\protect\citeauthoryear{{Hobbs} et~al.,}{{Hobbs} et~al.}{2020}]{hob20}
{Hobbs} G.,  et~al., 2020, \mn@doi [\mnras] {10.1093/mnras/stz3071}, \href
  {https://ui.adsabs.harvard.edu/abs/2020MNRAS.491.5951H} {491, 5951}

\bibitem[\protect\citeauthoryear{Hunter}{Hunter}{2007}]{hun07}
Hunter J.~D.,  2007, \mn@doi [Computing in Science Engineering]
  {10.1109/MCSE.2007.55}, 9, 90

\bibitem[\protect\citeauthoryear{{Jaffe} \& {Backer}}{{Jaffe} \&
  {Backer}}{2003}]{jaf03}
{Jaffe} A.~H.,  {Backer} D.~C.,  2003, \mn@doi [\apj] {10.1086/345443}, \href
  {https://ui.adsabs.harvard.edu/abs/2003ApJ...583..616J} {583, 616}

\bibitem[\protect\citeauthoryear{Jeffreys}{Jeffreys}{1961}]{jef61}
Jeffreys H.,  1961, Theory of Probability, third edn.
Oxford, Oxford, England

\bibitem[\protect\citeauthoryear{Jenet et~al.,}{Jenet et~al.}{2006}]{jen06}
Jenet F.~A.,  et~al., 2006, \mn@doi [The Astrophysical Journal]
  {10.1086/508702}, 653, 1571–1576

\bibitem[\protect\citeauthoryear{Kass \& Raftery}{Kass \&
  Raftery}{1995}]{kas95}
Kass R.~E.,  Raftery A.~E.,  1995, \mn@doi [Journal of the American Statistical
  Association] {10.1080/01621459.1995.10476572}, 90, 773

\bibitem[\protect\citeauthoryear{{Keith} et~al.,}{{Keith} et~al.}{2013}]{kei13}
{Keith} M.~J.,  et~al., 2013, \mn@doi [\mnras] {10.1093/mnras/sts486}, \href
  {https://ui.adsabs.harvard.edu/abs/2013MNRAS.429.2161K} {429, 2161}

\bibitem[\protect\citeauthoryear{{Kerr} et~al.,}{{Kerr} et~al.}{2020}]{ker20}
{Kerr} M.,  et~al., 2020, \mn@doi [\pasa] {10.1017/pasa.2020.11}, \href
  {https://ui.adsabs.harvard.edu/abs/2020PASA...37...20K} {37, e020}

\bibitem[\protect\citeauthoryear{Kullback}{Kullback}{1959}]{kul59}
Kullback S.,  1959, Information Theory and Statistics.
Wiley, New York

\bibitem[\protect\citeauthoryear{{Lam} et~al.,}{{Lam} et~al.}{2016}]{lcc+16}
{Lam} M.~T.,  et~al., 2016, \mn@doi [ApJ] {10.3847/0004-637X/819/2/155}, \href
  {https://ui.adsabs.harvard.edu/abs/2016ApJ...819..155L} {819, 155}

\bibitem[\protect\citeauthoryear{Lam et~al.,}{Lam et~al.}{2018}]{lam18}
Lam M.~T.,  et~al., 2018, \mn@doi [The Astrophysical Journal]
  {10.3847/1538-4357/aac770}, 861, 132

\bibitem[\protect\citeauthoryear{Lee, Jenet  \& Price}{Lee
  et~al.}{2008}]{lee08}
Lee K.~J.,  Jenet F.~A.,   Price R.~H.,  2008, \mn@doi [The Astrophysical
  Journal] {10.1086/591080}, 685, 1304

\bibitem[\protect\citeauthoryear{Lentati, Alexander, Hobson, Taylor, Gair,
  Balan  \& van Haasteren}{Lentati et~al.}{2013}]{len13}
Lentati L.,  Alexander P.,  Hobson M.~P.,  Taylor S.,  Gair J.,  Balan S.~T.,
  van Haasteren R.,  2013, \mn@doi [Physical Review D]
  {10.1103/physrevd.87.104021}, 87

\bibitem[\protect\citeauthoryear{Lentati et~al.,}{Lentati et~al.}{2015}]{len15}
Lentati L.,  et~al., 2015, \mn@doi [{Monthly Notices of the Royal Astronomical
  Society}] {10.1093/mnras/stv1538}, 453, 2576

\bibitem[\protect\citeauthoryear{{Lentati} et~al.,}{{Lentati}
  et~al.}{2016}]{len16}
{Lentati} L.,  et~al., 2016, \mn@doi [\mnras] {10.1093/mnras/stw395}, \href
  {https://ui.adsabs.harvard.edu/abs/2016MNRAS.458.2161L} {458, 2161}

\bibitem[\protect\citeauthoryear{{Liu}, {Keane}, {Lee}, {Kramer}, {Cordes}  \&
  {Purver}}{{Liu} et~al.}{2012}]{lkl+11}
{Liu} K.,  {Keane} E.~F.,  {Lee} K.~J.,  {Kramer} M.,  {Cordes} J.~M.,
  {Purver} M.~B.,  2012, \mn@doi [MNRAS] {10.1111/j.1365-2966.2011.20041.x},
  \href {http://adsabs.harvard.edu/abs/2012MNRAS.420..361L} {420, 361}

\bibitem[\protect\citeauthoryear{{Lorimer} \& {Kramer}}{{Lorimer} \&
  {Kramer}}{2004}]{lor04}
{Lorimer} D.~R.,  {Kramer} M.,  2004, Handbook of Pulsar Astronomy.
Cambridge University Press

\bibitem[\protect\citeauthoryear{{Lyne}, {Hobbs}, {Kramer}, {Stairs}  \&
  {Stappers}}{{Lyne} et~al.}{2010}]{lyn10}
{Lyne} A.,  {Hobbs} G.,  {Kramer} M.,  {Stairs} I.,   {Stappers} B.,  2010,
  \mn@doi [Science] {10.1126/science.1186683}, \href
  {https://ui.adsabs.harvard.edu/abs/2010Sci...329..408L} {329, 408}

\bibitem[\protect\citeauthoryear{{Maggiore}}{{Maggiore}}{2000}]{maj2000}
{Maggiore} M.,  2000, \mn@doi [\physrep] {10.1016/S0370-1573(99)00102-7}, \href
  {http://cdsads.u-strasbg.fr/abs/2000PhR...331..283M} {331, 283}

\bibitem[\protect\citeauthoryear{Main et~al.,}{Main et~al.}{2020}]{mai20}
Main R.~A.,  et~al., 2020, \mn@doi [Monthly Notices of the Royal Astronomical
  Society] {10.1093/mnras/staa2955}, 499, 1468

\bibitem[\protect\citeauthoryear{Manning \& Sch{\"u}tze}{Manning \&
  Sch{\"u}tze}{1999}]{man99}
Manning C.~D.,  Sch{\"u}tze H.,  1999, Foundations of Statistical Natural
  Language Processing.
The {MIT} Press, Cambridge, Massachusetts, \url
  {http://nlp.stanford.edu/fsnlp/}

\bibitem[\protect\citeauthoryear{{Matsakis}, {Taylor}  \& {Eubanks}}{{Matsakis}
  et~al.}{1997}]{mat97}
{Matsakis} D.~N.,  {Taylor} J.~H.,   {Eubanks} T.~M.,  1997, \aap, \href
  {https://ui.adsabs.harvard.edu/abs/1997A&A...326..924M} {326, 924}

\bibitem[\protect\citeauthoryear{{Melatos} \& {Link}}{{Melatos} \&
  {Link}}{2014}]{mel14}
{Melatos} A.,  {Link} B.,  2014, \mn@doi [\mnras] {10.1093/mnras/stt1828},
  \href {https://ui.adsabs.harvard.edu/abs/2014MNRAS.437...21M} {437, 21}

\bibitem[\protect\citeauthoryear{{Metropolis}, {Rosenbluth}, {Rosenbluth},
  {Teller}  \& {Teller}}{{Metropolis} et~al.}{1953}]{met53}
{Metropolis} N.,  {Rosenbluth} A.~W.,  {Rosenbluth} M.~N.,  {Teller} A.~H.,
  {Teller} E.,  1953, \mn@doi [\jcp] {10.1063/1.1699114}, \href
  {https://ui.adsabs.harvard.edu/abs/1953JChPh..21.1087M} {21, 1087}

\bibitem[\protect\citeauthoryear{Oliphant}{Oliphant}{2007}]{oli07}
Oliphant T.~E.,  2007, \mn@doi [Computing in Science Engineering]
  {10.1109/MCSE.2007.58}, 9, 10

\bibitem[\protect\citeauthoryear{{Perera} et~al.,}{{Perera}
  et~al.}{2018}]{per18}
{Perera} B.~B.~P.,  et~al., 2018, \mn@doi [\mnras] {10.1093/mnras/sty1116},
  \href {https://ui.adsabs.harvard.edu/abs/2018MNRAS.478..218P} {478, 218}

\bibitem[\protect\citeauthoryear{{Perera} et~al.,}{{Perera}
  et~al.}{2019}]{per19}
{Perera} B.~B.~P.,  et~al., 2019, \mn@doi [\mnras] {10.1093/mnras/stz2857},
  \href {https://ui.adsabs.harvard.edu/abs/2019MNRAS.490.4666P} {490, 4666}

\bibitem[\protect\citeauthoryear{{Rajagopal} \& {Romani}}{{Rajagopal} \&
  {Romani}}{1995}]{raj95}
{Rajagopal} M.,  {Romani} R.~W.,  1995, \mn@doi [\apj] {10.1086/175813}, \href
  {https://ui.adsabs.harvard.edu/abs/1995ApJ...446..543R} {446, 543}

\bibitem[\protect\citeauthoryear{{Reardon} et~al.,}{{Reardon}
  et~al.}{2016}]{rea16}
{Reardon} D.~J.,  et~al., 2016, \mn@doi [\mnras] {10.1093/mnras/stv2395}, \href
  {https://ui.adsabs.harvard.edu/abs/2016MNRAS.455.1751R} {455, 1751}

\bibitem[\protect\citeauthoryear{{Sazhin}}{{Sazhin}}{1978}]{saz78}
{Sazhin} M.~V.,  1978, Soviet Astronomy, \href
  {http://cdsads.u-strasbg.fr/abs/1978SvA....22...36S} {22, 36}

\bibitem[\protect\citeauthoryear{Sesana, Haardt, Madau  \& Volonteri}{Sesana
  et~al.}{2004}]{ses04}
Sesana A.,  Haardt F.,  Madau P.,   Volonteri M.,  2004, \mn@doi [The
  Astrophysical Journal] {10.1086/422185}, 611, 623–632

\bibitem[\protect\citeauthoryear{Sesana, Vecchio  \& Volonteri}{Sesana
  et~al.}{2009}]{ses09}
Sesana A.,  Vecchio A.,   Volonteri M.,  2009, \mn@doi [Monthly Notices of the
  Royal Astronomical Society] {10.1111/j.1365-2966.2009.14499.x}, 394, 2255

\bibitem[\protect\citeauthoryear{Shannon \& Cordes}{Shannon \&
  Cordes}{2010}]{sha10}
Shannon R.~M.,  Cordes J.~M.,  2010, \mn@doi [The Astrophysical Journal]
  {10.1088/0004-637x/725/2/1607}, 725, 1607

\bibitem[\protect\citeauthoryear{Shannon et~al.,}{Shannon et~al.}{2014}]{sha14}
Shannon R.~M.,  et~al., 2014, \mn@doi [Monthly Notices of the Royal
  Astronomical Society] {10.1093/mnras/stu1213}, 443, 1463

\bibitem[\protect\citeauthoryear{Sivia \& Skilling}{Sivia \&
  Skilling}{2006}]{siv06}
Sivia D.~S.,  Skilling J.,  2006, {Data Analysis - A Bayesian Tutorial}, 2nd
  edn.
Oxford Science Publications, Oxford University Press

\bibitem[\protect\citeauthoryear{{Skilling}}{{Skilling}}{2004}]{ski04}
{Skilling} J.,  2004, in {Fischer} R.,  {Preuss} R.,   {Toussaint} U.~V.,  eds,
   American Institute of Physics Conference Series Vol. 735, Bayesian Inference
  and Maximum Entropy Methods in Science and Engineering: 24th International
  Workshop on Bayesian Inference and Maximum Entropy Methods in Science and
  Engineering. pp 395--405, \mn@doi{10.1063/1.1835238}

\bibitem[\protect\citeauthoryear{Skilling}{Skilling}{2006}]{ski06}
Skilling J.,  2006, \mn@doi [Bayesian Analysis] {10.1214/06-BA127}, 1, 833

\bibitem[\protect\citeauthoryear{{Speagle}}{{Speagle}}{2020}]{spe20}
{Speagle} J.~S.,  2020, \mn@doi [\mnras] {10.1093/mnras/staa278}, \href
  {https://ui.adsabs.harvard.edu/abs/2020MNRAS.493.3132S} {493, 3132}

\bibitem[\protect\citeauthoryear{{Taylor}}{{Taylor}}{1992}]{tay92}
{Taylor} J.~H.,  1992, \mn@doi [Philosophical Transactions of the Royal Society
  of London Series A] {10.1098/rsta.1992.0088}, \href
  {https://ui.adsabs.harvard.edu/abs/1992RSPTA.341..117T} {341, 117}

\bibitem[\protect\citeauthoryear{Taylor, van Haasteren  \& Sesana}{Taylor
  et~al.}{2020}]{tay20}
Taylor S.~R.,  van Haasteren R.,   Sesana A.,  2020, \mn@doi [Physical Review
  D] {10.1103/physrevd.102.084039}, 102

\bibitem[\protect\citeauthoryear{Tiburzi et~al.,}{Tiburzi et~al.}{2015}]{tib15}
Tiburzi C.,  et~al., 2015, \mn@doi [Monthly Notices of the Royal Astronomical
  Society] {10.1093/mnras/stv2143}, 455, 4339

\bibitem[\protect\citeauthoryear{{Tsang} \& {Gourgouliatos}}{{Tsang} \&
  {Gourgouliatos}}{2013}]{ts13}
{Tsang} D.,  {Gourgouliatos} K.~N.,  2013, \mn@doi [\apjl]
  {10.1088/2041-8205/773/1/L17}, \href
  {https://ui.adsabs.harvard.edu/abs/2013ApJ...773L..17T} {773, L17}

\bibitem[\protect\citeauthoryear{{Vallisneri}}{{Vallisneri}}{2020}]{val20}
{Vallisneri} M.,  2020, {libstempo: Python wrapper for Tempo2} (\mn@eprint
  {ascl} {2002.017})

\bibitem[\protect\citeauthoryear{{Verbiest} et~al.,}{{Verbiest}
  et~al.}{2009}]{ver09}
{Verbiest} J.~P.~W.,  et~al., 2009, \mn@doi [\mnras]
  {10.1111/j.1365-2966.2009.15508.x}, \href
  {https://ui.adsabs.harvard.edu/abs/2009MNRAS.400..951V} {400, 951}

\bibitem[\protect\citeauthoryear{Verbiest et~al.,}{Verbiest
  et~al.}{2016}]{ver16}
Verbiest J. P.~W.,  et~al., 2016, \mn@doi [Monthly Notices of the Royal
  Astronomical Society] {10.1093/mnras/stw347}, 458, 1267

\bibitem[\protect\citeauthoryear{Wyithe \& Loeb}{Wyithe \& Loeb}{2003}]{wyi03}
Wyithe J. S.~B.,  Loeb A.,  2003, \mn@doi [The Astrophysical Journal]
  {10.1086/375187}, 590, 691–706

\bibitem[\protect\citeauthoryear{You et~al.,}{You et~al.}{2007}]{you07}
You X.~P.,  et~al., 2007, \mn@doi [Monthly Notices of the Royal Astronomical
  Society] {10.1111/j.1365-2966.2007.11617.x}, 378, 493

\bibitem[\protect\citeauthoryear{Zhu et~al.,}{Zhu et~al.}{2015}]{zhu15}
Zhu W.~W.,  et~al., 2015, \mn@doi [The Astrophysical Journal]
  {10.1088/0004-637x/809/1/41}, 809, 41

\bibitem[\protect\citeauthoryear{van Haasteren \& Levin}{van Haasteren \&
  Levin}{2012}]{van12}
van Haasteren R.,  Levin Y.,  2012, \mn@doi [Monthly Notices of the Royal
  Astronomical Society] {10.1093/mnras/sts097}, 428, 1147

\bibitem[\protect\citeauthoryear{van Haasteren \& Vallisneri}{van Haasteren \&
  Vallisneri}{2014}]{van14}
van Haasteren R.,  Vallisneri M.,  2014, \mn@doi [Phys. Rev. D]
  {10.1103/PhysRevD.90.104012}, 90, 104012

\bibitem[\protect\citeauthoryear{van Haasteren \& Vallisneri}{van Haasteren \&
  Vallisneri}{2015}]{van15}
van Haasteren R.,  Vallisneri M.,  2015, \mn@doi [Monthly Notices of the Royal
  Astronomical Society] {10.1093/mnras/stu2157}, 446, 1170

\bibitem[\protect\citeauthoryear{van Haasteren, Levin, McDonald  \& Lu}{van
  Haasteren et~al.}{2009}]{van09}
van Haasteren R.,  Levin Y.,  McDonald P.,   Lu T.,  2009, \mn@doi [Monthly
  Notices of the Royal Astronomical Society]
  {10.1111/j.1365-2966.2009.14590.x}, 395, 1005

\bibitem[\protect\citeauthoryear{van Straten}{van Straten}{2013}]{van13}
van Straten W.,  2013, \mn@doi [The Astrophysical Journal Supplement Series]
  {10.1088/0067-0049/204/1/13}, 204, 13

\bibitem[\protect\citeauthoryear{van~der Walt, Colbert  \& Varoquaux}{van~der
  Walt et~al.}{2011}]{van11}
van~der Walt S.,  Colbert S.~C.,   Varoquaux G.,  2011, \mn@doi [Computing in
  Science Engineering] {10.1109/MCSE.2011.37}, 13, 22

\makeatother
\end{thebibliography}



\ \\
$^{1}$Universit{\'e} de Paris, CNRS, Astroparticule et Cosmologie, 75013 Paris, France\\
$^{2}$Laboratoire de Physique et Chimie de l'Environnement et de l'Espace LPC2E UMR7328, Universit{\'e} d'Orl{\'e}ans, CNRS, 45071 Orl{\'e}ans, France\\
$^{3}$Station de Radioastronomie de Nan\c{c}ay, Observatoire de Paris, PSL University, CNRS, Universit{\'e} d'Orl{\'e}ans, 18330 Nan\c{c}ay, France\\
$^{4}$Moscow Institute of Physics and Technology, Dolgoprudny, Moscow region, Russia\\
$^{5}$IRFU, CEA, Universit\'e Paris-Saclay, F-91191, Gif-sur-Yvette, France\\
$^{6}$Dipartimento di Fisica ``G. Occhialini'', Universit{\'a} degli Studi di Milano-Bicocca, Piazza della Scienza 3, 20126 Milano, Italy\\
$^{7}$Kavli Institute for Astronomy and Astrophysics, Peking University, Beijing 100871, P.R.China\\
$^{8}$Laboratoire Univers et Th{\'e}ories LUTh, Observatoire de Paris, Universit{\'e} PSL, CNRS, Universit{\'e} de Paris, 92190 Meudon, France\\
$^{9}$Max-Planck-Institut f{\"u}r Radioastronomie, Auf dem H{\"u}gel 69, 53121 Bonn, Germany\\
$^{10}$LESIA, Observatoire de Paris, Universit{\'e} PSL, CNRS, Sorbonne Universit{\'e}, Universit{\'e} de Paris, 5 place Jules Janssen, 92195 Meudon, France\\
$^{11}$INFN, Sezione di Milano-Bicocca, Piazza della Scienza 3, 20126 Milano, Italy\\
$^{12}$ASTRON, Netherlands Institute for Radio Astronomy, Oude Hoogeveensedijk 4, 7991 PD, Dwingeloo, The Netherlands\\
$^{13}$Department of Astrophysics/IMAPP, Radboud University Nijmegen, P.O. Box 9010, 6500 GL Nijmegen, The Netherlands\\
$^{14}$Institute of Astrophysics, FORTH, N. Plastira 100, 70013, Heraklion, Greece\\
$^{15}$Argelander Institut für Astronomie, Auf dem H{\"u}gel 71, 53117, Bonn, Germany\\
$^{16}$Fakult{\"a}t für Physik, Universit{\"a}t Bielefeld, Postfach 100131, 33501 Bielefeld, Germany\\
$^{17}$INAF - Osservatorio Astronomico di Cagliari, via della Scienza 5, 09047 Selargius (CA), Italy\\
$^{18}$School of Physics, Faculty of Science, University of East Anglia, Norwich NR4 7TJ, UK\\
$^{19}$Max-Planck-Institut f{\"u}r Gravitationsphysik (Albert Einstein Institut), Am M{\"u}hlenberg 1, 14476 Golm, Germany\\
$^{20}$Jodrell Bank Centre for Astrophysics, Department of Physics and Astronomy, University of Manchester, Manchester M13 9PL, UK\\
$^{21}$National Astronomical Observatories, Chinese Academy of Sciences, Beijing, 100101, P. R. China\\
$^{22}$Laboratory of Gravitational Waves and Cosmology, Advanced Institute of Natural Sciences, Beijing Normal University at Zhuhai 519087, P.R.China\\
$^{23}$Canadian Institute for Theoretical Astrophysics, University of Toronto, 60 St. George Street, Toronto, ON M5S 3H8, Canada\\
$^{24}$Arecibo Observatory, HC3 Box 53995, Arecibo, PR 00612, USA\\
$^{25}$Max-Planck-Institut f{\"u}r Gravitationsphysik, Albert Einstein Institut, Am M{\"u}hlenberg 1, 14476 Golm, Germany\\
$^{26}$Institute for Gravitational Wave Astronomy and School of Physics and Astronomy, University of Birmingham, Edgbaston, Birmingham B15 2TT, UK\\
$^{27}$Department of Astronomy, Peking University, Beijing 100871, P. R. China\\



\appendix

\section{Achromatic red-noise properties}

Here we provide additional plots which demonstrate our main findings about the RN in each pulsar. \par

\begin{figure*}
	\centering
	\caption{(left panels) $68\%$ confidence interval of $100$ time-domain random realizations of the achromatic red-noise included in the default base (\textit{RN30\_DM100}) model (red) and the "custom" selected models (light grey) (cf. Table \ref{tab:finalMS}) for each of the six pulsars. (right panels) Achromatic red-noise spectrum for the corresponding pulsar (noted in left panel) included in the default base model and described with (i) a free-spectrum PSD (grey violins); or (ii) a broken power-law (blue), here showing $1000$ random realizations drawn from the posterior distributions. }
	\label{fig:timedom_freespec}
	\begin{subfigure}{.32\textwidth}
		\centering
		\hspace*{-3.5cm}\includegraphics[keepaspectratio=true,scale=0.24]{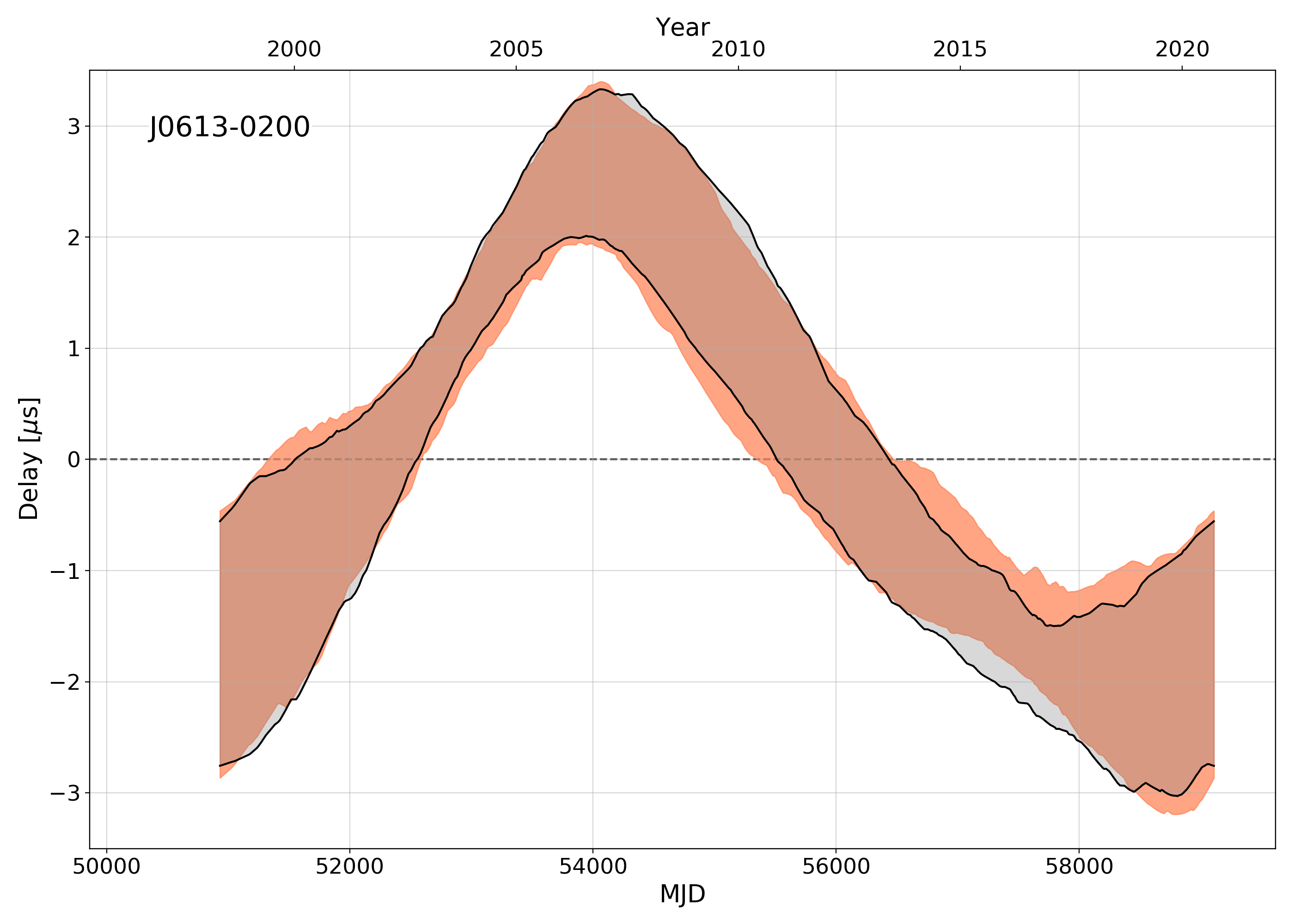}
	\end{subfigure}
	\begin{subfigure}{.32\textwidth}
		\centering
		\vspace*{.4cm}\includegraphics[keepaspectratio=true,scale=0.24]{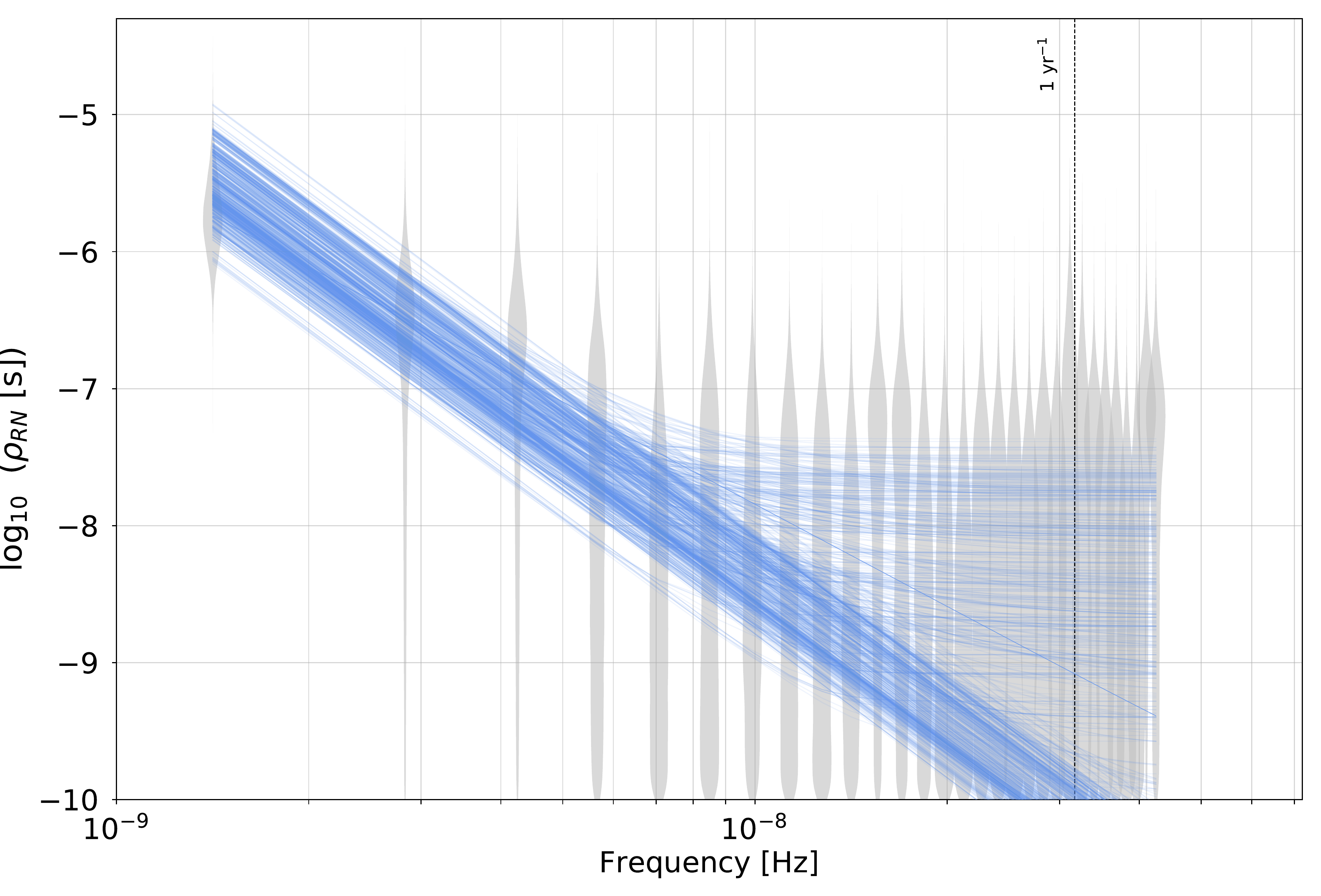}
	\end{subfigure}\\
	
	\begin{subfigure}{.32\textwidth}
		\centering
		\hspace*{-3.5cm}\includegraphics[keepaspectratio=true,scale=0.24]{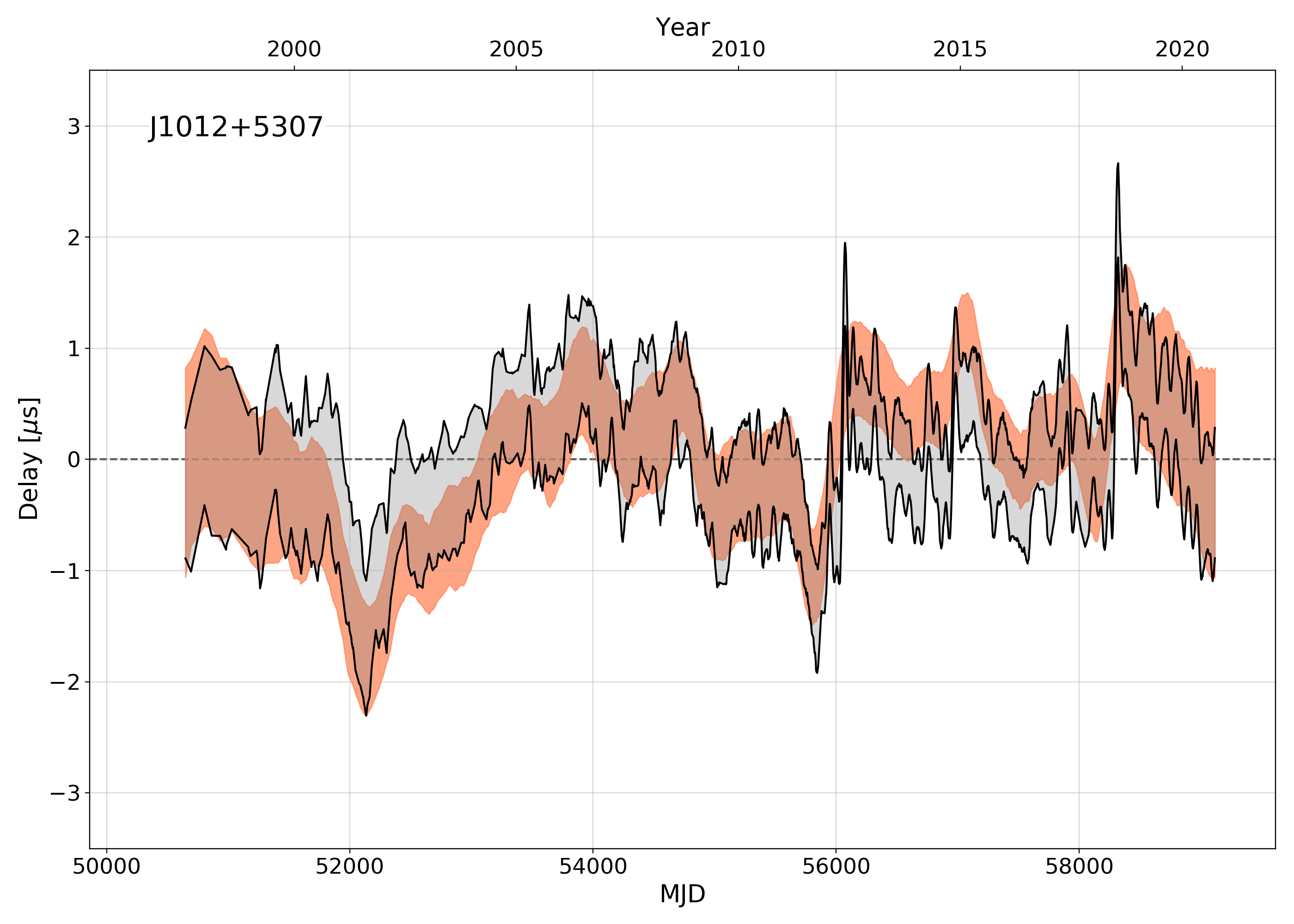}
	\end{subfigure}
	\begin{subfigure}{.32\textwidth}
		\centering
		\vspace*{.4cm}\includegraphics[keepaspectratio=true,scale=0.24]{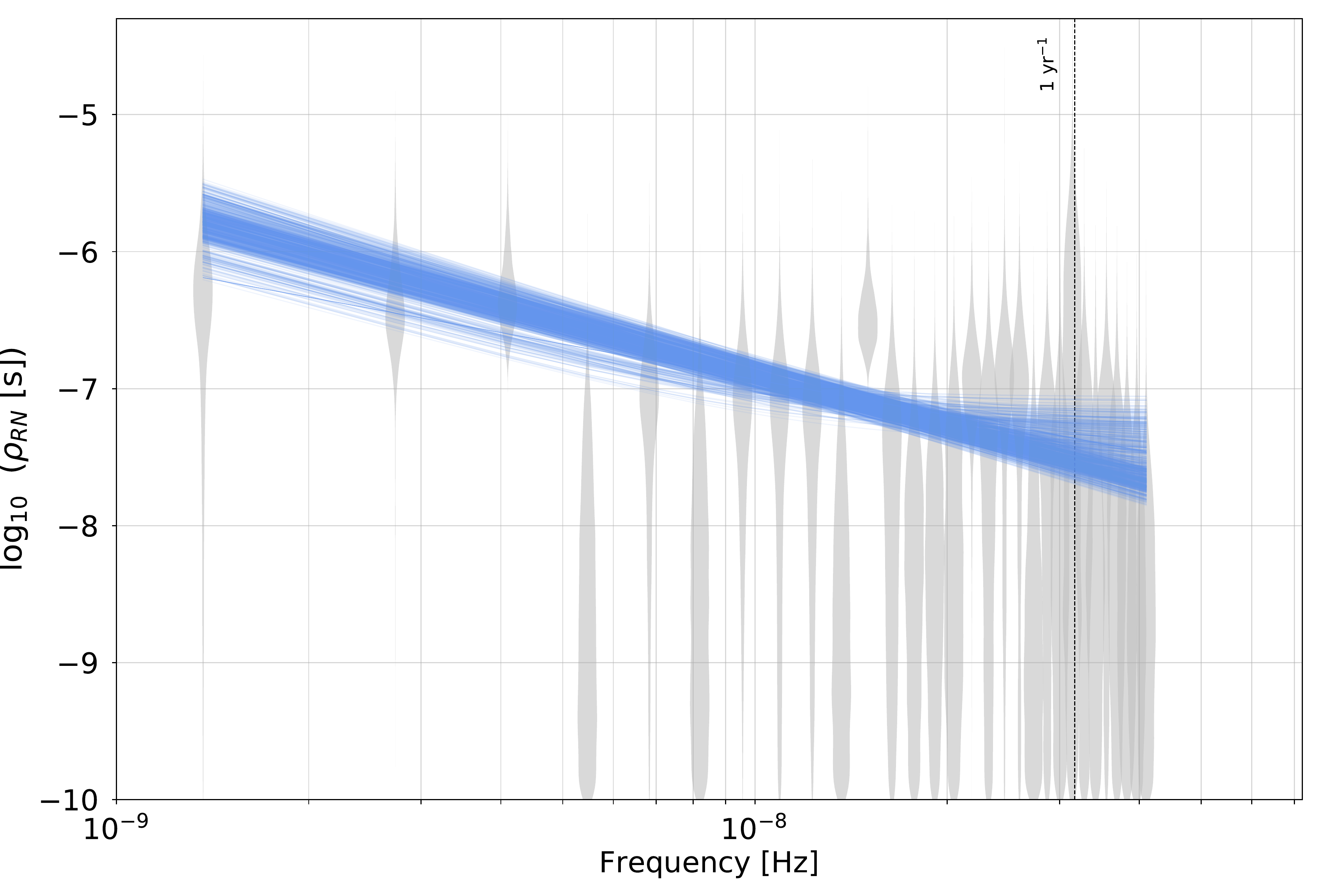}
	\end{subfigure}\\
	
	\begin{subfigure}{.32\textwidth}
		\centering
		\hspace*{-3.5cm}\includegraphics[keepaspectratio=true,scale=0.24]{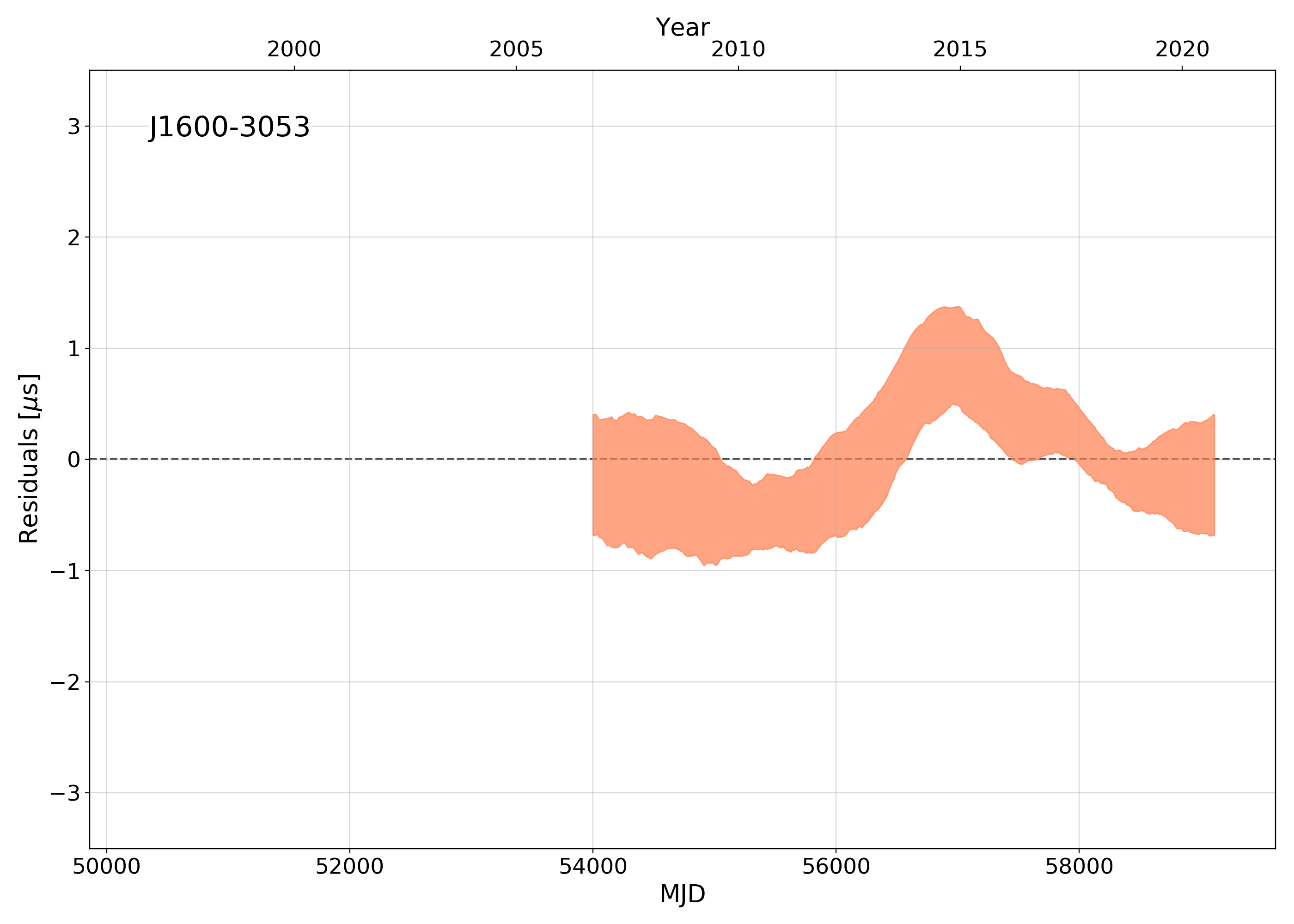}
	\end{subfigure}
	\begin{subfigure}{.32\textwidth}
		\centering
		\vspace*{.4cm}\includegraphics[keepaspectratio=true,scale=0.24]{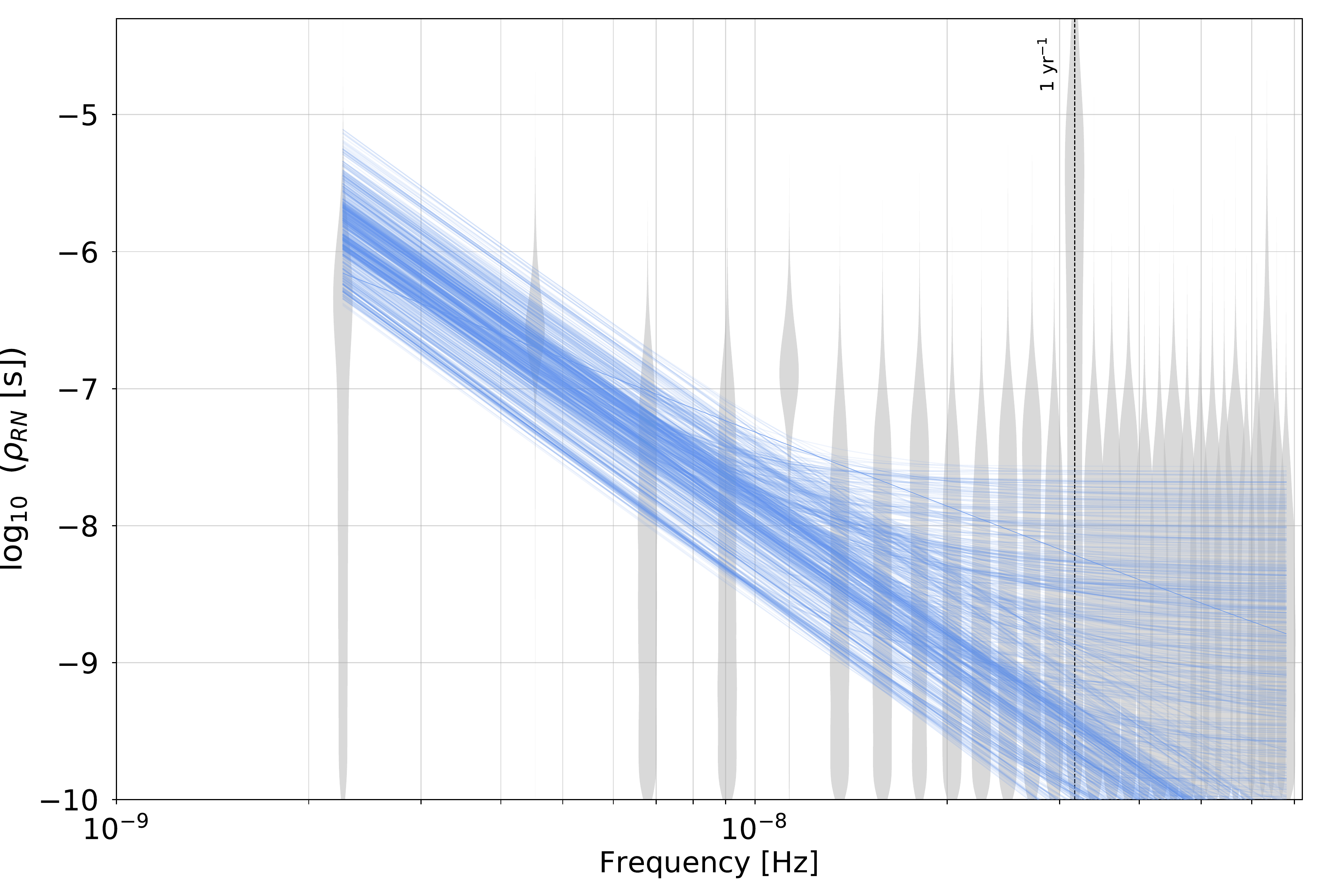}
	\end{subfigure}\\
\end{figure*}

We reconstruct time-domain Gaussian process realizations of this signal modelled with a power-law PSD. The left panels of Fig. \ref{fig:timedom_freespec} display the $68\%$ confidence interval of $100$ random realizations drawn from the posterior distributions of the RN amplitude and spectral index included in two different single-pulsar noise models: the default base model, \textit{RN30\_DMv100} (red), and the "custom" model shown in Table~\ref{tab:finalMS} (grey). The peculiar high frequency red noise is clearly seen in  PSR~J1012+5307 (see Section \ref{ssec:Stoch_det} for a detailed discussion).
The RN in PSR~J1744$-$1134 is considerably reduced in the "custom" model (see discussion at the end of Section \ref{ssec:Stoch_det}). Note that we did not include RN in the custom model for PSR~J1600$-$3053. The RN in the default and custom models are quite similar for PSRs J0613$-$0200, J1713+0747 and J1909$-$3053.

The right panels of Fig. \ref{fig:timedom_freespec} display the spectrum of the achromatic red-noise (using \textit{RN30\_DMv100}) for each pulsar computed with (i) a free-spectrum PSD (grey violins); and (ii) a broken-power law PSD (blue), where we give  $1000$ realizations randomly drawn from the posterior distributions.

\begin{figure*}
	\begin{subfigure}{.32\textwidth}
		\centering
		\hspace*{-3.5cm}\includegraphics[keepaspectratio=true,scale=0.24]{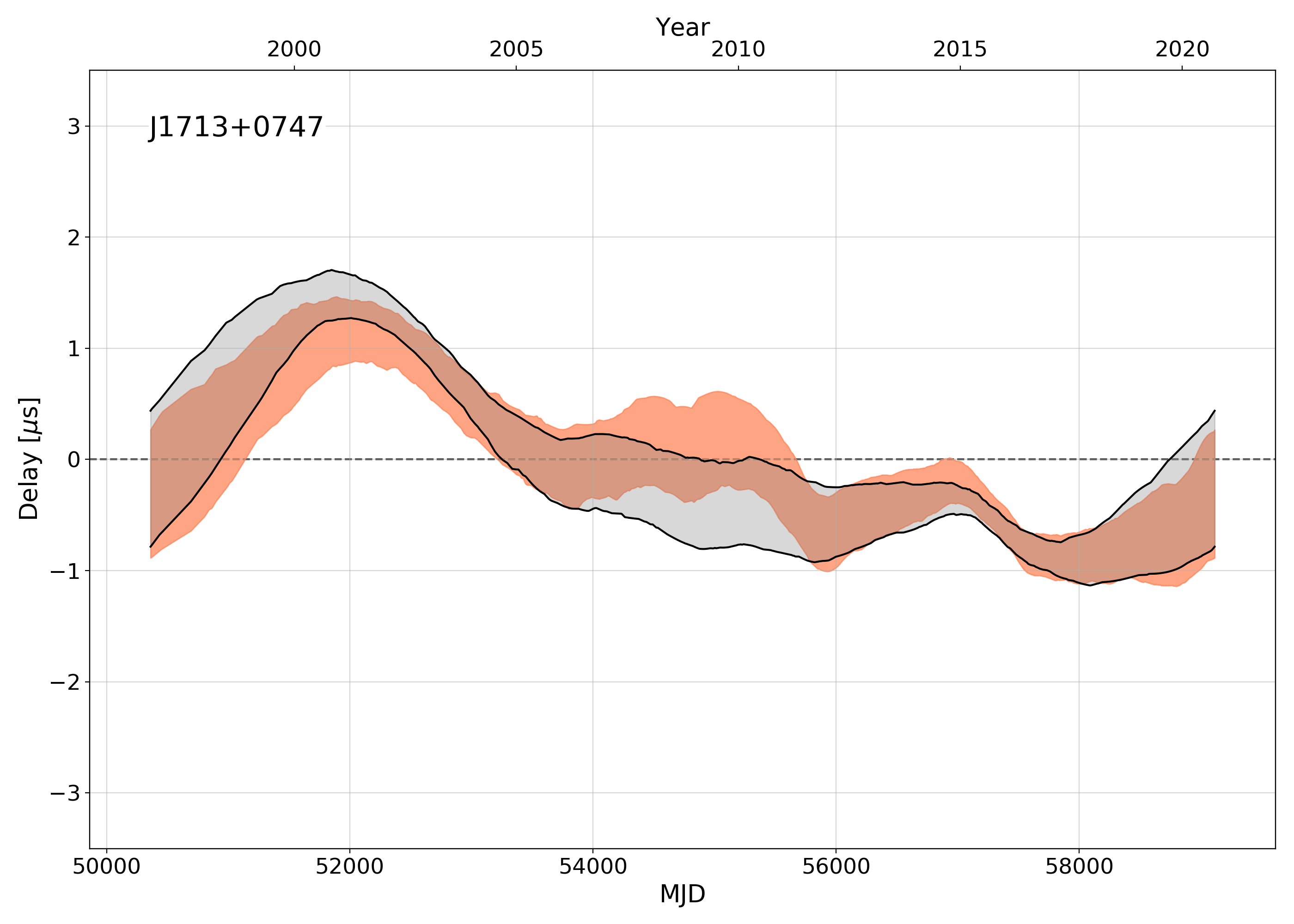}
	\end{subfigure}
	\begin{subfigure}{.32\textwidth}
		\centering
		\vspace*{.4cm}\includegraphics[keepaspectratio=true,scale=0.24]{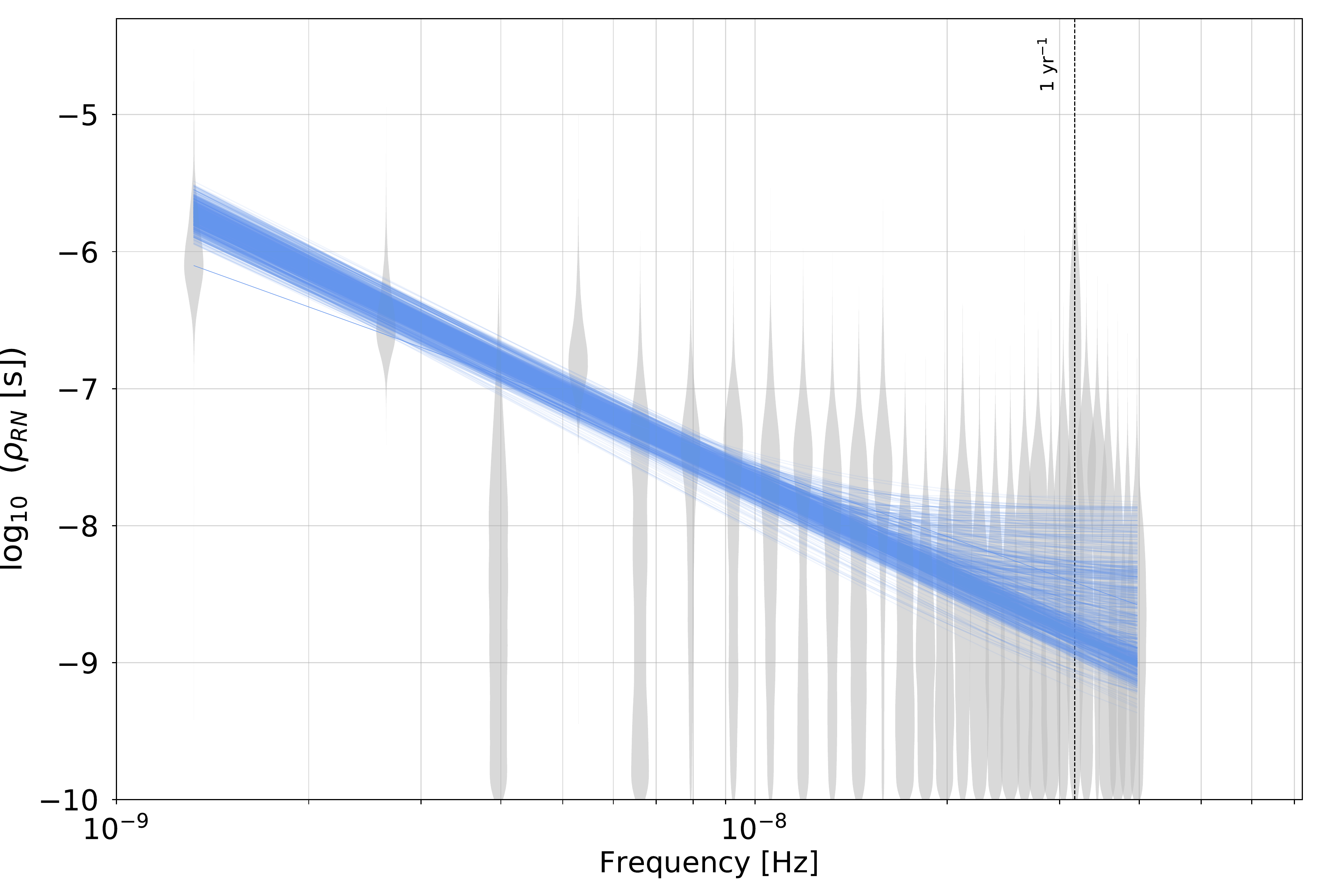}
	\end{subfigure}\\
	
	\begin{subfigure}{.32\textwidth}
		\centering
		\hspace*{-3.5cm}\includegraphics[keepaspectratio=true,scale=0.24]{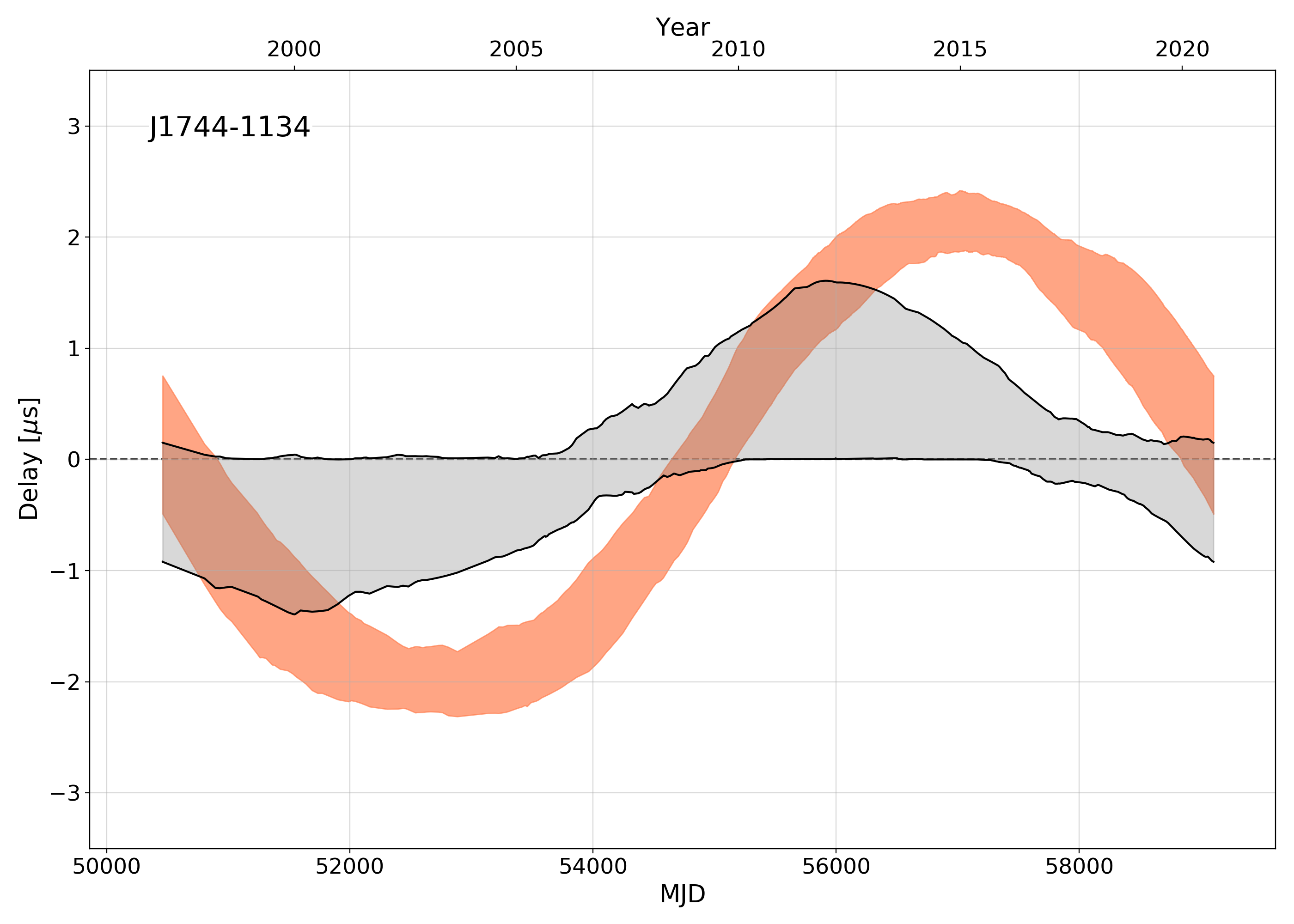}
	\end{subfigure}
	\begin{subfigure}{.32\textwidth}
		\centering
		\vspace*{.4cm}\includegraphics[keepaspectratio=true,scale=0.24]{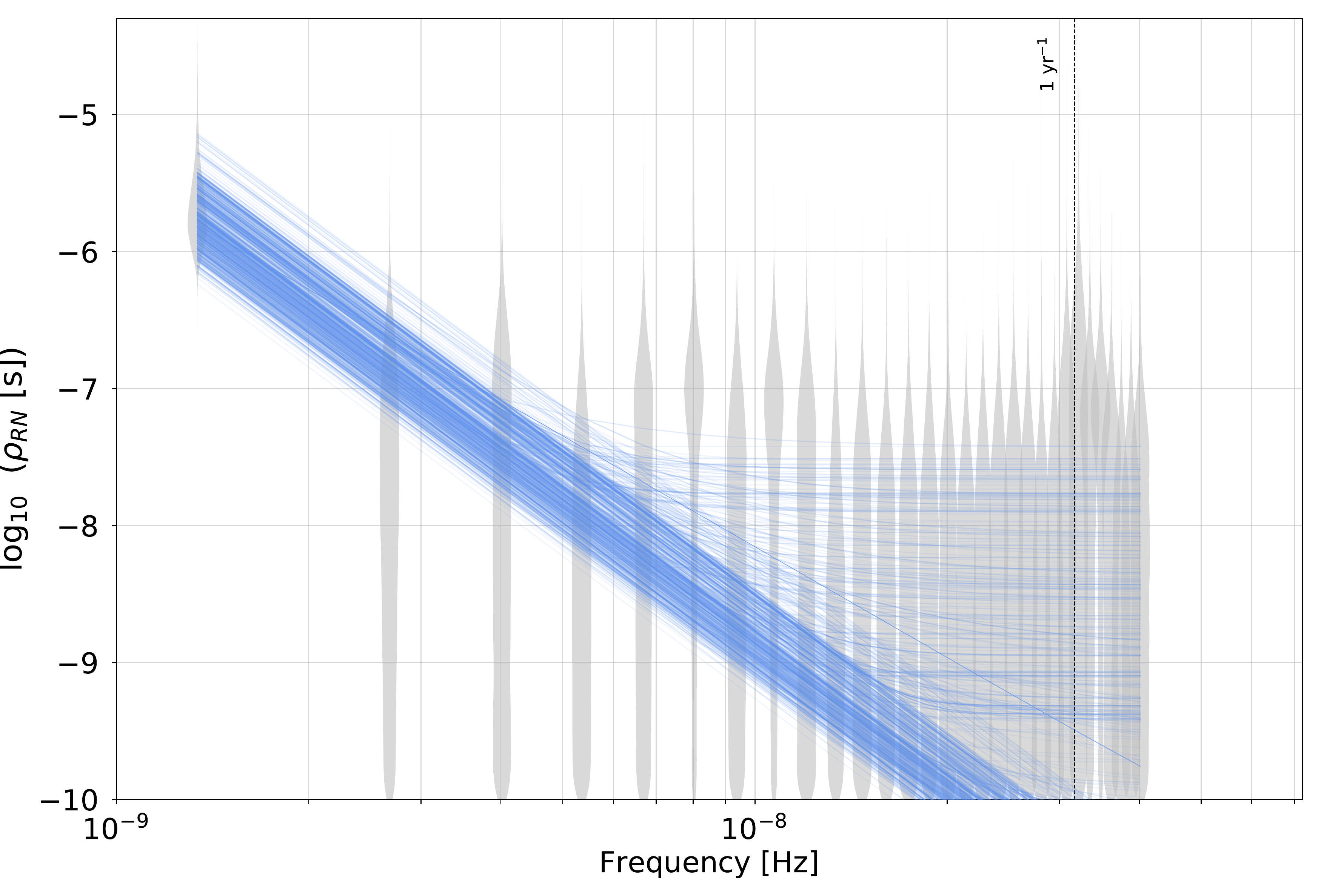}
	\end{subfigure}\\
	
	\begin{subfigure}{.32\textwidth}
		\centering
		\hspace*{-3.5cm}\includegraphics[keepaspectratio=true,scale=0.24]{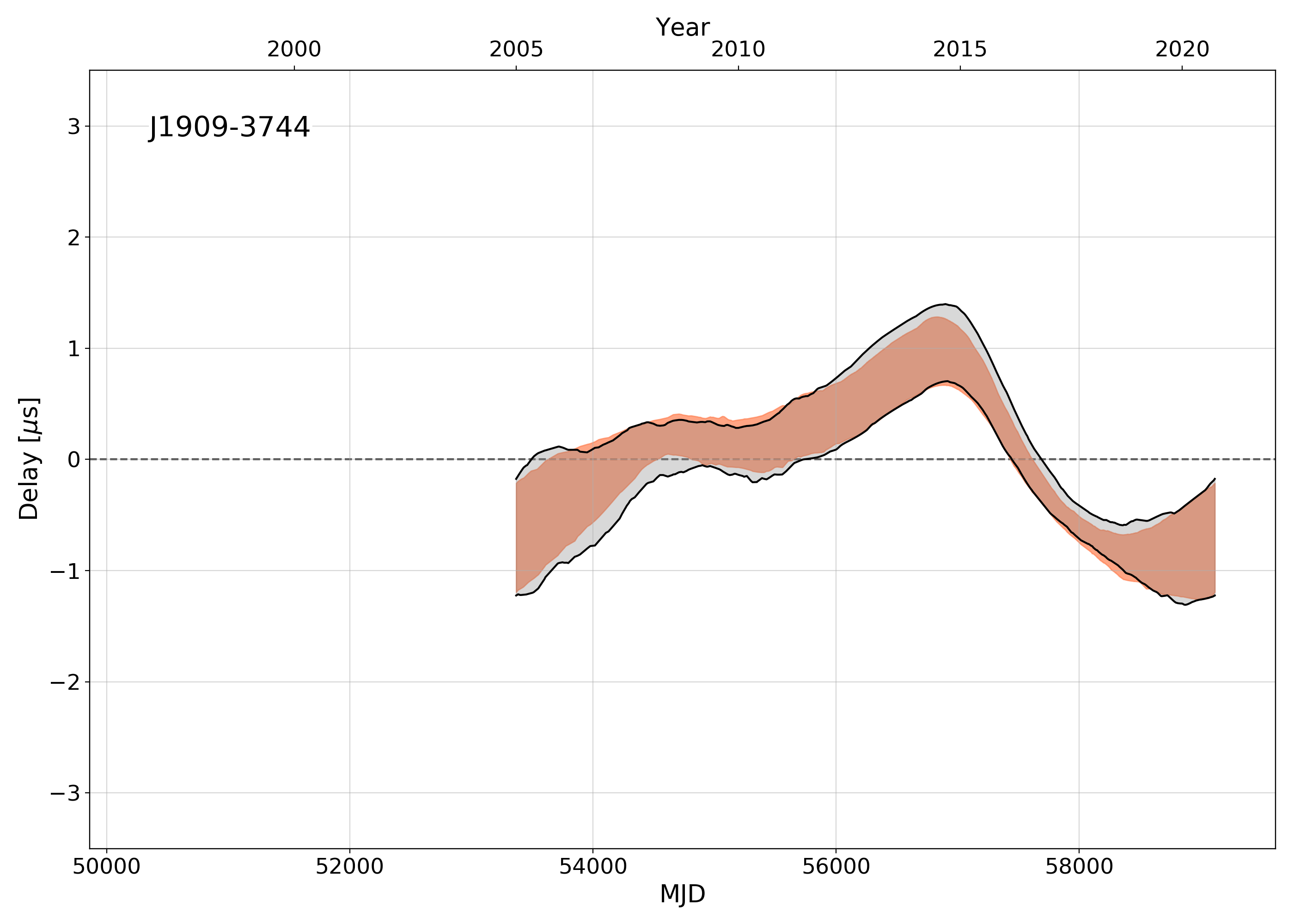}
	\end{subfigure}
	\begin{subfigure}{.32\textwidth}
		\centering
		\vspace*{.4cm}\includegraphics[keepaspectratio=true,scale=0.24]{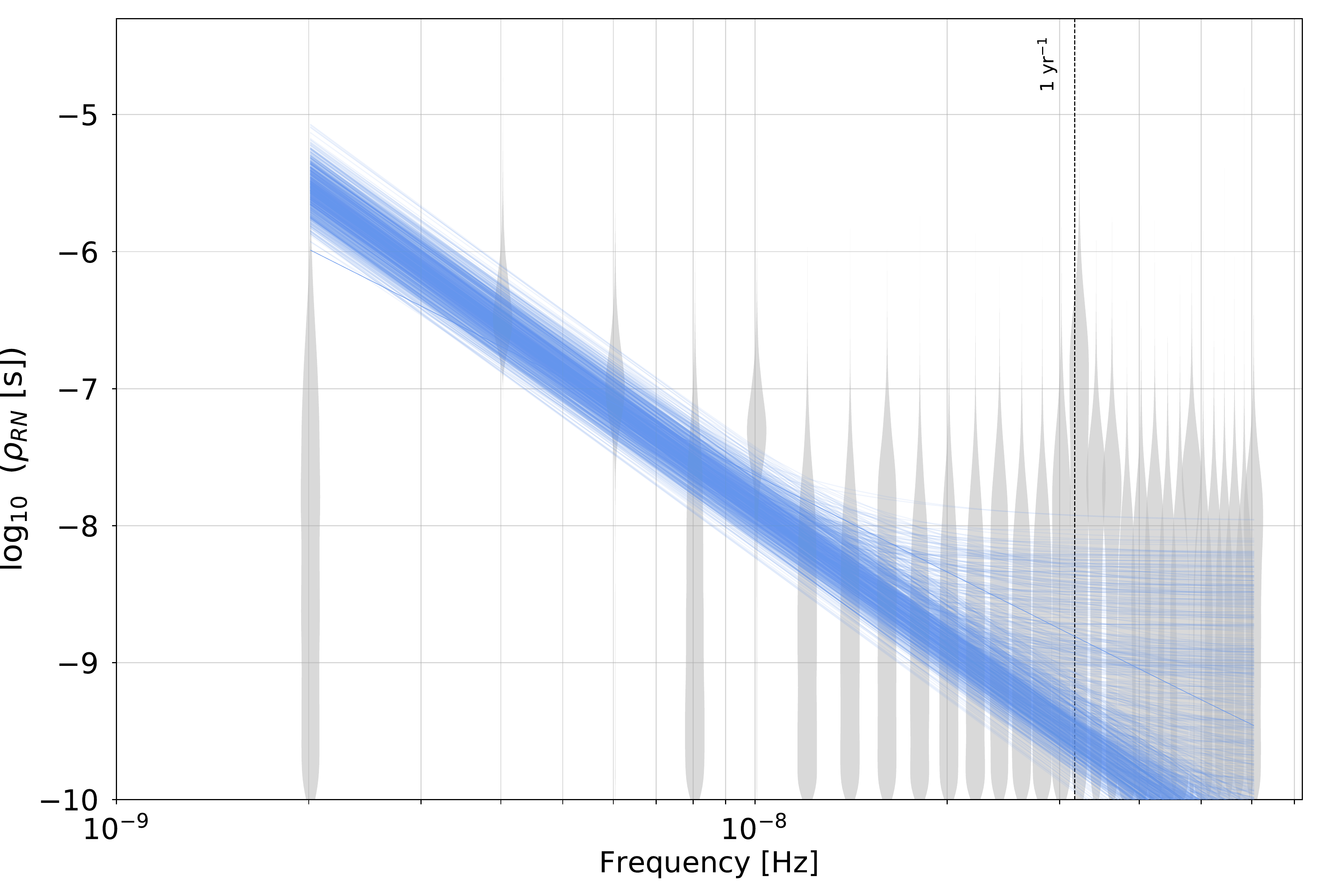}
	\end{subfigure}\\
\end{figure*}


\section{Single-pulsar noise model parameters}

In Table~\ref{tab:NoiseParams} we  give median values for each noise component with 68\% confidence interval in the custom model of each pulsar.

\begin{table*}
	\centering
	\caption{Medians and $16-84\%$ credible intervals of the $1$D marginalized posterior distributions of each single-pulsar noise model parameters for (i) the default base models used in \protect\cite{DR2_Chen+}; and (ii) the final custom models. Note that \textit{DMv} amplitudes are given with \textsc{Enterprise} normalization set for the radio frequency at $1.4$ GHz (see Eq. \ref{eq:RN_powerlaw}).}
	\label{tab:NoiseParams}
	\begin{tabular}{ccccccccc}
		\hline
		Model & Signal & parameter & J0613$-$0200 & J1012+5307 & J1600$-$3053 & J1713+0747 & J1744$-$1134 & J1909$-$3744 \\
		\hline\hline
		\multirow{5.5}{*}{Default} & \multirow{2.5}{*}{\textit{RN}} & $\mathrm{log}_{10} A$ & $-14.72^{+0.56}_{-0.60}$ & $-13.12^{+0.08}_{-0.08}$ & $-14.05^{+0.33}_{-0.53}$ & $-14.13^{+0.18}_{-0.19}$ & $-15.16^{+0.69}_{-0.72}$ & $-14.65^{+0.32}_{-0.37}$ \\
		\cmidrule(l){3-9}
		& & $\gamma$ & $4.76^{+1.09}_{-1.04}$ & $1.66^{+0.32}_{-0.30}$ & $3.50^{+1.22}_{-0.89}$ & $3.29^{+0.54}_{-0.47}$ & $5.19^{+1.10}_{-1.09}$ & $4.65^{+0.96}_{-0.83}$ \\
		\cmidrule(l){2-9}\morecmidrules\cmidrule(l){2-9}
		& \multirow{2.5}{*}{\textit{DMv}} & $\mathrm{log}_{10} A$ & $-13.75^{+0.21}_{-0.26}$ & $-13.36^{+0.05}_{-0.05}$ & $-13.09^{+0.04}_{-0.04}$ & $-13.47^{+0.04}_{-0.04}$ & $-13.33^{+0.05}_{-0.05}$ & $-13.56^{+0.04}_{-0.04}$ \\
		\cmidrule(l){3-9}
		& & $\gamma$ & $2.89^{+0.66}_{-0.59}$ & $1.20^{+0.18}_{-0.16}$ & $2.08^{+0.12}_{-0.11}$ & $1.49^{+0.20}_{-0.19}$ & $1.23^{+0.21}_{-0.21}$ & $1.53^{+0.15}_{-0.14}$ \\
		\hline\hline
		\multirow{36}{*}{Custom} & \multirow{2.5}{*}{\textit{RN}} & $\mathrm{log}_{10} A$ & $-14.82^{+0.64}_{-0.67}$ & $-13.02^{+0.03}_{-0.03}$ & - & $-14.48^{+0.25}_{-0.30}$ & $-15.25^{+1.31}_{-1.39}$ & $-14.46^{+0.36}_{-0.41}$ \\
		\cmidrule(l){3-9}
		& & $\gamma$ & $4.81^{+1.17}_{-1.15}$ & $1.19^{+0.13}_{-0.13}$ & - & $3.95^{+0.65}_{-0.56}$ & $3.67^{+2.17}_{-2.22}$ & $4.24^{+1.03}_{-0.88}$ \\
		\cmidrule(l){2-9}\morecmidrules\cmidrule(l){2-9}
		& \multirow{2.5}{*}{\textit{DMv}} & $\mathrm{log}_{10} A$ & $-13.58^{+0.16}_{-0.21}$ & $-13.66^{+0.12}_{-0.14}$ & $-14.16^{+0.32}_{-0.41}$ & $-13.78^{+0.05}_{-0.06}$ & $-13.45^{+0.07}_{-0.07}$ & $-13.92^{+0.27}_{-1.21}$ \\
		\cmidrule(l){3-9}
		& & $\gamma$ & $2.47^{+0.55}_{-0.47}$ & $2.09^{+0.42}_{-0.39}$ & $4.69^{+0.92}_{-0.76}$ & $1.16^{+0.20}_{-0.21}$ & $0.46^{+0.36}_{-0.30}$ & $2.64^{+2.67}_{-0.93}$ \\
		\cmidrule(l){2-9}\morecmidrules\cmidrule(l){2-9}
		& \multirow{2.5}{*}{\textit{Sv}} & $\mathrm{log}_{10} A$ & - & - & $-13.26^{+0.04}_{-0.04}$ & - & - & $-13.84^{+0.10}_{-0.19}$ \\
		\cmidrule(l){3-9}
		& & $\gamma$ & - & - & $1.48^{+0.14}_{-0.13}$ & - & - & $0.78^{+0.31}_{-0.34}$ \\
		\cmidrule(l){2-9}\morecmidrules\cmidrule(l){2-9}
		& \multirow{4}{*}{\textit{Exp. dip 1}} & $\mathrm{log}_{10} A$ [s] & - & - & - & $-5.54^{+0.05}_{-0.04}$ & - & - \\
		\cmidrule(l){3-9}
		& & $\mathrm{log}_{10} \tau$ [day] & - & - & - & $1.54^{+0.07}_{-0.07}$ & - & - \\
		\cmidrule(l){3-9}
		& & $t_0$ [MJD] & - & - & - & $54752.49^{+3.24}_{-3.17}$ & - & - \\
		\cmidrule(l){2-9}\morecmidrules\cmidrule(l){2-9}
		& \multirow{4}{*}{\textit{Exp. dip 2}} & $\mathrm{log}_{10} A$ [s] & - & - & - & $-5.89^{+0.05}_{-0.05}$ & - & - \\
		\cmidrule(l){3-9}
		& & $\mathrm{log}_{10} \tau$ [day] & - & - & - & $1.51^{+0.09}_{-0.09}$ & - & - \\
		\cmidrule(l){3-9}
		& & $t_0$ [MJD] & - & - & - & $57510.65^{+2.17}_{-2.20}$ & - & - \\
		\cmidrule(l){2-9}\morecmidrules\cmidrule(l){2-9}
		& \multirow{2.5}{*}{\textit{SN\_BON\_2.0}} & $\mathrm{log}_{10} A$ & - & - & - & $-14.78^{+1.03}_{-0.97}$ & - & - \\
		\cmidrule(l){3-9}
		& & $\gamma$ & - & - & - & $4.36^{+1.59}_{-1.80}$ & - & - \\
		\cmidrule(l){2-9}\morecmidrules\cmidrule(l){2-9}
		& \multirow{2.5}{*}{\textit{SN\_JBO\_1.5}} & $\mathrm{log}_{10} A$ & - & - & - & $-13.10^{+0.21}_{-0.22}$ & - & - \\
		\cmidrule(l){3-9}
		& & $\gamma$ & - & - & - & $1.47^{+0.68}_{-0.65}$ & - & - \\
		\cmidrule(l){2-9}\morecmidrules\cmidrule(l){2-9}
		& \multirow{2.5}{*}{\textit{SN\_LEAP\_1.4}} & $\mathrm{log}_{10} A$ & - & - & $-14.97^{+0.94}_{-0.66}$ & $-13.42^{+0.20}_{-0.21}$ & - & - \\
		\cmidrule(l){3-9}
		& & $\gamma$ & - & - & $5.16^{+1.19}_{-1.75}$ & $1.70^{+0.60}_{-0.59}$ & - & - \\
		\cmidrule(l){2-9}\morecmidrules\cmidrule(l){2-9}
		& \multirow{2.5}{*}{\textit{DMv-SN\_NUP\_1.4}} & $\mathrm{log}_{10} A$ & $-13.52^{+0.29}_{-0.33}$ & $-14.10^{+0.72}_{-0.91}$ & - & $-14.05^{+0.63}_{-0.64}$ & $-14.65^{+0.99}_{-1.03}$ & - \\
		\cmidrule(l){3-9}
		& & $\gamma$ & $2.88^{+0.95}_{-0.85}$ & $3.17^{+1.60}_{-1.31}$ & - & $2.75^{+1.20}_{-1.14}$ & $4.23^{+1.68}_{-1.75}$ & - \\
		\cmidrule(l){2-9}\morecmidrules\cmidrule(l){2-9}
		& \multirow{2.5}{*}{\textit{SN\_NUP\_2.5}} & $\mathrm{log}_{10} A$ & - & $-13.22^{+0.73}_{-1.03}$ & - & - & - & - \\
		\cmidrule(l){3-9}
		& & $\gamma$ & - & $2.50^{+1.84}_{-1.44}$ & - & - & - & - \\
		\cmidrule(l){2-9}\morecmidrules\cmidrule(l){2-9}
		& \multirow{2.5}{*}{\textit{BN\_Band.2}} & $\mathrm{log}_{10} A$ & - & - & - & - & $-13.84^{+0.35}_{-0.42}$ & - \\
		\cmidrule(l){3-9}
		& & $\gamma$ & - & - & - & - & $3.02^{+0.86}_{-0.77}$ & - \\
		\cmidrule(l){2-9}\morecmidrules\cmidrule(l){2-9}
		& \multirow{2.5}{*}{\textit{BN\_Band.3}} &$\mathrm{log}_{10} A$ & - & - & - & $-14.32^{+0.37}_{-0.48}$ & - & - \\
		\cmidrule(l){3-9}
		& & $\gamma$ & - & - & - & $2.68^{+1.02}_{-0.87}$ & - & - \\
		\hline
	\end{tabular}
\end{table*}


\bsp	
\label{lastpage}
\end{document}